\documentclass[3p]{elsarticle}

\usepackage[utf8]{inputenc}
\usepackage{hyperref}
\usepackage{theorem}
\usepackage{amsmath}
\usepackage{amssymb}

\newcommand\comment[1]{}

\newcommand\hsp[1][3ex]{\hspace*{#1}}

\newcommand\vide{\emptyset}
\newcommand\eg{{\em e.g.} }
\newcommand\ie{{\em i.e.} }

\newcommand\ala{{\em \`a la} }
\newcommand\furl[1]{\footnote{\url{http://#1}}}

\newcommand\glb{\mr{glb}}

\newcommand\dom{\mr{dom}}

\newcommand\FV{\mr{FV}}
\newcommand\Pos{\mr{Pos}}
\newcommand\lex{\mr{lex}}
\newcommand\mul{\mr{mul}}
\renewcommand\prod{\mr{prod}}
\newcommand\stat{\mr{stat}}

\renewcommand\a{\rightarrow}
\newcommand\A{\Rightarrow}
\renewcommand\aa{\leftrightarrow}

\newcommand\la{\leftarrow}

\newcommand\ad{\downarrow}

\renewcommand\to{\mapsto}

\newcommand\ab{\a_\b}

\renewcommand\ae{\a_\eta}
\newcommand\abe{\a_{\b\eta}}

\newcommand\ar{\a_\cR}

\newcommand\I[1]{[\![#1]\!]}

\newcommand\N[1]{\|#1\|_\infty}

\newcommand\ex{\exists}
\newcommand\all{\forall}
\newcommand\ou{\vee}

\newcommand\et{\wedge}

\newcommand\sle{\subseteq}
\newcommand\sge{\supseteq}

\newcommand\tle{\unlhd}
\newcommand\tge{\unrhd}
\newcommand\tlt{\lhd}
\newcommand\tgt{\rhd}

\newcommand\h[1]{{\widehat{#1}}}
\renewcommand\o[1]{{\overline{#1}}}

\newcommand\al{\alpha}
\renewcommand\b{\beta}

\newcommand\vep{\varepsilon}

\renewcommand\t{\theta}

\renewcommand\l{\lambda}
\renewcommand\L{\Lambda}
\renewcommand\r{\rho}
\newcommand\s{\sigma}
\renewcommand\S{\Sigma}

\newcommand\vphi{\varphi}
\newcommand\w{\omega}

\newcommand\mc{\mathcal}

\newcommand\mr{\mathrm}
\newcommand\mb{\mathbb}
\newcommand\mg{\mathbf}
\newcommand\mk{\mathfrak}
\newcommand\ms{\mathsf}

\newcommand\bM{\mb{M}}
\newcommand\bN{\mb{N}}

\newcommand\cB{\mc{B}}

\newcommand\cD{\mc{D}}
\newcommand\cE{\mc{E}}
\newcommand\cF{\mc{F}}

\newcommand\cL{\mc{L}}
\newcommand\cM{\mc{M}}

\newcommand\cP{\mc{P}}

\newcommand\cR{\mc{R}}
\newcommand\cS{\mc{S}}
\newcommand\cT{\mc{T}}

\newcommand\cX{\mc{X}}

\newcommand\ka{\mk{a}}

\newcommand\kl{\mk{l}}

\newcommand\sa{\ms{a}}

\renewcommand\sc{\ms{c}}
\newcommand\sd{\ms{d}}

\renewcommand\sf{\ms{f}}
\newcommand\sg{\ms{g}}
\newcommand\sh{\ms{h}}
\newcommand\si{\ms{i}}

\newcommand\sk{\ms{k}}

\newcommand\so{\ms{o}}

\let\latexss\ss
\renewcommand\ss{\ms{s}}

\newcommand\sz{\ms{z}}

\newcommand\sA{\ms{A}}
\newcommand\sB{\ms{B}}
\newcommand\sC{\ms{C}}
\newcommand\sD{\ms{D}}

\newcommand\sF{\ms{F}}

\newcommand\sK{\ms{K}}
\newcommand\sL{\ms{L}}

\newcommand\sN{\ms{N}}
\newcommand\sO{\ms{O}}

\newcommand\sR{\ms{R}}

\newcommand\sT{\ms{T}}

\newcommand\va{{\vec{a}}}
\newcommand\vb{{\vec{b}}}

\newcommand\vk{{\vec{k}}}
\newcommand\vl{{\vec{l}}}
\newcommand\vm{{\vec{m}}}

\newcommand\vp{{\vec{p}}}
\newcommand\vq{{\vec{q}}}

\newcommand\vs{{\vec{s}}}
\newcommand\vt{{\vec{t}}}
\newcommand\vu{{\vec{u}}}
\newcommand\vv{{\vec{v}}}
\newcommand\vw{{\vec{w}}}
\newcommand\vx{{\vec{x}}}
\newcommand\vy{{\vec{y}}}

\newcommand\vM{{\vec{M}}}

\newcommand\vT{{\vec{T}}}
\newcommand\vU{{\vec{U}}}
\newcommand\vV{{\vec{V}}}

\newenvironment{rul}
  {$\begin{array}{rcl}}
  {\end{array}$}
\newenvironment{rulc}
  {\begin{center}\begin{rul}}
  {\end{rul}\end{center}}

\newenvironment{rew}[1][~~\a~~]
  {$\begin{array}{r@{#1}l}}
  {\end{array}$}
\newenvironment{rewc}[1][~~\a~~]
  {\begin{center}\begin{rew}[#1]}
  {\end{rew}\end{center}}

\comment{
\newcounter{counter}

{\theorembodyfont{\rmfamily} % theorem
  \newtheorem{dfn}[counter]{Definition}
  \newtheorem{lem}[counter]{Lemma}
  \newtheorem{thm}[counter]{Theorem}
  \newtheorem{cor}[counter]{Corollary}

  \newtheorem{expl}[explnum]{Example}
}}

\newcommand\cqfd{\hfill$\blacksquare$}

\newenvironment{lstgeneric}[2]
  {\begin{list}{#1}{\topsep=.5mm\itemsep=.5mm\parsep=0mm%
    \itemindent=-3ex\labelsep=1ex\labelwidth=0ex #2}}
  {\end{list}}

\newenvironment{enumii}[1]
  {\begin{lstgeneric}{}{\usecounter{enumii}%
    }}
  {\end{lstgeneric}}

\newtheorem{thm}{Theorem}
\newtheorem{lem}{Lemma}
\newtheorem{cor}{Corollary}
\newdefinition{dfn}{Definition}
\newdefinition{expl}{Example}
\newproof{prf}{Proof}
\comment{
\textwidth=15.92cm
\oddsidemargin=0cm
\evensidemargin=0cm
\newcounter{dfncnt}
\newcounter{lemcnt}
\newcounter{thmcnt}
\newcounter{corcnt}

{\theorembodyfont{\rmfamily} % theorem
  \newtheorem{dfn}[dfncnt]{Definition}
  \newtheorem{lem}[lemcnt]{Lemma}
  \newtheorem{thm}[thmcnt]{Theorem}
  \newtheorem{cor}[corcnt]{Corollary}
  }
\newenvironment{keyword}{{\bf Keywords:}}{}
\newcommand\sep{, }
\newenvironment{prf}{{\bf Proof.}}{}
\newcommand\qed\cqfd
}

\newcommand\CC{\mr{CC}}

\newcommand\Acc{\mr{Acc}}
\newcommand\card{\mr{card}}

\newcommand\Aliens{\mr{Aliens}}
\newcommand\rk{\mr{rk}}

\newcommand\SN{\mr{SN}}
\newcommand\BV{\mr{BV}}

\newcommand\rec{\ms{rec}}
\renewcommand\sup{\ms{sup}}
\newcommand\suc{\ms{suc}}
\renewcommand\lim{\ms{lim}}
\newcommand\zero{\ms{zero}}
\newcommand\nil{\ms{nil}}
\newcommand\cons{\ms{cons}}

\newcommand\sub{\ms{sub}}
\newcommand\ack{\ms{ack}}
\newcommand\arrow{\ms{arrow}}
\newcommand\pred{\ms{pred}}
\newcommand\height{\ms{height}}
\newcommand\node{\ms{node}}
\newcommand\leaf{\ms{leaf}}
\newcommand\fex{\ms{ex}}
\renewcommand\max{\ms{max}}
\renewcommand\sin{\ms{sin}}
\renewcommand\cos{\ms{cos}}
\newcommand\val{\ms{val}}

\newcommand\Cand{\mg{Cand}}
\newcommand\Red{\mg{Red}}
\newcommand\Sat{\mg{Sat}}
\newcommand\SatInd{\mg{SatInd}}
\newcommand\Bi{\mg{Bi}}

\newcommand\af{{\al_\sf}}
\newcommand\ag{{\al_\sg}}

\newcommand\mset[1]{\{\!|#1|\!\}}
\newcommand\at[2][2.5]{\stackrel{\hsp[-#1mm]#2}}
\newcommand\be{{\b\eta}}

\begin{document}

\title{Termination of rewrite relations on $\l$-terms\\based on Girard's notion of reducibility}

% with elsarticle style
%\comment{
\author[fb]{Frédéric Blanqui\fnref{fn1}}
\address[fb]{Institut National de Recherche en Informatique et Automatique (INRIA), France}
\fntext[fn1]{Hosted from July 2012 to August 2013 by the Institute of Software of the Chinese Academy of Sciences, Beijing, China.}
%}

% with article style
\comment{
\author{Frédéric Blanqui\thanks{Hosted from July 2012 to August 2013 by the Institute of Software of the Chinese Academy of Sciences, Beijing, China.}\\[2mm]
INRIA, France}
\date{29 August 2015}
\maketitle
}

\begin{abstract}
  In this paper, we show how to extend the notion of reducibility
  introduced by Girard for proving the termination of $\b$-reduction
  in the polymorphic $\l$-calculus, to prove the termination of
  various kinds of rewrite relations on $\l$-terms, including
  rewriting modulo some equational theory and rewriting with matching
  modulo $\b\eta$, by using the notion of {\em computability
    closure}. This provides a powerful termination criterion for
  various higher-order rewriting frameworks, including Klop's
  Combinatory Reductions Systems with simple types and Nipkow's
  Higher-order Rewrite Systems.
\end{abstract}


\maketitle % with elsarticle style

\section{Introduction}
\label{sec-intro}

This paper addresses the problem of checking the termination of
various kinds of rewrite relations on simply typed $\l$-terms.

\medskip

First-order rewriting \cite{knuth67,dershowitz90chapter} and
$\l$-calculus \cite{church40jsl,barendregt84book} are two general
(Turing-complete) computational frameworks with different strengths
and limitations.

The $\l$-calculus is a language for expressing arbitrary functions
based on a few primitives (abstraction over some variable and
application of a function to an argument). Computation is done by
repeatedly substituting formal arguments by actual ones
($\b$-reduction) \cite{church40jsl}.

In first-order rewriting, one considers a fixed set of function
symbols and a fixed set of term transformation rules. Computation is
done by repeatedly substituting the left-hand side of a rule by the
corresponding right-hand side \cite{knuth67}.

Hence, in $\l$-calculus, there is only one computation rule and it is
unconditional while, in rewriting, a computation step occurs only if a
term {\em matches} a pattern (possibly modulo some equational theory).

But first-order rewriting cannot express in a simple way anonymous
functions or patterns with bound variables. See for instance the works
on Combinatory Logic \cite{curry58book}, first-order definitions of a
substitution operation compatible with $\al$-equivalence
\cite{debruijn78tr,abadi91jfp,kesner07csl} (to cite just a few, for the
amount of publications on this subject is very important), or
first-order encodings of higher-order rewriting \cite{bonelli05jlc}.

Rewriting on $\l$-terms, or higher-order rewriting, aims at unifying
these two languages. Several approaches exist like Klop's Combinatory
Reduction Systems (CRSs) \cite{klop80phd,klop93tcs}, Khasidashvili's
Expression Reduction Systems (ERSs)
\cite{khasidashvili90tr,glauert05lncs}, Nipkow's Higher-order Rewrite
Systems (HRSs) \cite{nipkow91lics,mayr98tcs}, or Jouannaud and Okada's
higher-order algebraic specification languages (HALs)
\cite{jouannaud91lics,jouannaud97tcs}. Van Oostrom and van Raamsdonk
studied the relations between CRSs and HRSs \cite{oostrom93hoa} and
developed a general framework (HORSs) that subsumes most of the
previous approaches \cite{oostrom94phd,raamsdonk96phd}.

In another direction, some researchers introduced calculi where
patterns are first-class citizens: van Oostrom's pattern calculus
\cite{oostrom90tr,klop08tcs}, Cirstea and Kirchner's $\r$-calculus
\cite{cirstea01jigpal1,cirstea01jigpal2}, Jay and Kesner's pattern calculus
\cite{jay04toplas,jay09jfp}, or some extensions of ML or Haskell
\cite{erwig96ifl,tullsen00padl}.

In this paper, I will consider HALs with curried symbols (\ie all
symbols are of arity $0$), that is, arbitrary simply typed $\l$-terms
with curried symbols defined by the combination of rewrite rules and
$\b$-reduction. But, as we will see in Section \ref{sec-hrs}, our
results easily apply to HRSs and simply typed CRSs as well.

My goal is to develop techniques for proving the termination of such a
system, \ie the combination of $\b$-reduction and arbitrary
user-defined rewrite rules.

\medskip

For proving the termination of rewrite relations on $\l$-terms, one
can try to extend to $\l$-calculus techniques developed for
first-order rewriting (\eg
\cite{loria92ctrs,vandepol96phd,sakai01ieice,jones04rta,fuhs12rta})
or, vice versa, adapt to rewriting techniques developed for
$\l$-calculus (\eg
\cite{jouannaud91lics,blanqui04rta,blanqui06lpar-sbt}).

Since $\b$-reduction does not terminate in general, one usually
restricts his attention to some strict subset of the set of all
$\l$-terms, like the set of $\l$-terms typable in some type system
\cite{barendregt92chapter} (types were first introduced by logicians as
an alternative to the restriction of the comprehension axiom in set
theory, and later found important applications in programming
languages and compilers).

To prove the termination of $\b$-reduction in typed $\l$-calculi,
there are essentially three techniques:

\begin{description}
\item[Direct proof.] In the simply-typed $\l$-calculus, it is possible
  to prove the termination of $\b$-reduction by induction on the size
  of the type of the substituted variable
  \cite{sanchis67ndj,vandaalen80phd}. For instance, in the reduction
  sequence $(\l x^{A\A B}xy)(\l y^Az)\ab (\l y^Az)y\ab z$, the type in
  the first reduction step of the substituted variable $x$ is $A\A B$
  while, in the second reduction step (which is generated by the first
  one), the type of the substituted variable $y$ is $A$.

But this technique extends neither to polymorphic types nor to
rewriting since, in both cases, the type of the substituted variables
may increase:

\begin{itemize}
\item
With polymorphic types, consider the reduction sequence $(\l
x^{(\all\al)\al\A B}xYy)(\L\al\l y^\al z)\ab (\L\al\l y^\al z)Yy$ $\ab
(\l y^Y z)y\ab z$. In the first reduction step, the type of the
substituted variable $x$ is $(\all\al)\al\A B$ while, in the last
reduction step, the type of the substituted variable $y$ is the
arbitrary type $Y$.
\item
With the rule $\sK~x~\sa\ar x$ where $\sK:T\a\sA\a T$, consider the
reduction sequence $(\l z\sK xz)\sa\ab \sK x\sa\ar x$. In the first
reduction step, the type of the substituted variable $z$ is $\sA$ while,
in the second reduction step which is generated by the first one, the
type of the substituted variable $x$ is the arbitrary type $T$.
\end{itemize}

\item[Interpretation.]
For the simply-typed $\l$-calculus again, Gandy showed that $\l
I$-terms ($\l$-terms where, in every subterm $\l xt$, $x$ has at least one free occurrence in $t$), can be interpreted by hereditarily monotone functionals
on $\bN$ \cite{gandy80chapter}. Then, van de Pol showed that there is a
transformation from $\l$-terms to $\l I$-terms that strictly decreases
when there is a $\b$-reduction, and extended this to higher-order
rewriting and other domains than $\bN$ \cite{vandepol96phd}. Finally,
Hamana developed a categorical semantics for terms with bound
variables \cite{hamana06hosc} based on the work of Fiore, Plotkin and
Turi \cite{fiore99lics}, that is complete for termination (which is
not the case of van de Pol's interpretations), and extended to
higher-order terms the technique of semantic labeling
\cite{hamana07ppdp} introduced for first-order terms by Zantema
\cite{zantema95fi}. However, Roux showed that its application to
$\b$-reduction itself is not immediate since the interpretation of
$\b$-reduction is not $\b$-reduction \cite{blanqui09csl,roux11phd}.

\item[Computability.]  The last technique, not limited to simply-typed
  $\l$-calculus, is based on Tait and Girard's notions of
  computability\footnote{In fact, Tait speaks of ``convertibility''
    and Girard of ``reducibility''. To the best of my knowledge, the
    expression ``computability'' is due to Troelstra
    \cite{troelstra73chapter} although Troelstra himself invokes
    Tait. This notion of computability has to be distinguished from
    the one of Turing and Church \cite{turing37jsl}. However, given a
    Tait-computable $\l$-term $t:U\A V$, the function that maps every
    Tait-computable $\l$-term $u:U$ to the normal form of $tu$ is
    indeed Turing-computable.} introduced by Tait \cite{tait67jsl} for
  the weak normalization of the simply-typed $\l$-calculus, and
  extended by Girard to polymorphic types \cite{girard71sls} and
  strong normalization \cite{girard72phd}.
\end{description}

There are however relations between these techniques. For instance,
van de Pol proved that his interpretations on $\bN$ can be obtained
from a computability proof by adding information on the length of
reductions \cite{vandepol96phd}. Conversely, the author and Roux
proved that size-based termination
\cite{gimenez98icalp,abel04ita,barthe04mscs,blanqui04rta}, which is a
refinement of computability, can to some extent be seen as an instance
of Hamana's higher-order semantic labeling technique
\cite{blanqui09csl}.

\medskip

In this paper, we will consider a technique based on computability.

Computability has been first used for proving the termination of the
combination of $\b$-reduction, in the simply typed or polymorphic
$\l$-calculus, together with a first-order rewrite system that is
terminating on first-order terms, by Tannen and Gallier
\cite{tannen89icalp,tannen91tcs} and Okada \cite{okada89issac}
independently. It was noticed later by Dougherty that, with
first-order rewriting, a proof can be given that is independent of the
proof of termination of $\b$-reduction
\cite{dougherty91rta,dougherty92ic}, because first-order rewriting
cannot create $\b$-redexes (but just duplicate them). But this does
not extend to higher-order rewriting or to function symbols with
polymorphic types.

In \cite{jouannaud91lics,jouannaud97tcs}, Jouannaud and Okada extended
computability to higher-order rewrite rules following a schema
extending G\"odel' system T recursion schema on Peano integers
\cite{godel58} to arbitrary first-order data types. This work was then
extended to Coquand and Huet's Calculus of Constructions
\cite{coquand84,coquand88ic} in a series of papers culminating in
\cite{barbanera97jfp}.

In \cite{jouannaud97draft}, Jouannaud and Okada reformulated this
general schema as an inductively defined set called {\em computability
  closure}. This notion was then extended with the author to strictly
positive inductive types \cite{blanqui99rta,blanqui02tcs} and to the
Calculus of Algebraic Constructions, that is an extension of the
Calculus of Constructions where types equivalent modulo user-defined
rewrite rules are identified and function symbols can be given
polymorphic and dependent types \cite{blanqui05mscs}.

In this paper, we provide a new presentation of the notion of
computability closure for standard rewriting and show how to extend it
for dealing with rewriting modulo some equational theory and
higher-order pattern-matching, by providing detailed proofs of results
sketched in \cite{blanqui03rta,blanqui07jpbirthday}. We do it in a
progressive way by showing, step by step, how the notion of
computability closure can be extended to cope with new term
constructions or new rewriting mechanisms. To avoid unnecessary
technicalities related to the type discipline, we do it in the simply
typed $\l$-calculus but this work could be conducted in the Calculus
of Algebraic Constructions as well, following the lines of
\cite{blanqui05mscs}.

\medskip

The paper is organized as follows. In Section \ref{sec-term}, we
define the set of terms that will be considered, introduce our notations
and recall some general results on well-founded relations. In Section
\ref{sec-comp}, we present the different definitions of computability
introduced so far and discuss their relations and applicability to
rewriting. In Section \ref{sec-core}, we show how Girard's definition
of computability can be extended to deal with rewriting with matching
modulo $\al$-equivalence by introducing the notion of computability
closure, and provide a first core definition of such a computability
closure. Then follows a number of subsections and sections showing how
to extend this core definition to deal with new constructions or more
general notions of rewriting: abstraction and bound variables, basic
subterms, recursive functions, higher-order subterms, matching on
defined symbols, rewriting modulo an equational theory and rewriting
with matching modulo $\b\eta$. We finally explain why our results
apply to HRSs and simply typed CRSs as well.

Parts of this work have already been formalized in the Coq proof
assistant \cite{blanqui13coq-cc}. See the conclusion for more details
about that.

\section{Definitions and notations}
\label{sec-term}

We first recall some definitions and notations about simply-typed
$\l$-terms, rewriting and well-founded relations. See for instance
\cite{dershowitz90chapter,barendregt92chapter,terese03book} for more
details.

%%%%%%%%%%%%%%%%%%%%%%%%%%%%%%%%%%%%%%%%%%%%%%%%%%%%%%%%%%%%%%%%%%%%%%%%%%%%%%
\subsection{Notations for sequences}
\label{sec-words}

Given a set $A$, let $A^*$ be the free monoid generated from $A$, \ie
the set of finite sequences of elements of $A$ or {\em words} on
$A$. We denote the empty word by $\vep$, word concatenation by
juxtaposition, and the length of a word $w$ by $|w|$. We often denote
a word $a_1\ldots a_n$ by $\va$. A word $p$ is a {\em prefix} of a
word $q$, written $p\le q$, if there is $r$ such that $q=pr$. The
prefix relation is a partial ordering. We write $p\#q$ if $p$ and $q$
are not comparable or {\em disjoint}.

%%%%%%%%%%%%%%%%%%%%%%%%%%%%%%%%%%%%%%%%%%%%%%%%%%%%%%%%%%%%%%%%%%%%%%%%%%%%%%
\subsection{Simple types}
\label{sec-types}

We assume given a set $\cB$ of {\em type constants}. As usual, the set
$\cT$ of (simple) {\em types} is defined recursively as follows
\cite{church40jsl}:

\begin{itemize}
\item
a type constant $B\in\cB$ is a type;
\item
if $T$ and $U$ are types, then $T\A U$ is a type.
\end{itemize}

%%%%%%%%%%%%%%%%%%%%%%%%%%%%%%%%%%%%%%%%%%%%%%%%%%%%%%%%%%%%%%%%%%%%%%%%%%%%%%
\subsection{Terms}
\label{sec-terms}

All over the paper, we only consider simply typed $\l$-terms (terms
are always well-typed).

We follow Pottinger's approach \cite{pottinger78ndj}, that is, we
assume that every variable or function symbol comes equipped with a
fixed (simple) type and that $\al$-equivalence replaces a variable by
another variable of the same type only (assuming an infinite set of
variable for each type). Hence, we do not have to consider untyped
terms and introduce typing environments (finite map from variables to
types) for terms and rules: it is like working in a fixed infinite
typing environment.

Let $\cX$ be an infinite set of {\em variables} and $\cF$ be a set of
{\em function symbols} disjoint from $\cX$, and assume that each
variable or function symbol $s$ is equipped with a type $\tau(s)$ so
that there is an infinite number of variable of each type.

The family $(L^T)_{T\in\cT}$ of the sets of {\em raw terms of type
  $T$} is inductively defined as follows:

\begin{itemize}
\item
if $s\in\cX\cup\cF$, then $s\in L^{\tau(s)}$;
\item
if $x\in\cX$, $T\in\cT$ and $t\in L^T$, then $\l xt\in L^{\tau(x)\A T}$;
\item
if $U,V\in\cT$, $t\in L^{U\A V}$ and $u\in L^U$, then $tu\in L^V$.
\end{itemize}

\noindent
For every type $T$, the set $\cL^T$ of {\em terms of type $T$} is the
quotient of $L^T$ by type-preserving $\al$-equivalence, that is, $\l
xt=_\al\l yu$ only if $\tau(x)=\tau(y)$ \cite{pottinger78ndj}. Let
$\cL=\bigcup_{T\in\cT}\cL^T$ be the set of all (typed) terms. We write
$t:T$ or $\tau(t)=T$ if the $\al$-equivalence class of $t$ belongs to
$\cL^T$. A relation $R$ on terms {\em preserves types} if
$\tau(t)=\tau(u)$ whenever $(t,u)\in R$, written $tRu$ (\eg
$\al$-equivalence).

Note that function symbols do not have to be applied to any argument
nor any fixed number of arguments. Hence, $\sf$, $\sf x$, $\sf xy$,
etc. are legal terms (as long as they are well-typed). However, in
some examples, for convenience, we may use infix notations, like $x+y$
for denoting $+xy$.

Let $\FV(t)$ be the set of variables having a free occurrence in $t$
(\ie not bound by a $\l$), and $\BV(t)$ be the set of binding
variables of a raw term $t$ (\eg $\BV(\l xy)=\{x\}$).

A term is {\em linear} if no variable has more than one {\em free}
occurrence in it.

A term is {\em algebraic} if it contains no subterm of the form $\l
xt$ or $xt$.

A type-preserving relation $R$ on terms is {\em monotone} if, for all
$t,u,v,x$ such that $tRu$, one has $(\l xt)R(\l xu)$, $(tv)R(uv)$
whenever $tv$ is well-typed, and $(vt)R(vu)$ whenever $vt$ is
well-typed.

%%%%%%%%%%%%%%%%%%%%%%%%%%%%%%%%%%%%%%%%%%%%%%%%%%%%%%%%%%%%%%%%%%%%%%%%%%%%%%
\subsection{Substitution}
\label{sec-subs}

A substitution $\s$ is a map from $\cX$ to $\cL$ such that (1) for all
$x\in\cX$, $\tau(\s(x))=\tau(x)$, and (2) its {\em domain}
$\dom(\s)=\{x\in\cX\mid\s(x)\neq x\}$ is finite. In particular, we
write $_x^u$ for the substitution $\s$ such that $\s(x)=u$ and
$\s(y)=y$ if $y\neq x$. Let $\FV(\s)=\bigcup\{\FV(\s(x))\mid
x\in\dom(\s)\}$. A substitution $\s$ is {\em away from $X\sle\cX$} if
$(\dom(\s)\cup\FV(\s))\cap X=\vide$.

Given a term $t$ and a substitution $\s$, we denote by $t\s$ the term
obtained by replacing in $t$ each free occurrence of a variable $x$ by
$\s(x)$ by renaming, if necessary, variables bound in $t$ so that no
variable free in $\s(x)$ becomes bound \cite{curry58book}. Note that
substitution preserves typing: $\tau(t\s)=\tau(t)$.

A relation $R$ is {\em stable by substitution} (away from $X$) if
$(t\s)R(u\s)$ whenever $tRu$ (and $\s$ is away from $X$). It is a {\em
  congruence} if it is an equivalence relation that is monotone and
stable by substitution.

%%%%%%%%%%%%%%%%%%%%%%%%%%%%%%%%%%%%%%%%%%%%%%%%%%%%%%%%%%%%%%%%%%%%%%%%%%%%%%
\subsection{Stable subterm ordering}
\label{sec-stable-subterm-ord}

The notion of sub-raw-term is not compatible with
$\al$-equivalence. Instead, we consider the notion of {\em stable
  subterm}: $t\tle_\ss u$ if $t$ is a sub-raw-term of $u$ and
$\FV(t)\sle\FV(u)$. The relation $\tle_\ss$ is a partial ordering
stable by substitution. Let $\tlt_\ss$ be its strict part and
$\tge_\ss$ (resp. $\tgt_\ss$) be the inverse of $\tle_\ss$
(resp. $\tlt_\ss$).

%%%%%%%%%%%%%%%%%%%%%%%%%%%%%%%%%%%%%%%%%%%%%%%%%%%%%%%%%%%%%%%%%%%%%%%%%%%%%%
\subsection{Positions}
\label{sec-pos}

The set of {\em positions} in a (raw) term $t$, $\Pos(t)$, is the
subset of $\{0,1\}^*$ such that:

\begin{itemize}
\item
$\Pos(x)=\Pos(\sf)=\{\vep\}$ if $x\in\cX$ and $\sf\in\cF$
\item
$\Pos(tu)=\{\vep\}\cup\{0w\mid w\in\Pos(t)\}\cup\{1w\mid w\in\Pos(u)\}$
\item
$\Pos(\l xt)=\{\vep\}\cup\{0w\mid w\in\Pos(t)\}$
\end{itemize}

Given a (raw) term $t$, we denote by $t|_p$ its sub-raw-term at
position $p\in\Pos(t)$, and by $t[u]_p$ the (raw) term obtained by
replacing it by $u$.

A term $t$ is {\em $\eta$-long} if every variable or function symbol
occurring in it is maximally applied, that is, for all $p\in\Pos(t)$,
if $t|_p\in\cX\cup\cF$ and $t|_p:\vT\A\sA$, then there are
$q\in\Pos(t)$ and $\vt$ such that $p=q0^{|\vT|}$ and $t|_q=t|_p\vt$
\cite{huet76hdr}.

Given a term $t$ and $p\in\Pos(t)$, the set $\BV(t,p)$ of binding
variables above $t|_p$ is defined as follows:

\begin{itemize}
\item $\BV(t,\vep)=\vide$
\item $\BV(tu,0p)=\BV(t,p)$
\item $\BV(tu,1p)=\BV(u,p)$
\item $\BV(\l xt,0p)=\{x\}\cup\BV(t,p)$
\end{itemize}

For instance, $\Pos(\l x\sf x)=\{\vep,0,00,01\}$ and $\BV(\l x\sf x,0)=\{x\}$.

%%%%%%%%%%%%%%%%%%%%%%%%%%%%%%%%%%%%%%%%%%%%%%%%%%%%%%%%%%%%%%%%%%%%%%%%%%%%%%
\subsection{Rewriting}
\label{sec-rewriting}

The relation of {\em $\b$-reduction} (resp. {\em $\eta$-reduction}),
$\ab$ (resp. $\ae$), is the monotone closure of $\{((\l
xt)u,t_x^u)\mid t,u\in\cL,x\in\cX\}$ (resp. $\{(\l x(tx),t)\mid
t\in\cL,x\in\cX,x\notin\FV(t)\}$). We write $t\at{p}\ab u$ to indicate
that $t|_p=(\l xa)b$ and $u=t[a_x^b]_p$, and similarly for $t\at{p}\ae
u$. Note that the relation ${\abe}={\ab\cup\ae}$ preserves typing: if
$t:T$ and $t\abe t'$, then $t':T$.

An {\em equation} is a pair of terms $(l,r)$, written $l=r$, such that
$\tau(l)=\tau(r)$. A {\em (rewrite) rule} is a pair of terms $(l,r)$,
written $l\a r$, such that $\tau(l)=\tau(r)$, $l$ is of the form
$\sf\vl$ and $\FV(r)\sle\FV(l)$.

We assume neither that, if $\sf\vl\a r$ is a rule, then every
occurrence of $\sf$ in $r$ comes applied to $|\vl|$ arguments, nor
that, if $\sf\vl\a r$ and $\sf\vm\a s$ are two distinct rules, then
$|\vl|=|\vm|$. And, indeed, we will give examples of systems that do
not satisfy these constraints in Section \ref{sec-ho-subterm}
(function $\ms{ex}$) and Section \ref{sec-hopm} (after Lemma
\ref{lem-eta}). Such systems are necessary for dealing with matching
modulo $\be$ because we use curried symbols. In contrast, in HRSs
\cite{nipkow91lics}, function symbols are always maximally applied
(wrt their types) since terms are in $\eta$-long form and rules are of
the form $\sf\vl\a r$ with $\sf\vl$ of base type. Note however that,
in \cite{vandepol96phd}, van de Pol considers rules not necessarily in
$\eta$-long form nor of base type.

The {\em rewriting relation} generated by a set of rules $\cR$,
written $\ar$, is the closure by monotony and substitution of
$\cR$. Hence, $t\ar u$ if there are $p\in\Pos(t)$, $l\a r\in\cR$ and
$\s$ such that $t|_p=l\s$ and $u=t[r\s]_p$. For instance, with
$\cR=\{\sf x\a x\}$, we have $\l x\sf xy\ar \l xxy$. Note that
rewriting preserves typing: if $t:T$ and $t\ar t'$, then $t':T$.

Given a set of rules $\cR$, let $\cD(\cR)=\{\sf\in\cF\mid\ex\vl,\ex
r,\sf\vl\a r\in\cR\}$ be the subset of symbols {\em defined} by $\cR$,
and $\af=\sup\{|\vl|\mid\ex r,\sf\vl\a r\in\cR\}$. Note that $\af$ is
finite even if $\cR$ is infinite for $\sf\vl$ is {\em simply} typed by
assumption.\footnote{However, with polymorphic types, or dependent
  types together with type-level rewriting (\eg strong elimination),
  $\af$ may be infinite if $\cR$ is infinite.}

%%%%%%%%%%%%%%%%%%%%%%%%%%%%%%%%%%%%%%%%%%%%%%%%%%%%%%%%%%%%%%%%%%%%%%%%%%%%%%
\subsection{Notations for relations}
\label{sec-rel}

Given a relation $R$ on a set $A$, let $R(t)=\{u\in A\mid tRu\}$ be
the set of reducts or successors of $t$. An element $t$ such that
$R(t)=\vide$ is said to be in {\em normal form} or irreducible.

Given a relation $R$, let $R^=$ be the reflexive closure of $R$, $R^+$
its transitive closure, $R^*$ its reflexive and transitive closure,
and $R^{-1}$ its inverse ($xR^{-1}y$ iff $yRx$).

However, we will denote by $\la_\b$, $\la_\eta$ and $\la_\cR$ the
inverse relations of $\ab$, $\a_\eta$ and $\ar$ respectively; by
$\aa_\b$, $\aa_\eta$ and $\aa_\be$ the {\em symmetric closures} of
$\ab$, $\ae$ and $\abe$ respectively (\ie ${\ab}\cup{\la_\b}$, etc.);
and by $=_\eta$ and $=_\be$ the reflexive and transitive closures of
$\aa_\eta$ and $\aa_\be$ respectively.

Given two relations $R$ and $S$, we denote their composition by
juxtaposition and say that $R$ {\em commutes} with $S$ if
${RS}\sle{SR}$. For instance, if $R$ is monotone, then $\tgt_\ss$
commutes with $R$.

A relation $R$ is {\em strongly confluent} if
$R^{-1}R\sle(R^=)(R^=)^{-1}$, {\em locally confluent} if $R^{-1}R\sle
R^*(R^{-1})^*$, and {\em confluent} if $(R^{-1})^*R^*\sle
R^*(R^{-1})^*$. For instance, the relations $\ae$, $\ab$ and their
union $\abe$ are all confluent \cite{pottinger78ndj}.

%%%%%%%%%%%%%%%%%%%%%%%%%%%%%%%%%%%%%%%%%%%%%%%%%%%%%%%%%%%%%%%%%%%%%%%%%%%%%%
\subsection{Notations for quasi-orderings}
\label{sec-qo}

Given an equivalence relation $R$ on a set $A$, we denote by $[t]_R$
the equivalence class of an element $t$, and by $A/R$ the set of
equivalence classes modulo $R$.

Given a quasi-ordering $\ge$ on a set $A$ (transitive and reflexive
relation), let ${\simeq}={\ge\cap\ge^{-1}}$ be its {\em associated
  equivalence relation} and ${>}={\ge-\ge^{-1}}$ be its {\em strict
  part} (transitive and irreflexive relation).

%%%%%%%%%%%%%%%%%%%%%%%%%%%%%%%%%%%%%%%%%%%%%%%%%%%%%%%%%%%%%%%%%%%%%%%%%%%%%%
\subsection{Well-founded relations}
\label{sec-wf}

Given a set $A$, an element $a\in A$ is {\em strongly normalizing} wrt
a relation $R$ on $A$ if there is no infinite sequence
$a=a_0Ra_1R\ldots$ The relation $R$ {\em terminates} (or is {\em
  noetherian} or {\em well-founded}\footnote{In contrast with the
  mathematical tradition where a relation $R$ is said {\em
    well-founded} if there is no infinite descending chain
  $a_0R^{-1}a_1R^{-1}\ldots$}) on $A$ if every element of $A$ is
strongly normalizing wrt $R$. Let $\SN(R)$ be the set of elements of
$A$ that are strongly normalizing wrt $R$. By abuse of language, we
sometimes say that a quasi-ordering $\ge$ is well-founded when its
strict part so is.

If $R$ terminates (resp. is confluent) then every element has at least
(resp. at most) one normal form. In particular, we will denote by
${t\!\!\ad_\eta}$ the unique normal form of $t$ wrt $\ae$.

Note that, if $R$ is monotone, then $R\cup\tgt_\ss$ terminates iff $R$
terminates.

In this paper, we are interested in the termination of the relation
$\ab\cup\ar$, or variants thereof. Note that $\ab$ terminates on
well-typed terms \cite{sanchis67ndj}. However, since termination is
not a modular property (already in the first-order case)
\cite{toyama87ipl}, the termination of $\ar$ is generally not
sufficient to guarantee the termination of $\ab\cup\ar$. Moreover,
considering $\ar$ alone does not make sense when, in a right-hand side
of a rule, a free variable is applied to a term. This is not the case
in CRSs and HRSs since, in these systems, the definition of rewriting
includes some $\b$-reductions after a rule application
\cite{oostrom93hoa}.

%%%%%%%%%%%%%%%%%%%%%%%%%%%%%%%%%%%%%%%%%%%%%%%%%%%%%%%%%%%%%%%%%%%%%%%%%%%%%%
\subsection{Product quasi-ordering}
\label{sec-prod}

The {\em product} of $n$ relations $R_1,\ldots,R_n$ on the sets
$A_1,\ldots,A_n$ respectively is the relation $(R_1,\ldots,R_n)_\prod$
on $A_1\times\ldots\times A_n$ such that
$\vx\,(R_1,\ldots,R_n)_\prod\,\vy$ if, for all $i\in[1,n]$,
$x_iR_iy_i$.

If each $R_i$ is a quasi-ordering, then $(R_1,\ldots,R_n)_\prod$ is a
quasi-ordering too. If, moreover, the strict parts of $R_1,\ldots,R_n$
are well-founded, then the strict part of $(R_1,\ldots,R_n)_\prod$ is
well-founded too.

Given a quasi-ordering $\ge$ on a set $A$, let also $\ge_\prod$ denote
the product quasi-ordering on $A^n$ with each component ordered by
$\ge$.

%%%%%%%%%%%%%%%%%%%%%%%%%%%%%%%%%%%%%%%%%%%%%%%%%%%%%%%%%%%%%%%%%%%%%%%%%%%%%%
\subsection{Multiset quasi-ordering}
\label{sec-mul}

Given a set $A$, let $\cM=\bM(A)$ be the set of {\em finite multisets}
on $A$ (functions from $A$ to $\bN$ with finite support)
\cite{dershowitz79cacm}. Given a quasi-ordering $\ge_A$ on $A$, the
     {\em extension} of $\ge_A$ on finite multisets is the smallest
     quasi-ordering $\ge_\cM$ containing ${>^1_\cM}\cup{\simeq_\cM}$
     where $\simeq_\cM$ and $>^1_\cM$ are defined as follows
     \cite{comon03ln}:

\begin{itemize}
\item
$\vide\simeq_\cM\vide$, and $M+\mset{x}\simeq_\cM N+\mset{y}$ if
  $M\simeq_\cM N$ and $x\simeq_A y$;\footnote{Here, $A+B$ is the
  multiset union of the multisets $A$ and $B$, and
  $\mset{y_1,\ldots,y_n}$ the multiset made of $y_1,\ldots,y_n$.}
\item
$M+\mset{x}>^1_\cM M+\mset{y_1,\ldots,y_n}$ ($n\ge 0$) if, for every
  $i\in[0,n]$, $x>_A y_i$;
\end{itemize}
\noindent
where $\simeq_A$ (resp. $>_A$) is the equivalence relation associated
to (resp. strict part of) $\ge_A$.

Its associated equivalence relation is $\simeq_\cM$. Its strict part
${>_\cM}$ is ${{(>^1_\cM)^+}\simeq_\cM}$. It is well-founded if $>_A$
is well-founded.

Finally, let $\ge_\mul$ be the quasi-ordering on $A^*$ such that
$\vx\ge_\mul\vy$ if $\mset{\vx}\ge_\cM\mset{\vy}$.

%%%%%%%%%%%%%%%%%%%%%%%%%%%%%%%%%%%%%%%%%%%%%%%%%%%%%%%%%%%%%%%%%%%%%%%%%%%%%%
\subsection{Lexicographic quasi-ordering}
\label{sec-lex}

Given quasi-orderings $\ge_1,\ldots,\ge_n$ on sets $A_1,\ldots,A_n$,
the {\em lexicographic quasi-ordering} on $A_1\times\ldots\times A_n$,
written $(\ge_1,\ldots,\ge_n)_\lex$, is the union of the following two
relations:

\begin{itemize}
\item $(\simeq_1,\ldots,\simeq_n)_\prod$;
\item $\vx>\vy$ if there is $i\in[1,n]$ such that $x_i>_iy_i$ and,
  for all $j<i$, $x_j\simeq_jy_j$;
\end{itemize}

\noindent
where $\simeq_i$ (resp. $>_i$) is the equivalence relation associated
to (resp. strict part of) $\ge_i$. If $>_1,\ldots,>_n$ are
well-founded, then $>$ is well-founded too.

Given a quasi-ordering $\ge$ on a set $A$, let $\ge_\lex$ also denote
the lexicographic quasi-ordering on $A^n$ with each component ordered
by $\ge$.

%%%%%%%%%%%%%%%%%%%%%%%%%%%%%%%%%%%%%%%%%%%%%%%%%%%%%%%%%%%%%%%%%%%%%%%%%%%%%%
\subsection{Dependent lexicographic quasi-ordering}
\label{sec-dlqo}

Given two sets $A$ and $B$ and, for each $x\in A$, a set $B_x\sle B$,
the {\em dependent product} of $A$ and $(B_x)_{x\in A}$ is the set
$\S_{x\in A}B_x$ of pairs $(x,y)\in A\times B$ such that $y\in
B_x$. In the following, we use in many places a generalization to
dependent products of the lexicographic quasi-ordering (generalizing
to quasi-orderings Paulson's lexicographic ordering on dependent pairs
\cite{paulson86jsc}):

\begin{dfn}[Dependent lexicographic quasi-ordering]
\label{def-dlqo}
The {\em dependent lexicographic quasi-ordering}\\(DLQO) on a dependent
product $\S_{x\in A}B_x$ associated to:
\begin{itemize}
\item a quasi-ordering $\ge_A$ on $A$;
\item for each equivalence class $E$ modulo $\simeq_A$, a
set $C_E$ equipped with a quasi-ordering $\ge_E$;
\item for each $x\in A$, a {\em partial} function
  $\psi_x:B_x\a C_{[x]_{\simeq_A}}$;
\end{itemize}
is the union of the following two relations:
\begin{itemize}
\item
$(x,y)\simeq(x',y')$ if $x\simeq_A
  x'\et\psi_x(y)\simeq_{[x]_{\simeq_A}}\psi_{x'}(y')$;
\item
$(x,y)>(x',y')$ if $x>_Ax'\ou(x\simeq_A
  x'\et\psi_x(y)>_{[x]_{\simeq_A}}\psi_{x'}(y'))$;
\end{itemize}
\noindent
where $\simeq_A$ (resp. $\simeq_E$) is the equivalence relation
associated to $\ge_A$ (resp. $\ge_E$), and $>_A$ (resp. $>_E$) the
strict part of $\ge_A$ (resp. $\ge_E$).
\end{dfn}

If $>_A$ and each $>_E$ are well-founded, then $>$ is well-founded
too. Various examples of DLQOs will be given and used in the paper (in
particular, in Sections \ref{sec-f-qo} and \ref{sec-f-qo-mod}).

\section{Computability}
\label{sec-comp}

The computability method was introduced by Tait to prove the weak
normalization of (\ie the existence of a normal form wrt)
$\b$-reduction in some extensions of the simply typed $\l$-calculus
\cite{tait67jsl}, and was later extended by Girard for dealing with
polymorphic types \cite{girard71sls} and strong normalization
\cite{girard72phd,girard88book}. This method consists of:

\begin{enumerate}
\item
defining a domain $\Cand\sle\cP(\SN(\ab))$ of {\em computability
  candidates} for interpreting types;\footnote{In the following, like
  Girard in \cite{girard72phd,girard88book}, we will in fact consider
  a domain $\Cand^T\sle\cP(\cL^T)$ for each type $T$, but this is not
  relevant in this section.}
\item
interpreting each type $T$ by a candidate $\I{T}\in\Cand$;
\item
proving that each term of type $T$ is {\em computable}, \ie belongs to
$\I{T}$, from which it follows that every typed term is strongly
normalizing wrt $\ab$.
\end{enumerate}

In this section, we will see the various definitions that have been
proposed for $\Cand$ so far, and discuss which ones are best suited
for extension to arbitrary, and in particular
non-orthogonal\footnote{A rewrite system is orthogonal if it is
  left-linear and non-ambiguous (\ie has no critical pair). This is in
  particular the case of ML-like programs. An important property of
  orthogonal systems is their confluence
  \cite{huet80jacm,klop80phd,oostrom94phd}.}, rewrite
systems. However, all those definitions satisfy the following
properties:

\begin{itemize}
\item
variables are computable: for every $P\in\Cand$, $\cX\sle P$;
\item
$\Cand$ is stable by the operation
  $\propto:\cP(\cL)\times\cP(\cL)\a\cP(\cL)$ defined
  by: $$\propto\!(P,Q)=\{v\in\cL\mid\all t\in P,vt\in Q\}$$ \ie if
  $P,Q\in\Cand$, then $\propto\!(P,Q)\in\Cand$;
\item
$\Cand$ is stable by arbitrary\footnote{Finite or infinite.} non-empty
  intersection:\\if $(A_i)_{i\in I}$ is a non-empty family of
  candidates, then $\bigcap_{i\in I}A_i\in\Cand$;
\item
$\Cand$ contains $\SN(\ab)$.
\end{itemize}

The last two conditions imply that $\Cand$ has a structure of complete
lattice for inclusion,\footnote{An inf-complete lattice $L$ that has a
  biggest element is complete. The supremum of a set $P\sle L$ is
  indeed $\glb(\mr{ub}(P))$ where $\glb$ is the greatest lower bound
  and $\mr{ub}(P)$ is the {\em non-empty} set of all the upper bounds
  of $P$.} the greatest lower bound of a set $P\sle\Cand$ being given
by the intersection $\bigcap P$ if $P\neq\vide$, and $\SN(\ab)$ if
$P=\vide$. However, its lowest upper bound (the smallest candidate
containing the union) is {\em not} necessarily the union
\cite{riba07fossacs}.

The intersection allows one to interpret quantification on types
(polymorphism) or inductive types (see Section \ref{sec-ho-subterm}),
while $\propto$ allows one to interpret $\A$ so that, by definition,
$vt\in\I{V}$ if $v\in\I{T\A U}$ and $t\in\I{T}$, which is the main
problem when trying to prove the termination of $\b$-reduction.

\bigskip

We now see every definition we are aware of:

\begin{itemize}
\item $\Red$: Girard' set of {\em reducibility candidates}
  \cite{girard72phd,girard88book}. A set $P$ belongs to $\Red$ if the
  following conditions are satisfied:

\begin{enumii}{R}
\item $P\sle\SN(\ab)$;
\item $P$ is stable by reduction: if $t\in P$ and $t\ab u$, then $u\in
  P$;
\item if $t$ is a {\em neutral}\footnote{Called ``simple'' in
    \cite{girard72phd} and ``neutral'' in \cite{girard88book}.} term
  and $\ab\!(t)\sle P$, then $t\in P$.
\end{enumii}

In $\l$-calculus with no function symbols, a term is neutral if it is
not an abstraction. Neutral terms satisfy the following key property:
if $t$ is neutral then, for all terms $u$, $\ab\!(tu)=\{t'u\mid t\ab
t'\}\cup\{tu'\mid u\ab u'\}$, that is, the application of $t$ cannot
create new redexes.

\item $\Sat$: Tait' set of {\em saturated\footnote{This expression
    seems due to Gallier \cite{gallier90chapter}.} sets}
  \cite{tait72lc}. A set $P$ belongs to $\Sat$ if the following
  conditions are satisfied:

\begin{enumii}{S}
\item $P\sle\SN(\ab)$;
\item $P$ contains all the strongly normalizable terms of the form
  $x\vt$;
\item if $t_x^u\vv\in P$ and $u\in\SN(\ab)$, then $(\l xt)u\vv\in P$.
\end{enumii}

\item $\SatInd$: Parigot' smallest subset of $\Sat$ containing
  $\SN(\ab)$ and stable by $\propto$ and $\bigcap$
  \cite{parigot97jsl}. As Parigot remarked, for $\b$-reduction, it is
  not necessary to consider all saturated sets but only those that can
  be obtained from $\SN(\ab)$ by $\propto$ and intersection.

\item $\Bi$: Parigot' set of {\em bi-orthogonals}\footnote{Parigot did
  not use the expression ``bi-orthogonal''. To my knowledge, this
  expression first appears in \cite{vouillon04popl}. See
  \cite{abel06phd}, p. 67, for a discussion about the origin of this
  expression. Anyway, Parigot computability predicates are indeed
  bi-orthogonals wrt the orthogonality relation $\bot$ between
  $\cP(\SN(\ab))$ and $\cP(\SN(\ab)^*)$ such that $P\bot E$ if $\all
  v\in P,\all\vt\in E,v\vt\in\SN(\ab)$. The (right) orthogonal of
  $P\sle\SN(\ab)$ is $P^\bot=\{\vt\in\SN(\ab)^*\mid\all v\in
  P,v\vt\in\SN(\ab)\}$, while the (left) orthogonal of
  $E\sle\SN(\ab)^*$ is ${^\bot E}={\propto^*\!(E,\SN(\ab))}$. One can
  then see that $\Bi=\{P\sle\SN(\ab)\mid P\neq\vide\et
  {}^\bot(P^\bot)=P\}$.}  \cite{parigot93lics,parigot97jsl} is the set
  $\{\propto^*\!(E,\SN(\ab))\mid\vide\neq E\sle\SN(\ab)^*\}$ where
  $\SN(\ab)^*$ is the set of finite sequences of elements of
  $\SN(\ab)$ and $\propto^*:\cP(\cL^*)\times\cP(\cL)\a\cP(\cL)$
  extends $\propto$ as
  follows: $$\propto^*\!(E,Q)=\{v\in\cL\mid\all\vt\in E,v\vt\in Q\}$$

Note that a sequence $\vt\in\cL^*$ can be seen as the context
$[]\vt$. Hence, $$\propto^*\!(E,Q)=\{v\in\cL\mid\all e\in E,e[v]\in
Q\}.$$
\end{itemize}

Reducibility candidates and saturated sets are studied in
\cite{gallier90chapter}. In particular, every reducibility candidate is a
saturated set: $\Red\sle\Sat$. The converse does not hold in general
since a saturated set does not need to be stable by reduction: for
instance, the smallest saturated set containing $\l x(\l yy)x$ does
not contain $\l xx$. However, Riba showed that every saturated set
stable by reduction is a reducibility candidate
\cite{riba07fossacs}. Hence,
$\Red=\Sat_\a=\{P\in\Sat\mid{\ab\!(P)}\sle{P}\}$. In
\cite{parigot97jsl}, Parigot showed that every element of $\SatInd$ is
a bi-orthogonal: $\SatInd\sle\Bi$. Finally, Riba showed that every
bi-orthogonal is a reducibility candidate \cite{riba07lics}:
$\Bi\sle\Red$. In particular, bi-orthogonals are stable by
reduction. On the other hand, I don't know whether $\SatInd$, $\Bi$
and $\Red$ are distinct. In conclusion, we currently have the
following relations:
\[\SatInd\sle\Bi\sle\Red=\Sat_\a\subsetneq\Sat\]

A natural question is then to know to which extent each one of these
sets can be used to handle rewriting, and if a set allows to show the
termination of more systems than the others. All these definitions
rely on the form of {\em redexes} (reducible expressions): $\Red$ uses
the notion of neutral term, a set $P\in\Sat$ has to be stable by
head-expansion (inverse relation of head-reduction), and $\Bi$ is
defined as the set of bi-orthogonals wrt a relation between terms and
contexts that allows one to build redexes.

\begin{itemize}
\item $\Bi$ being exclusively based on the notion of context, it does
  not seem possible to extend it to non-orthogonal rewrite relations.

\item The saturated sets could perhaps be extended by adding:

\hsp[5mm](S4) if $l\a r\in\cR$, $r\s\vt\in P$ and
$\s\in\SN(\ab\cup\ar)$, then $l\s\vt\in P$.

In order to have $\SN(\ab\cup\ar)\in\Sat_\cR$, one has then to prove
that $l\s\vt\in\SN(\ab\cup\ar)$ if $r\s\vt\in\SN(\ab\cup\ar)$ and
$\s\in\SN(\ab\cup\ar)$, which is generally not the case if $\cR$ is
not orthogonal. This problem could perhaps be solved by considering
{\em all} the head-reducts of $l\s$, but then we would arrive at a
condition similar to (R3).

\item In contrast to the previous domains, Girard's reducibility
  candidates seem easy to extend to arbitrary rewrite relations. This
  is therefore the notion of computability that we will use in the
  following.
\end{itemize}

\section{Rewriting with matching modulo $\al$-equivalence}
\label{sec-core}

In this section, we provide a survey on the notion of computability
closure for standard rewriting (that is in fact rewriting modulo
$\al$-equivalence, because terms are defined modulo $\al$-equivalence)
first introduced in \cite{blanqui99rta,blanqui02tcs}. We present the
computability closure progressively by showing at each step how it has
to be extended to handle new term constructions. Omitted proofs can be
found in \cite{blanqui02tcs,blanqui05mscs}. For dealing with recursive
function definitions (Section \ref{sec-rec} below), we introduce a new
more general rule based on the notion of $\cF$-quasi-ordering
compatible with application (Definition \ref{def-f-quasi-ord}) and
provide various examples of such $\cF$-quasi-orderings in Section
\ref{sec-f-qo} (and later in Section \ref{sec-f-qo-mod}).

\newcommand\compsn{(R1)}
\newcommand\compred{(R2)}
\newcommand\compneutral{(R3)}

%%%%%%%%%%%%%%%%%%%%%%%%%%%%%%%%%%%%%%%%%%%%%%%%%%%%%%%%%%%%%%%%%%%%%%%%%%%%%%
\subsection{Definition of computability}
\label{sec-comp-core}

To extend to rewriting Girard's definition of computability predicates
\cite{girard88book}, we first have to define the set of neutral
terms. By analogy with abstractions, a term of the form $\sf\vt$ with
$\sf\in\cD(\cR)$ should be neutral only if $\sf$ is applied to enough
arguments wrt $\cR$, \ie $|\vt|\ge\af$. Otherwise, $\sf\vt u$ could be
head-reducible and the key property of neutral terms, that
$\a\!(tu)=\{t'u\mid t\a t'\}\cup\{tu'\mid u\a u'\}$ whenever $t$ is
neutral, would not hold. Now, what about terms of the form $\sf\vt$
with $\sf\in\cF-\cD(\cR)$ (undefined symbols)? We could {\em a priori}
consider them as neutral. However, for dealing with higher-order
subterms in Sections \ref{sec-ho-subterm} and \ref{sec-match-def},
we will consider type interpretations for which it seems difficult to
prove \compneutral\ if such terms are neutral. We therefore exclude
them from neutral terms:

\begin{dfn}[Computability candidates]
\label{def-comp-core}
Given a set $\cR$ of rewrite rules of the form $\sf\vl\a r$, a term is
{\em neutral}\footnote{We will give a more general definition in
  Definition \ref{def-neutral-acc}.} if it is of the form $x\vv$, $(\l
xt)u\vv$ or $\sf\vv$ with $\sf\in\cD(\cR)$ and
$|\vv|\ge\af=sup\{|\vl|\mid\ex r,\sf\vl\a r\in\cR\}$.

Given a type $T$, let $\Red_\cR^T$ be the set of all the sets
$P\sle\cL^T$ such that:

\begin{enumii}{R}
\item
$P\sle\SN(\a)$ where ${\a}={\ab\cup\ar}$;
\item
$P$ is stable by reduction: if $t\in P$ and $t\a u$, then $u\in P$;
\item
if $t:T$ is neutral and $\a\!(t)\sle P$, then $t\in P$.
\end{enumii}

Given $P\in\Red_\cR^T$ and $Q\in\Red_\cR^U$, let
$\propto\!(P,Q)=\{v:T\A U\mid\all t\in P,vt\in Q\}$.
\end{dfn}

Note that computability predicates are sets of well-typed terms and
that all the elements of a computability predicate have the same type.

For the sake of simplicity, in all the remaining of the paper, we
write $\SN$ instead of $\SN(\a)$, but $\a$ will have different
meanings in sections \ref{sec-rew-mod} and \ref{sec-hopm}.

We now check that the family $(\Red_\cR^T)_{T\in\cT}$ has the
properties described in Section \ref{sec-comp}:

\begin{lem}
\label{lem-cand}
For every type $T$, $\Red_\cR^T$ is stable by non-empty intersection
and admits $\SN^T=\{t:T\mid t\in\SN\}$ as greatest element. Moreover,
for all $T,U\in\cT$, $P\in\Red_\cR^T$ and $Q\in\Red_\cR^U$,
$\propto\!(P,Q)\in\Red_\cR^{T\A U}$.
\end{lem}

\begin{prf}
The fact that $\SN^T\in\Red_\cR$ and the stability by non-empty
intersection are easily proved. We only detail the stability by
$\propto$. Let $T,U\in\cT$, $P\in\Red_\cR^T$ and
$Q\in\Red_\cR^U$. Every element of $\propto\!(P,Q)$ is of type $T\A
U$.

\begin{enumii}{R}
\item
Let $v\in{\propto\!(P,Q)}$. Let $x$ be a variable of type $T$. By
\compneutral, $x\in P$. By definition of $\propto$, $vx\in Q$. By
\compsn, $vx\in\SN$. Thus, $v\in\SN$.
\item
Let $v\in{\propto\!(P,Q)}$, ${v'}\in{\a\!(v)}$ and $t\in P$. By
definition of $\propto$, $vt\in Q$. By \compred, $v't\in Q$.
\item
  Let $v:T\A U$ neutral such that ${\a\!(v)}\sle{\propto\!(P,Q)}$, and
  $t\in P$. We show that $vt\in Q$ by well-founded induction on $t$
  with $\a$ as well-founded relation ($t\in\SN$ by \compsn). Since $v$
  is neutral, $vt$ is neutral too. Hence, by \compneutral, it is
  sufficient to show that $\a\!(vt)\sle Q$. Let $w\in{\a\!(vt)}$. We
  first prove (a): either $w=v't$ with $v\a v'$, or $w=vt'$ with $t\a
  t'$. We proceed by case on $vt\a w$:
\begin{itemize}
\item
$vt\ab w$. Since $v$ is neutral, $v$ is not an abstraction and (a) is
  satisfied.
\item
$vt\ar w$. If there are $\sf\vl l\a r\in\cR$ and $\s$ such that
  $vt=\sf\vl\s l\s$ and $w=r\s$, then $v=\sf\vl\s$ and $|\vl|<|\vl
  l|\le\af$. Since $v$ is neutral, this is not possible. Thus (a) is
  verified.
\end{itemize}
We now show that $w\in Q$.
\begin{itemize}
\item Case $w=v't$ with $v\a v'$. By assumption,
  $v'\in{\propto\!(P,Q)}$. As $t\in P$, we have $w\in Q$.
\item Case $w=vt'$ with $t\a t'$. By \compred, $t'\in P$. Thus, by the
  induction hypothesis, $w\in Q$.\cqfd
\end{itemize}
\end{enumii}
\end{prf}

Therefore, as already mentioned in Section \ref{sec-comp}, every
$\Red_\cR^T$ is a complete lattice for inclusion.

Now, one can easily check Tait's property (S3) described in the
previous section (implying that elements of $\Red_\cR^T$ are Tait
saturated sets with $\SN(\ab)$ replaced by $\SN(\ab\cup\ar)$):

\begin{lem}
\label{lem-comp-beta}
Given $T\in\cT$ and $P\in\Red_\cR^T$, $(\l xt)u\vv\in P$ iff $(\l
xt)u\vv:T$, $t_x^u\vv\in P$ and $u\in\SN$.
\end{lem}

\begin{prf}
  Assume that $(\l xt)u\vv\in P$. Then, $(\l xt)u\vv:T$. By \compred,
  $t_x^u\vv\in P$. By \compsn, $(\l xt)u\vv\in\SN$. Therefore,
  $u\in\SN$.

  Assume now that $(\l xt)u\vv:T$, $t_x^u\vv\in P$ and $u\in\SN$. By
  \compsn, $t_x^u\vv\in\SN$. Therefore, $\vv\in\SN$, $t_x^u\in\SN$ and
  $t\in\SN$. We now prove that, for all $t,u,\vv\in\SN$, $(\l
  xt)u\vv\in P$, by induction on $\a_\prod$. Since $(\l xt)u\vv:T$ and
  $(\l xt)u\vv$ is neutral, by \compneutral, it suffices to prove that
  every reduct $w$ of $(\l xt)u\vv$ belongs to $P$. Since rules are of
  the form $\sf\vl\a r$, there are two possible cases:
\begin{itemize}
\item $w=t_x^u\vv$. Then, $w\in P$ by assumption.
\item $w=(\l xt')u'\vv'$ and $tu\vv\a_\prod t'u'\vv'$. Then, $w\in P$
  by the induction hypothesis.\cqfd
\end{itemize}
\end{prf}

\begin{cor}
\label{cor-comp-lam}
Given $T,U\in\cT$, $P\in\Red_\cR^T$ and $Q\in\Red_\cR^U$, ${\l
  xt}\in{\propto\!(Q,P)}$ iff $\l xt:U\A T$ and, for all $u\in Q$,
$t_x^u\in P$.
\end{cor}

\begin{prf}
  Assume that ${\l xt}\in{\propto\!(Q,P)}$ and $u\in Q$. Then, by
  definition of $\propto$, $\l xt:U\A T$ and $(\l xt)u\in
  P$. Therefore, by \compred, $t_x^u\in P$. Assume now that $\l xt:U\A
  T$ and, for all $u\in P$, $t_x^u\in P$. By definition, ${\l
    xt}\in{\propto\!(Q,P)}$ if, for all $u\in Q$, $(\l xt)u\in P$. So,
  let $u\in Q$. By \compsn, $u\in\SN$. Therefore, by Lemma
  \ref{lem-comp-beta}, $(\l xt)u\in P$.\cqfd
\end{prf}

Given two sets $A$ and $B$ and, for each $x\in A$, a set $B_x\sle B$,
let $\Pi_{x\in A}B_x=\cP(\S_{x\in A}B_x)$ be the set of {\em partial}
functions $f:A\a B$ such that, for all $x\in\dom(f)$, $f(x)\in B_x$.

Given an interpretation of type constants
$I\in\Pi_{\sB\in\cB}\Red_\cR^\sB$, the interpretation of types
$\I{\_}^I\in\Pi_{T\in\cT}\Red_\cR^\cT$ is defined as follows:

\begin{itemize}
\item
$\I\sB^I=I(\sB)$ if $\sB\in\cB$,
\item
${\I{T\A U}^I}={\propto\!(\I{T}^I,\I{U}^I)}$.
\end{itemize}

We say that a type constant $\sB$ is {\em basic} if its interpretation
is $\SN^\sB$, and that a symbol $\sf:\vT\A\sB$ is {\em basic} if $\sB$
is basic. Let the {\em basic interpretation} be the interpretation $I$
such that $I(\sB)=\SN^\sB$ for all $\sB\in\cB$.

We say that a term $t:T$ is {\em computable} wrt a base type
interpretation $I$ if $t\in\I{T}^I$. A substitution $\s$ is computable
wrt a base type interpretation $I$ if, for all $x\in\cX$,
$x\s\in\I{\tau(x)}^I$. Note that, by \compneutral, variables are
computable. Therefore, the identity substitution is always computable.

By definition of the interpretation of arrow types, a symbol
$\sf:\vT\A U$ is computable wrt a base type interpretation $I$ if, for
all $\vt\in\I\vT^I$, $\sf\vt\in\I{U}^I$. So, let $\S^I$ be the set of
pairs $(\sf,\vt)$ such that $\sf:\vT\A U$ and $\vt\in\I\vT^I$ ($\sf$
may be partially applied in $\sf\vt$), and let $\S_\max^I$ be the
subset of $\S^I$ made of the pairs $(\sf,\vt)$ such that $U\in\cB$,
that is, when $\sf$ is maximally applied.

In the following, we may drop the exponent $I$ when it is clear from
the context.

\begin{thm}
\label{thm-comp}
The relation ${\ab}\cup{\ar}$ terminates on well-typed terms if there
is $I\in\Pi_{\sB\in\cB}\Red_\cR^\sB$ such that every non-basic
undefined symbol and every defined symbol is computable.
\end{thm}

\begin{prf}
  It suffices to prove that every well-typed term is computable. For
  dealing with abstraction, we prove the more general statement that,
  for all $t:T$ and computable $\s$, $t\s\in\I{T}$, by induction on
  $t$. This indeed implies that every well-typed term is computable
  since the identity substitution is computable by \compneutral. We
  proceed by case on $t$:
\begin{itemize}
\item $t=x\in\cX$. Then, $t\s=x\s\in\I{T}$ since $\s$ is computable.
\item $t=uv$. By the induction hypothesis, $u\s\in\I{\tau(v)\A T}$ and
  $v\s\in\I{\tau(v)}$. Therefore, by definition of $\I{\_}$,
  $t\s=(u\s)(v\s)\in\I{T}$.
\item $t=\sf:\vT\A\sA$. If $\sf$ is a defined symbol or a non-basic
  undefined symbol, then $t\s=\sf$ is computable by
  assumption. Otherwise, $\sf$ is a basic undefined symbol. By
  definition, it is computable if, for all $\vt\in\I\vT$,
  $\sf\vt\in\I\sA$. Since $\sf$ is basic, $\I\sA=\SN$. Now, one can
  easily prove that $\sf\vt\in\SN$, by induction on $\vt$ with
  $\a_\prod$ as well-founded relation ($\vt\in\SN$ by \compsn).
\item $t=\l xu$. Wlog we can assume that $\s$ is away from
  $\{x\}$. Hence, $t\s=\l x(u\s)$. By Corollary \ref{cor-comp-lam},
  $\l x(u\s)\in\I{T}=\I{\tau(x)\A\tau(u)}$ if $\l
  x(u\s):\tau(x)\A\tau(U)$ and $(u\s)_x^v\in\I{\tau(u)}$ for all
  $v\in\I{\tau(x)}$. Since $\s$ is away from $\{x\}$, we have
  $(u\s)_x^v=u\t$ where $x\t=v$ and $y\t=y\s$ if $y\neq x$. Since $\t$
  is computable, by induction hypothesis, $u\t\in\I{\tau(u)}$.\cqfd
\end{itemize}
\end{prf}

Note that, with the basic interpretation, there is no
non-basic undefined symbol. In Section \ref{sec-basic-subterm}, we
will see another interpretation with which non-basic
undefined symbols are computable.

%%%%%%%%%%%%%%%%%%%%%%%%%%%%%%%%%%%%%%%%%%%%%%%%%%%%%%%%%%%%%%%%%%%%%%%%%%%%%%
\subsection{Core computability closure}

The next step consists then in proving that every defined symbol is
computable, \ie $\sf\in\I{\tau(\sf)}^I$ for all
$\sf\in\cD(\cR)$. Assume that $\tau(\sf)=\vT\A\sA$. As just seen
above, $\sf$ is computable if, for all $\vt\in\I\vT^I$, $\sf\vt\in
I(\sA)$. Since $\sf\in\cD(\cR)$ and $|\vt|\ge\al_\sf$, $\sf\vt$ is
neutral and, by \compneutral, belongs to $I(\sA)$ if all its reducts
so do. The notion of {\em computability closure} enforces this
property.

\begin{dfn}[Computability closure]
\label{def-cc-core}
A {\em computability closure} is a function $\CC$ mapping every
$\sf\in\cD(\cR)$ and $\vl\in\cL^*$ such that $\sf\vl$ is well-typed to
a set of well-typed terms.
\end{dfn}

\begin{dfn}[Valid computability closure -- first definition]\footnote{We give a more general definition in Definition \ref{def-cc-valid-2}.}
\label{def-cc-valid-1}
A computability closure $\CC$ is {\em valid} wrt a base type
interpretation $I$ if it satisfies the following properties:
\begin{itemize}\itemsep=0mm
\item it is {\em stable by substitution}: $t\s\in\CC_\sf(\vl\s)$
  whenever $t\in\CC_\sf(\vl)$;
\item it {\em preserves computability} wrt $I$: every element of
  $\CC_\sf(\vl)$ is computable whenever $\vl$ so are.
\end{itemize}
\end{dfn}

\begin{thm}
\label{thm-cc}
Given $I\in\Pi_{\sB\in\cB}\Red_\cR^\sB$, every defined symbol is
computable if there is a valid computability closure $\CC$ such that,
for every rule $\sf\vl\a r\in\cR$, we have $r\in\CC_\sf(\vl)$.
\end{thm}

\begin{prf}
  As just explained, it is sufficient to prove that, for all
  $(\sf,\vt)\in\S_\max$ with $\sf\in\cD(\cR)$ and $\sf:\vT\A\sA$,
  every reduct $t$ of $\sf\vt$ belongs to $\I\sA$. We proceed by
  well-founded induction on $\vt$ with $\a_\prod$ as well-founded
  relation ($\vt\in\SN$ by \compsn). There are two possible cases:

\begin{itemize}
\item There is $\vu$ such that $t=\sf\vu$ and $\vt\a_\prod\vu$. By
  \compred, $\vu\in\I\vT$. Therefore, by the induction hypothesis,
  $\sf\vu\in\I\sA$.

\item There are $\vw$, $\sf\vl\a r\in\cR$ and $\s$ such
  that $\vt=\vl\s\vw$ and $t=r\s\vw$. Since $r\in\CC_\sf(\vl)$ and
  $\CC$ is stable by substitution, we have
  $r\s\in\CC_\sf(\vl\s)$. Since $\vl\s$ are computable and $\CC$
  preserves computability, we have $r\s$ computable. Finally, since
  $\vw$ is computable, we have $t$ computable.\cqfd
\end{itemize}
\end{prf}

Hence, the termination of $\ab\cup\ar$ can be reduced to finding
computability-preserving operations to define a computability
closure. Among such operations, one can consider the ones of Figure
\ref{fig-cc-core} that directly follow from the definition or
properties of computability.

\begin{figure}[ht]
\caption{Computability closure operations I\label{fig-cc-core}}
\begin{center}
\fbox{\begin{tabular}{rl}
(arg)&$\{\vl\}\sle\CC_\sf(\vl)$\\

(app)&if $t\in\CC_\sf(\vl)$, $t:U\A V$, $u\in\CC_\sf(\vl)$ and $u:U$,
then $tu\in\CC_\sf(\vl)$\\

(red)&if $t\in\CC_\sf(\vl)$ and $t\a u$, then $u\in\CC_\sf(\vl)$\\

(undef-basic)&if $\sg\in\cF-\cD(\cR)$, $\sg:\vT\A\sB$, $\sB\in\cB$
and $\I\sB=\SN^\sB$, then $\sg\in\CC_\sf(\vl)$
\end{tabular}}
\end{center}
\end{figure}

\begin{thm}
\label{thm-cc-core}
For all $I\in\Pi_{\sB\in\cB}\Red_\cR^\sB$, the smallest computability
closure closed by the operations I is valid.
\end{thm}

\begin{prf}
\begin{itemize}
\item Stability by substitution. We prove that, for all
  $t\in\CC_\sf(\vl)$, we have $t\s\in\CC_\sf(\vl\s)$, by induction on
  the definition of $\CC_\sf(\vl)$.
\begin{description}
\item[(arg)] By (arg), $\{\vl\s\}\sle\CC_\sf(\vl\s)$.
\item[(app)] By the induction hypothesis,
  $\{t\s,u\s\}\sle\CC_\sf(\vl\s)$. Therefore, by (app),
  $(tu)\s=(t\s)(u\s)\in\CC_\sf(\vl\s)$.
\item[(red)] By the induction hypothesis, $t\s\in\CC_\sf(\vl\s)$. Since
  $\a$ is stable by substitution, $t\s\a u\s$. Therefore, by (red),
  $u\s\in\CC_\sf(\vl\s)$.
\item[(undef-basic)] By (undef-basic), $\sg\s=\sg\in\CC_\sf(\vl\s)$.
\end{description}

\item Preservation of computability. Assume that $\vl$ are
  computable. We prove that, for all $t\in\CC_\sf(\vl)$, $t$ is
  computable, by induction on the definition of $\CC_\sf(\vl)$.
\begin{description}
\item[(arg)] $\vl$ are computable by assumption.
\item[(app)] By the induction hypothesis, $t$ and $u$ are
  computable. Therefore, by definition of $\I{U\A V}$, $tu$ is
  computable.
\item[(red)] By the induction hypothesis, $t$ is
  computable. Therefore, by \compred, $u$ is computable.
\item[(undef-basic)] After the proof of Theorem \ref{thm-comp}, $\sg$
  is computable.\cqfd
\end{description}
\end{itemize}
\end{prf}

Therefore, using for $I$ the basic interpretation, we get:

\begin{cor}
\label{cor-cc-core}
The relation $\ab\cup\ar$ terminates on well-typed terms if, for every
rule $\sf\vl\a r\in\cR$, we have $r\in\CC_\sf(\vl)$, where $\CC$ is
the smallest computability closure closed by the operations I.
\end{cor}

%%%%%%%%%%%%%%%%%%%%%%%%%%%%%%%%%%%%%%%%%%%%%%%%%%%%%%%%%%%%%%%%%%%%%%%%%%%%%%
\subsection{Handling abstractions and bound variables}
\label{sec-bv}

Consider now the following symbol definition, where $T$, $U$ and $V$
are any type:

\begin{center}
$\so:(U\A V)\A(T\A U)\A(T\A V)$\\[2mm]
\begin{rew}
f~\so~g&\l x~f~(g~x)\\
\end{rew}
\end{center}

It cannot be handled by the operations I. By Corollary
\ref{cor-comp-lam}, a term $\l xt$ is computable if, for all
$u\in\I{\tau(x)}$, $t_x^u$ is computable. We can therefore extend the
previous core definition of $\CC$ by the rules of Figure
\ref{fig-cc-bv} if we generalize validity as follows:

\begin{dfn}[Valid computability closure -- extended definition]\footnote{This definition replaces the one given in Definition \ref{def-cc-valid-1}.}
\label{def-cc-valid-2}
A computability closure $\CC$ is {\em valid} wrt a base type
interpretation $I$ if:
\begin{itemize}\itemsep=0mm
\item it is {\em stable by substitution}: $t\s\in\CC_\sf(\vl\s)$
  whenever $t\in\CC_\sf(\vl)$ and $\s$ is away from $\FV(t)-\FV(\vl)$;
\item it {\em preserves computability} wrt $I$: $t\t$ is computable
  whenever $t\in\CC_\sf(\vl)$, $\vl$ are computable, $\t$ is
  computable and $\dom(\t)\sle\FV(t)-\FV(\vl)$.
\end{itemize}
\end{dfn}

Note that, if $\FV(t)\sle\FV(\vl)$ as it is required for the
right-hand side of a rule, then the above conditions reduce to the
ones of Definition \ref{def-cc-valid-1}.

\begin{figure}[ht]
\caption{Computability closure operations II\label{fig-cc-bv}}
\begin{center}
\fbox{\begin{tabular}{rl}
(var)&$\cX-\FV(\vl)\sle\CC_\sf(\vl)$\\
(abs)&if $t\in\CC_\sf(\vl)$ and $x\in\cX-\FV(\vl)$,
then $\l xt\in\CC_\sf(\vl)$\\
\end{tabular}}
\end{center}
\end{figure}

\begin{thm}
\label{thm-cc-bv}
For all $I\in\Pi_{\sB\in\cB}\Red_\cR^\sB$, the smallest computability
closure $\CC$ closed by the operations I and II is valid.
\end{thm}

\begin{prf}
We proceed as in Theorem \ref{thm-cc-core} but only detail the new cases.
\begin{itemize}
\item Stability by substitution. We prove that, for all
  $t\in\CC_\sf(\vl)$ and $\s$ away from $\FV(t)-\FV(\vl)$, we have
  $t\s\in\CC_\sf(\vl\s)$, by induction on the definition of
  $\CC_\sf(\vl)$.
\begin{description}
\item[(var)] Let $x\in\cX-\FV(\vl)$. Since $\s$ is away from
  $\FV(x)-\FV(\vl)$, we have $x\s=x$ and
  $x\in\cX-\FV(\vl\s)$. Therefore, by (var), $x\in\CC_\sf(\vl\s)$.
\item[(abs)] Wlog we can assume that $\s$ is away from $\{x\}$. Hence,
  $(\l xt)\s=\l x(t\s)$ and, since $x\in\cX-\FV(\vl)$, we have
  $x\in\cX-\FV(\vl\s)$. Therefore, by (abs), $\l
  x(t\s)\in\CC_\sf(\vl\s)$ for, by the induction hypothesis,
  $t\s\in\CC_\sf(\vl\s)$.
\end{description}

\item Preservation of computability. Assume that $\vl$ are
  computable. We prove that, for all $t\in\CC_\sf(\vl)$ and computable
  $\t$ such that $\dom(\t)\sle\FV(t)-\FV(\vl)$, we have $t\t$
  computable, by induction on $\CC_\sf(\vl)$.
\begin{description}
\item[(var)] Let $x\in\cX-\FV(\vl)$. Then, $x\t$ is computable by
  assumption.
\item[(abs)] Wlog we can assume that $\t$ is away from $\{x\}$. Hence,
  $(\l xt)\t=\l x(t\t)$. Now, by Corollary \ref{cor-comp-lam}, $\l
  x(t\t)$ is computable if, for all computable $u:\tau(x)$,
  $(t\t)_x^u$ is computable. Since $\t$ is away from $\{x\}$,
  $(t\t)_x^u=t\s$ where $x\s=u$ and $y\t=y\s$ if $y\neq x$. Now, $\s$
  is computable and $\dom(\s)\sle\FV(t)-\FV(\vl)$ for
  $\dom(\t)\sle\FV(\l xt)-\FV(\vl)$. Therefore, by the induction
  hypothesis, $t\s$ is computable.\cqfd
\end{description}
\end{itemize}
\end{prf}

For instance, we have $\{f,g\}\sle\CC_\so$ by (arg), $x\in\CC_\so$ by
(var), $f(gx)\in\CC_\so(f,g)$ by (app) twice, and $\l
xf(gx)\in\CC_\so(f,g)$ by (abs).

%%%%%%%%%%%%%%%%%%%%%%%%%%%%%%%%%%%%%%%%%%%%%%%%%%%%%%%%%%%%%%%%%%%%%%%%%%%%%%
\subsection{Handling basic subterms}
\label{sec-basic-subterm}

Consider now the following definition on unary natural numbers (Peano
integers):

\begin{center}
$\sz:\sN;\quad
\ss:\sN\A\sN\quad$\\[2mm]
\begin{rew}
\pred~\sz&\sz\\
\pred~(\ss~x)&x\\
\end{rew}
\end{center}

In order to handle this definition, we need to extend the
computability closure with some subterm operation. Unfortunately,
$\tgt_\ss$ does not always preserve computability as shown by the
following example:\footnote{Note that the rule $\sf~(\sc~x)\a x$ means
  that $\sc$ is injective and thus that, in a set-theoretical
  interpretation of types, the cardinality of the function space
  $\sA\a\sB$ is smaller than or equal to the cardinality of $\sA$,
  which is hardly possible if $\sB$ is of cardinality greater than or
  equal to 2.}

\begin{center}
$\sf:\sA\A(\sA\A\sB);\quad\sc:(\sA\A\sB)\A\sA$\\[2mm]
\begin{rew}
\sf~(\sc~y)&y\\
\end{rew}
\end{center}

Indeed, with $w=\l x\sf xx$, we have $w(\sc w)\ab\sf(\sc w)(\sc w)\ar
w(\sc w)\ab\ldots$ \cite{mendler87phd}. Therefore, $\sc w\in\SN$ but
$w\notin{\propto\!(\SN,\SN)}$ since $w(\sc w)\notin\SN$.

On the other hand, $\tgt_\ss$ preserves termination. Hence, we can add
the operation of Figure \ref{fig-subterm-sn} for basic subterms (we
omit the proof).

\begin{figure}[ht]
\caption{Computability closure operations III\label{fig-subterm-sn}}
\begin{center}
\fbox{\begin{tabular}{rl}
(subterm-basic)&if $t\in\CC_\sf(\vl)$, $t\tgt_\ss u:\sB\in\cB$ and
  $\I\sB=\SN^\sB$, then $u\in\CC_\sf(\vl)$
\end{tabular}}
\end{center}
\end{figure}

Hence, to handle the above predecessor function definition, it is
enough to take $I(\sN)=\SN^\sN$.

In Sections \ref{sec-ho-subterm} and \ref{sec-hopm}, we will see
other computability-preserving subterm operations.

%%%%%%%%%%%%%%%%%%%%%%%%%%%%%%%%%%%%%%%%%%%%%%%%%%%%%%%%%%%%%%%%%%%%%%%%%%%%%%
\subsection{Handling recursive functions}
\label{sec-rec}

Consider now a simple recursive function definition:

\begin{center}
$+:\sN\A\sN\A\sN$\\[2mm]
\begin{rew}
\sz+y&y\\
(\ss~x)+y&\ss~(x+y)\\
\end{rew}
\end{center}

For handling such a recursive definition, and more generally mutually
recursively defined functions, we need to extend $\CC_\sf(\vl)$ with
terms of the form $\sg\vm$. In order to ensure termination, we can try
to use some well-founded DLQO $\ge$ on $\S$ (DLQOs are defined in
Definition \ref{def-dlqo} and $\S$ just before Theorem \ref{thm-comp})
so that $(\sf,\vl)>(\sg,\vm)$ and prove by induction on $>$ that $\CC$
preserves computability and defined function symbols are
computable. However, we cannot consider arbitrary DLQOs. Indeed, since
we consider curried symbols and, by (app), adding arguments preserves
termination, the number of arguments is not a valid termination
criterion as shown by the following example:

\begin{center}
$\val:(\sN\A\sN)\A\sN;\quad\sf:\sN\A(\sN\A\sN)$\\[2mm]
\begin{rew}
\val~x&x~\sz\\
\sf~x~\sz&\val~(\sf~x)\\
\end{rew}
\end{center}

\noindent
where $\sf~x~\sz\a\val~(\sf~x)\a\sf~x~\sz\a\ldots$

We therefore need to consider DLQOs compatible with application:

\begin{dfn}[$\cF$-quasi-ordering]
\label{def-f-quasi-ord}
An {\em $\cF$-quasi-ordering} is a DLQO on $\S$. A relation $R$ on $\S$ is:
\begin{itemize}\itemsep=0mm
\item {\em compatible with application} if, for all
  $(\sf,\vl\vv),(\sg,\vm\vw)\in\S_\max$, $(\sf,\vl\vv)R(\sg,\vm\vw)$
  whenever $(\sf,\vl)R(\sg,\vm)$;
\item {\em stable by substitution} if $(\sf,\vl\s)R(\sg,\vm\s)$
  whenever $(\sf,\vl)R(\sg,\vm)$, $\s$ is away from
  $\FV(\vm)-\FV(\vl)$ and $\vl\s,\vm\s$ are computable.
\end{itemize}
\end{dfn}

A simple way to get an $\cF$-quasi-ordering compatible with
application is to restrict comparisons to pairs $(\sf,\vt)$ such that
$\sf$ is maximally applied in $\sf\vt$. For instance, given a
quasi-ordering $\ge$ on terms, the DLQO associated to:
\begin{itemize}
\item the identity relation on $\cF$;
\item for each equivalence class $E$ modulo identity, the
  quasi-ordering $\ge_\prod$ (resp. $\ge_\mul$);
\item for each symbol $\sf$, the identity function $\psi_\sf(\vt)=\vt$
  if $\sf$ is maximally applied in $\sf\vt$;
\end{itemize}
is an $\cF$-quasi-ordering compatible with application, that is stable
by substitution if $\ge$ so is. For the sake of simplicity, we
also denote such a DLQO by $\ge_\prod$ (resp. $\ge_\mul$) and its
strict part by $>_\prod$ (resp. $>_\mul$). Hence, $\a^*_\prod$,
$(\tge_\ss)_\mul$ and, more generally,
$(\a^*\!\tge_\ss)_\mul$\footnote{$\a^*\!\tge_\ss$ is the smallest
  quasi-ordering containing both $\a$ and $\tgt_\ss$. Its strict part
  on $\SN$ is ${\a^+}\cup{\a^*\tgt_\ss}$.} are $\cF$-quasi-orderings
compatible with application and stable by substitution. For the sake
of simplicity, we will denote $(\a^+)_\prod$ by $\a_\prod$.

\begin{dfn}[Valid $\cF$-quasi-ordering]
  A quasi-ordering $\ge$ on terms is {\em compatible with reduction}
  if ${>}\cup{\a}$ is well-founded, where $>$ is the strict part of
  $\ge$. It is {\em valid} if, moreover, $>$ and the equivalence
  relation associated to $\ge$ are both stable by substitution.

  An $\cF$-quasi-ordering $\ge$ is {\em compatible with reduction} if
  ${>}\cup{\a_\prod}$ is well-founded on $\S_\max$, where $>$ is the
  strict part of $\ge$. It is {\em valid} if, moreover, $>$ and the
  equivalence relation associated to $\ge$ are both stable by
  substitution and compatible with application.
\end{dfn}

Note that ${>}\cup{\a_\prod}$ is required to be well-founded on
$\S_\max$, that is, on pairs $(\sf,\vt)$ such that $\sf:\vT\A\sA$,
$\sA\in\cB$ and $\vt\in\I\vT$. Note also that, by \compsn, $\a_\prod$
is well-founded on computable terms.

For instance, $\ge_\prod$ and $\ge_\mul$ are both valid if $\ge$ so
is. In particular, $\a^*_\prod$ and $(\a^*\!\tge_\ss)_\mul$ are both
valid ($\tgt_\ss$ commutes with $\a$ since $\a$ is monotone).

In the next subsection, we will give another example of valid
$\cF$-quasi-ordering.

With a valid $\cF$-quasi-ordering, we can add the operation of Figure
\ref{fig-cc-rec}.

\begin{figure}[ht]
\caption{Computability closure operations IV\label{fig-cc-rec}}
\begin{center}
\fbox{\begin{tabular}{rl}
(rec)&if $\sg:\vM\A U$, $\vm:\vM$, $\vm\in\CC_\sf(\vl)$
and $(\sf,\vl)>(\sg,\vm)$, then $\sg\vm\in\CC_\sf(\vl)$\\
\end{tabular}}
\end{center}
\end{figure}

\begin{lem}
\label{lem-cc}
Let $\ge$ be a valid $\cF$-quasi-ordering, $(\sf,\vt)\in\S_\max$ and
assume that, for all $(\sg,\vu)\in\S_\max$ such that
$(\sf,\vt)>(\sg,\vu)$, $\sg\vu$ is computable. Then, for all $\vl$,
$\vw$, $t$ and $\t$ such that $\vt=\vl\vw$, $\t$ is computable,
$\dom(\t)\sle\FV(t)-\FV(\vl)$ and $t\in\CC_\sf(\vl)$, where $\CC$ is
the smallest computability closure closed by the operations I to IV,
we have $t\t$ computable.
\end{lem}

\begin{prf}
  We proceed as for Theorem \ref{thm-cc-bv} by proving that, for all
  $t\in\CC_\sf(\vl)$ and computable $\t$ such that
  $\dom(\t)\sle\FV(t)-\FV(\vl)$, $t\t$ is computable, by induction on
  $\CC$, but only detail the new case:
  \begin{description}
  \item[(rec)] We have $\vm\t$ computable by the induction
    hypothesis. Since $\dom(\t)\sle\FV(\sg\vm)-\FV(\vl)$,
    $\vl\t=\vl$. Since $>$ is stable by substitution, we have
    $(\sf,\vl)>(\sg,\vm\t)$. Assume now that $U=\vV\A\sB$ and let
    $\vv\in\I\vV$. Since $>$ is compatible with application, we have
    $(\sf,\vl\vw)>(\sg,\vm\t\vv)$. Hence, by assumption, $\sg\vm\t\vv$
    is computable. Therefore, $\sg\vm\t$ is computable.\cqfd
  \end{description}
\end{prf}

\begin{thm}
\label{thm-cc-rec}
The relation ${\ab}\cup{\ar}$ terminates on well-typed terms if there
are $I\in\Pi_{\sB\in\cB}\Red_\cR^\sB$ and a valid $\cF$-quasi-ordering
$\ge$ such that:
\begin{itemize}
\item every non-basic undefined symbol is computable;
\item for every rule $\sf\vl\a r\in\cR$, we have $r\in\CC_\sf(\vl)$,
  where $\CC$ is the smallest computability closure closed by the
  operations I to IV.
\end{itemize}
\end{thm}

\begin{prf}
  We follow the proof of Theorem \ref{thm-cc} that, for all
  $(\sf,\vt)\in\S_\max$ with $\sf\in\cD(\cR)$, every reduct $t$ of
  $\sf\vt$ is computable, but proceed by induction on
  ${>}\cup{\a_\prod}$. There are two cases:

\begin{itemize}
\item There is $\vu$ such that $t=\sf\vu$ and $\vt\a_\prod\vu$. By
  \compred, $\vu$ is computable. Therefore, by the induction
  hypothesis, $\sf\vu$ is computable.

\item There are $\vw$, $\sf\vl\a r\in\cR$ and $\s$ such that
  $\vt=\vl\s\vw$ and $t=r\s\vw$. Since $r\in\CC_\sf(\vl)$ and $\CC$ is
  stable by substitution (for $>$ is stable by substitution), we have
  $r\s\in\CC_\sf(\vl\s)$. Thus, by Lemma \ref{lem-cc}, $r\s$ is
  computable since, for all $(\sg,\vu)\in\S_\max$, if
  $(\sf,\vt)>(\sg,\vu)$, then $\sg\vu$ is computable by the induction
  hypothesis.\cqfd
\end{itemize}
\end{prf}

As a consequence, by taking the basic interpretation for $I$, we get:

\begin{cor}
\label{cor-cc-rec}
The relation $\ab\cup\ar$ terminates on well-typed terms if, for every
rule $\sf\vl\a r\in\cR$, we have $r\in\CC_\sf(\vl)$, where $\CC$ is
the smallest computability closure closed by the operations I to IV,
and $\ge$ is any $\cF$-quasi-ordering valid wrt the basic
interpretation.
\end{cor}

In the first paper implicitly using the notion of computability
closure for higher-order rewriting
\cite{jouannaud91lics,jouannaud97tcs}, Jouannaud and Okada take the
basic interpretation for $I$ and define for $\CC$ a schema
generalizing G\"odel' system T recursion schema on Peano integers
\cite{godel58} to arbitrary first-order data types. This schema is
included in any computability closure closed by the operations I to IV
with the $\cF$-quasi-ordering $(\tge_\ss)_\stat$ defined in the next
subsection. The present inductive formulation first appeared in
\cite{blanqui99rta,blanqui02tcs}. In Section \ref{sec-cc-rec-ex}, we
provide various examples of systems that can be proved terminating by
using this corollary.

%%%%%%%%%%%%%%%%%%%%%%%%%%%%%%%%%%%%%%%%%%%%%%%%%%%%%%%%%%%%%%%%%%%%%%%%%%%%%%
\subsubsection{Examples of valid $\cF$-quasi-orderings}
\label{sec-f-qo}

We have seen that a simple way to get an $\cF$-quasi-ordering
compatible with application is to only compare terms of base
type. Another way is to always compare the same fixed subset of
arguments by using a particular case of {\em arguments filtering
  system} (AFS) \cite{arts00tcs}:

\begin{dfn}[Arguments filtering system]
  A {\em filter} is a word on $\bN-\{0\}$. The arity of a filter
  $\vphi=k_1\ldots k_n$ is
  $\N\vphi=\mathop{max}\{0,k_1,\ldots,k_n\}$. A word $w$ is compatible
  with a filter $\vphi$ if $|w|\ge\N\vphi$. We denote by $\vphi^A$ the
  function mapping every word $\va\in A^*$ compatible with
  $\vphi=k_1\ldots k_n$ to $a_{k_1}\ldots a_{k_n}$. An {\em arguments
    filtering system} (AFS) is a function $\vphi$ providing, for each
  $\sf\in\cD(\cR)$, a filter $\vphi_\sf$ of arity
  $\N{\vphi_\sf}\le\al_\sf$.
\end{dfn}

An AFS describes, for each symbol $\sf$, which arguments, and in which
order, these arguments must be compared. For instance, if
$\vphi_\sf=322$, then
$\vphi_\sf^\cL(t_1t_2t_3\ldots)=t_3t_2t_2$. Hence, when comparing
$(\sf,t_1t_2t_3)$ and $(\sf,u_1u_2u_3u_4)$, one in fact compares
$t_3t_2t_2$ and $u_3u_2u_2$ only.

Following \cite{dershowitz79focs,kamin80note}, here is an
$\cF$-quasi-ordering allowing both multiset and lexicographic
comparisons depending on a function $\stat:\cF\a\{\lex,\mul\}$:

\begin{dfn}[Status $\cF$-quasi-ordering]
\label{def-stat}
Given a quasi-ordering $\ge$ on terms, a quasi-ordering $\ge_\cF$ on
$\cF$, an AFS $\vphi$ and a function $\stat:\cF\a\{\lex,\mul\}$
compatible with $\simeq_\cF$, \ie such that:
\begin{itemize}
\item $\stat_\sf=\stat_\sg$ whenever $\sf\simeq_\cF\sg$;
\item $|\vphi_\sf|=|\vphi_\sg|$ whenever $\sf\simeq_\cF\sg$ and $\stat_\sf=\lex$;
\end{itemize}
\noindent
let $\ge_\stat$ be the DLQO associated to:
\begin{itemize}
\item the quasi-ordering $\ge_\cF$ on $\cF$;
\item for each equivalence class $E$ modulo $\simeq_\cF$ of status
  $\mul$ (resp. $\lex$), the quasi-ordering $\ge_\mul$
  (resp. $\ge_\lex$);
\item for each symbol $\sf$, the function
  $\psi_\sf(\vt)=\vphi_\sf^\cL(\vt)$.
\end{itemize}
\end{dfn}

The $\cF$-quasi-ordering $\ge_\stat$ is valid whenever $\ge$ so is. In
particular, $(\a^*\!\tge_\ss)_\stat$ is valid.

%%%%%%%%%%%%%%%%%%%%%%%%%%%%%%%%%%%%%%%%%%%%%%%%%%%%%%%%%%%%%%%%%%%%%%%%%%%%%%
\subsubsection{Examples of termination proofs based on computability closure}
\label{sec-cc-rec-ex}

With the above closure operations, one can already prove the
termination of a large class of rewrite systems including:

\begin{itemize}
\item G\"odel system T \cite{godel58}:

\begin{center}
$\rec_\sN^T:\sN\A T\A(\sN\A T\A T)\A T$, for every type $T$\\[2mm]
\begin{rew}
\rec_\sN^T~\sz~u~v&u\\
\rec_\sN^T~(\ss~x)~u~v&v~x~(\rec_\sN^T~x~u~v)\\
\end{rew}
\end{center}

To give an example, let us detail why the right-hand side of the
second rule is in the computability closure of the left-hand. We take
the identity relation on $\cF$ for $\ge_\cF$, $\vphi_{\rec_\sN^T}=1$
as AFS (only the first argument of $\rec_\sN^T$ will be used in
comparisons), and $\stat_{\rec_\sN^T}=\lex$. Then, we have
$\{\ss~x,u,v\}\sle\CC=\CC_{\rec_\sN^T}(\ss~x,u,v)$ by (arg),
$x\in\CC$ by (subterm-basic), $\rec_\sN^T~x~u~v\in\CC$ by (rec) for
$\ss~x\tgt_\ss x$, and $v~x~(\rec_\sN^T~x~u~v)\in\CC$ by (app) twice.

\item Ackermann's function:

\begin{center}
$\ack:\sN\A\sN\A\sN$\\[2mm]
\begin{rew}
\ack~\sz~n&\ss~n\\
\ack~(\ss~m)~\sz&\ack~m~(\ss~\sz)\\
\ack~(\ss~m)~(\ss~n)&\ack~m~(\ack~(\ss~m)~n)\\
\end{rew}
\end{center}

One can easily check that, for each rule, its right-hand side is in
the computability closure of its left-hand side by taking
$\vphi_\ack=12$ and $\stat_\ack=\lex$.

\item The following non-orthogonal set of rules for subtraction on
  unary natural numbers:

\begin{center}
$-:\sN\A\sN\A\sN$\\[2mm]
\begin{rew}
\sz-x&\sz\\
x-\sz&x\\
(\ss~x)-(\ss~y)&x-y\\
x-x&\sz\\
\end{rew}
\end{center}

can also be proved terminating by taking $\vphi_\sub=1$ and
$\stat_\sub=\lex$.

\item Here is an example of a rule for computing subtyping constraints
  on simple types that requires multiset comparisons (take
  $\vphi_\le=12$ and $\stat_\le=\mul$):

\begin{center}
$\arrow:\sT\A\sT\A\sT;\quad
\le\;:\sT\A\sT\A\sC;\quad
\et:\sC\A\sC\A\sC$\\[2mm]
\begin{rew}
\arrow~x~y\le \arrow~x'~y'&x'\le x\et y\le y'
\end{rew}
\end{center}

\item Here is an example of mutually defined functions requiring a
  true quasi-ordering on function symbols
  ($\height_\sT\simeq_\cF\height_\sF$):

\begin{center}
$\nil:\sF;\quad
\cons:\sT\A\sF\A\sF;\quad
\leaf:\sT;\quad
\node:\sF\A\sT;\quad
\height_\sT:\sT\A\sN;\quad
\height_\sF:\sF\A\sN$\\[2mm]
\begin{rew}
\height_\sF~\nil&\sz\\
\height_\sF~(\cons~t~f)&\max~(\height_\sT~t)~(\height_\sF~f)\\
\height_\sT~\leaf&\sz\\
\height_\sT~(\node~f)&\ss~(\height_\sF~f)\\
\end{rew}
\end{center}

\item Finally, here is an example showing that the operations I to IV
  can already handle rules with matching on basic {\em defined}
  symbols (we will see the case of {\em non-basic} defined symbols in
  Section \ref{sec-match-def}):

\begin{center}
$\times:\sN\A\sN\A\sN$\\[2mm]
\begin{rew}
\sz+y&y\\
(\ss~x)+y&x+(\ss~y)\\
(x+y)+z&x+(y+z)\\
\sz\times y&\sz\\
(\ss~x)\times y&(x\times y)+y\\
(x+y)\times z&(x\times z)+(y\times z)\\
\end{rew}
\end{center}
\end{itemize}

%%%%%%%%%%%%%%%%%%%%%%%%%%%%%%%%%%%%%%%%%%%%%%%%%%%%%%%%%%%%%%%%%%%%%%%%%%%%%%
\subsection{Handling higher-order subterms}
\label{sec-ho-subterm}

The closure operations presented so far do not enable us to deal with
functions defined by induction on higher-order inductive types, that
is, on inductive types with constructors taking functions as
arguments. Here are some examples:

\begin{itemize}
\item The ``addition'' on the following (type theoretic) ordinal
  notation \cite{coquand88colog}:
\begin{center}
$\zero:\sO\quad
\suc:\sO\A\sO\quad
\lim:(\sN\A\sO)\A\sO\quad
+:\sO\A\sO\A\sO$\\[2mm]
\begin{rew}
\zero+y&y\\
(\suc~x)+y&\suc~(x+y)\\
(\lim~x)+y&\lim~(\l n~(x~n)+y)\\
\end{rew}
\end{center}

\item The computation of the prenex normal form in the predicate
  calculus \cite{mayr98tcs}:

\begin{center}
$\bot,\top:\sF;\quad
\neg:\sF\A\sF;\quad
\et,\ou:\sF\A\sF\A\sF;\quad
\all,\ex:(\sT\A\sF)\A\sF$\\[2mm]
\begin{rew}
(\all~P)\et Q&\all~(\l x~(P~x)\et Q)\\
\neg~(\all~P)&\ex~(\l x~\neg~(P~x))\quad\ldots\\
\end{rew}
\end{center}

\item The list of labels of a tree in breadth-first order using
  continuations (we only give the definition of one of the functions)
  \cite{hofmann95note}:

\begin{center}
$\nil:\sL;\quad
\cons:\sN\A\sL\a\sL;\quad
\sd:\sC;\quad
\sc:((\sC\A\sL)\A\sL)\a\sC;\quad
\fex:\sC\A\sL$\\[2mm]
\begin{rew}
\fex~\sd&\nil\\
\fex~(\sc~x)&x~\fex\\
\end{rew}
\end{center}
\end{itemize}

Indeed, in all these examples, there are two problems. First, we need
the higher-order arguments of a computable function-headed term to be
computable, \eg $x$ in $(\lim~x)$. Second, we need to have a DLQO in
which $(\lim~x)$ is bigger than $(x~n)$, where $n$ is a bound
variable.

But we have already seen in Section \ref{sec-basic-subterm} that the
first property is not always satisfied. Fortunately, under some
conditions, it is possible to define an interpretation $I$ satisfying
this property by using the fact that $\Red_\cR$ is a complete lattice
(as seen in Section \ref{sec-comp}) on which, therefore, any monotone
function has a fixpoint \cite{tarski55pjm}. Following
\cite{matthes98phd}, two different definitions are possible that we
illustrate with the type $\sO$ of ordinals:\footnote{These definitions
  can be generalized to any {\em positive} inductive type (see
  Definition \ref{def-pos} just after)
  \cite{mendler87phd,blanqui02tcs}.}

\begin{itemize}
\item An {\em elimination-based} definition using recursor
  symbols. For instance, for $\sO$, one can define the family of
  recursor symbols $\rec_\sO^T$ indexed by $T\in\cT$ as follows:

\begin{center}
$\rec_\sO^T:\sO\A T\A(\sO\A T\A T)\A((\sN\A\sO)\A(\sN\A T)\A T)\A T$\\[2mm]
\begin{rew}
\rec_\sO^T~\zero~u~v~w&u\\
\rec_\sO^T~(\suc~x)~u~v~w&v~x~(\rec_\sO^T~x~u~v~w)\\
\rec_\sO^T~(\lim~x)~u~v~w&w~x~(\l n~\rec_\sO^T~(x~n)~u~v~w)\\
\end{rew}
\end{center}

\noindent
and define $I(\sO)$ as some fixpoint of the following monotone function:

\begin{center}
$F_\sO(X)=\{t\in\cL\mid\all T\in\cT,\all P\in\Red_\cR^T,\all u\in\I\sA^J,
\all v\in\I{\sO\A\sA\A\sA}^J,$\\
\hsp[26mm]$\all w\in\I{(\sN\A\sO)\A(\sN\A\sA)\A\sA}^J,
\rec_\sO^T~t~u~v~w\in\I\sA^J\}$\\[3mm]
where $\sA$ is a type constant distinct from $\sO$ and
$\sN$\footnote{We assume that $\cB$ is infinite. Alternatively, we
  could consider type variables.},\\$J(\sA)=P$, $J(\sO)=X$ and
$J(\sN)=I(\sN)$
\end{center}

\noindent
The computability of $\rec_\sO^T$ directly follows from the definition
of $I(\sO)$. And for proving that $x$ is computable if $(\lim\,x)$ so
is, it suffices to take $T=\sO$ and $w=\l x\l yx$ which is clearly
computable. Indeed, in this case, $\rec_\sO^\sO(\lim\,x)uvw\a wx(\l
n\rec_\sO^\sO(xn)uvw)\a x$ and we can conclude by \compred. Finally,
proving that constructors are computable is no more complicated.

\item An {\em introduction-based} definition using constructors
  only. In this approach, $I(\sO)$ is defined as some fixpoint of the
  following monotone function:

\begin{center}
$F_\sO(X)=\{t\in\SN\mid\all u,(t\a^*\suc~u\A u\in X)\et(t\a^*\lim~u\A
u\in\I{\sN\A\sO}^J)\}$\\[3mm]
where $J(\sO)=X$ and $J(\sN)=I(\sN)$
\end{center}

\noindent
In this case, the computability of constructor arguments directly
follows from the definition of $I(\sO)$.
\end{itemize}

In \cite{matthes98phd}, p. 116-117, Matthes proves that, when using
saturated sets, the introduction-based interpretation is included into
the elimination-based interpretation and provides an example of type
for which the two interpretations are distinct, by using the fact that
some saturated sets are not stable by reduction. It is not too
difficult to check that this cannot happen with reducibility
candidates.

Anyway, in both cases, the monotony of $F_\sO$ is due to the fact that
$\sO$ occurs only {\em positively} in the types of the arguments of
the constructors of $\sO$, knowing that $\sA$ occurs {\em positively}
in $\sB\A\sA$ and {\em negatively} in $\sA\A\sB$. More formally:

\begin{dfn}[Positive and negative positions]
\label{def-pos}
Given a type $T$, the {\em positive} (resp. {\em negative}) positions
of $T$, $\Pos^+(T)$ (resp. $\Pos^-(T)$), are the subsets of
$\{0,1\}^*$ defined as follows:

\begin{itemize}
\item
$\Pos^+(\sB)=\{\vep\}$
\item
$\Pos^-(\sB)=\vide$
\item
$\Pos^+(T\A U)=\{0w\mid w\in\Pos^-(T)\}\cup\{1w\mid w\in\Pos^+(U)\}$
\item
$\Pos^-(T\A U)=\{0w\mid w\in\Pos^+(T)\}\cup\{1w\mid w\in\Pos^-(U)\}$
\end{itemize}

\noindent
And the positions in a type $T$ of the occurrences of a type constant
$\sB$, $\Pos(\sB,T)$, are:

\begin{itemize}
\item
$\Pos(\sB,\sB)=\{\vep\}$
\item
$\Pos(\sB,\sC)=\vide$ if $\sB\neq\sC$
\item
$\Pos(\sB,T\A U)=\{0w\mid w\in\Pos(\sB,T)\}\cup\{1w\mid w\in\Pos(\sB,U)\}$
\end{itemize}
\end{dfn}

This leads to the following common restrictions one can for instance
find in the Calculus of Inductive Constructions (CIC)\footnote{In
  fact, in CIC, inductive types are even restricted to {\em
    strictly-positive} inductive types (see Definition
  \ref{def-struct}) for termination may be lost when considering some
  {\em polymorphic} non-strictly positive types
  \cite{coquand88colog}.} \cite{coquand88colog,werner94phd} and proof
assistants based on CIC like Agda \cite{bove09tphol}, Coq \cite{coq}
or Matita \cite{asperti11cade}:

\begin{dfn}[Standard inductive system]
\label{def-std-ind-sys}
  Given a set $\cR$ of rewrite rules, the set of type constants $\cB$
  and the set of undefined function symbols $\cF-\cD(\cR)$
  (constructors) form a {\em standard inductive system} if there is a
  well-founded quasi-ordering $\ge_\cB$ on $\cB$ such that, for all
  $\sB\in\cB$, $\sc\in\cF-\cD(\cR)$, $\sc:\vT\A\sB$, $i\in[1,|\vT|]$
  and $\sC$ occurring in $T_i$, either $\sC<_\cB\sB$ or else
  $\sC\simeq_\cB\sB$ and $\Pos(\sC,T_i)\sle\Pos^+(T_i)$.
\end{dfn}

Taking a quasi-ordering instead of an ordering allows us to deal with
mutually defined inductive types. However, in this case, one has to
reason on equivalence classes modulo $\simeq_\cB$ because, if
$\sB\simeq_\cB\sC$, then the interpretation of $\sB$ and the
interpretation of $\sC$ have to be defined at the same time.

In such a system, one can define $I\in\Pi_{\sB\in\cB}\Red_\cR^\sB$ by
induction on $>_\cB$ and, for each equivalence class $E$ modulo
$\simeq_\cB$, as some fixpoint $S_E$ of a monotone function $F_E$
(similar to the function $F_\sO$ above) on the complete lattice
$E\a\Red_\cR$ ordered point-wise by inclusion ($I\le J$ if, for all
$\sB\in E$, $I(\sB)\sle J(\sB)$). See Lemma 14 in \cite{blanqui02tcs}
or Section 6.3 in \cite{blanqui05mscs} for more details about
that. For each type constant $\sB$, $I(\sB)$ is then defined as
$S_{[\sB]_{\simeq_\cB}}(\sB)$.

With this interpretation, all the symbols $\sf\in\cF-\cD(\cR)$
(constructors) are computable and one can add to the computability
closure the operations of Figure \ref{fig-cc-std-ind}.

\begin{figure}[ht]
\caption{Computability closure operations V for standard inductive systems (Definition \ref{def-std-ind-sys})\label{fig-cc-std-ind}}
\begin{center}
\fbox{\begin{tabular}{rl}
(undef)&${\cF-\cD(\cR)}\sle\CC_\sf(\vl)$\\
(subterm-undef)&if $\sg\vt\in\CC_\sf(\vl)$, $\sg\vt:\sB$
and $\sg\in\cF-\cD(\cR)$, then $\{\vt\}\sle\CC_\sf(\vl)$\\
\end{tabular}}
\end{center}
\end{figure}

However, the stable subterm ordering is not sufficient to prove the
termination of the systems given above. For instance, for the addition
on $\sO$, starting from an argument of the form $(\lim~x)$, we have a
recursive call with an argument of the form $(x~n)$ where $n$ is a
bound variable. Although $x$ is a subterm of $(\lim~x)$, $(x~n)$ is
not. In the case of continuations, this is even worse: starting from
an argument of the form $(\sc~x)$, the function $\fex$ is applied to
no argument but is itself argument of $x$\ldots

If, for $S_E$, we take the {\em smallest} fixpoint of $F_E$ (the set
of fixpoints is itself a complete lattice \cite{tarski55pjm}), then it
can be obtained by transfinite iteration \cite{cousot79pjm}: there is
an ordinal $\ka$ such that, for all $\sB\in E$, $S_E=F_E^\ka(\bot_E)$
where $\bot_E$ is the smallest element of $E\a\Red_\cR$ and $F_E^\ka$
is defined by transfinite induction:

\begin{itemize}
\item
$F_E^0(X)=X$
\item
$F_E^{\ka+1}(X)=F_E(F_E^\ka(X))$
\item
$F_E^\kl(X)=\bigcup\{F_E^\ka(X)\mid\ka<\kl\}$ if $\kl$ is a limit ordinal
\end{itemize}

This provides us with a notion of rank to compare computable terms:

\begin{dfn}[Rank of a computable term]
\label{def-rank}
  The {\em rank} of a term $t\!\in\!I(\sB)$, $\rk_\sB(t)$, is the
  smallest ordinal $\ka$ such that $t\in
  F_{[\sB]_{\simeq_\cB}}^\ka\hsp[-1mm](\bot_{[\sB]_{\simeq_\cB}})(\sB)$. Let
  $\succeq_\sB$ be the quasi-ordering on $I(\sB)$ such that
  $t\succeq_\sB u$ if $\rk_\sB(t)\ge\rk_\sB(u)$.
\end{dfn}

Note that some terms may have a rank bigger than $\w$. For instance,
with $\si:\sN\A\sO$ defined by the rules $\si~\sz\a\zero$ and
$\si(\ss~n)\a\suc(\si~n)$, we have $\rk_\sO(\lim~\si)=\w+1$.

The relation $\succ_\sB$ is compatible with reduction since
$t\succeq_\sB u$ whenever $t\in\I\sB$ and $t\a u$ (reduction cannot
increase the rank of a term by \compred). However, it is not stable by
substitution. For instance, $\ss~\sz>_\sO y$ for
$\rk_\sO(\ss~\sz)=1$ and $\rk_\sO(y)=0$, but
$\ss~\sz<_\sO\ss~(\ss~\sz)$ for
$\rk_\sO(\ss~(\ss~\sz))=2$. Restricting $t\succ_\sB u$ to the
cases where $\FV(u)\sle\FV(t)$ is not a solution since, with the
addition on $\sO$, we have to compare $(\lim~x)$ and $(x~n)$. Instead,
we will consider a sub-quasi-ordering of $\succeq_\sB$ due to Coquand
\cite{coquand92types} that is valid (in a sense that will be precised
after the definition) and, in which, $(\lim~x)$ is bigger than
$(x~n)$:

\newcommand\tgtacc{\tgt_\ss^{\ms{acc}}}

\begin{dfn}[Structural subterm ordering]
\label{def-struct}
The $i$-th argument of $\sc:\vT\A\sB$ is {\em strictly positive} if
$T_i$ is of the form $\vU\A\sC$ with $\sC\simeq\sB$ and, for all $\sD$
occurring in $\vU$, $\sD<_\cB\sB$. Let $\tgtacc$ be the smallest
sub-ordering of $\tgt_\ss$ such that, for all $\sc:\vT\A\sB$,
$\vt:\vT$ and $i\in[1,|\vt|]$, we have $\sc\vt\tgtacc t_i$ if the
$i$-th argument of $\sc$ is strictly positive. Given a term of the
form $\sf\vl$, a term $t:T$ is {\em structurally bigger} than a term
$u:U$, written $t>_\ss^{\sf\vl}u$, if $T$ and $U$ are equivalent type
{\em constants} and there are $v$ and $\vx\in\cX-\FV(\vl)$ such that
$t\tgtacc v$ and $u=v\vx$.\footnote{We could improve this definition
  by taking $\vx\in\CC_\sf(\vl)$ instead of $\vx\in\cX-\FV(\vl)$ only
  \cite{blanqui02tcs,blanqui06tr}.} Finally, let $\ge_\ss^{\sf\vl}$ be
the reflexive closure of $>_\ss^{\sf\vl}$.
\end{dfn}

For instance, $\lim~x>_\ss^{(\lim~x)+y}x~n$ for $\lim~x:\sO$,
$x~n:\sO$, $\lim~x\tgtacc x$ and $n\in\cX-\FV((\lim~x)+y)$.

The relation $>_\ss^{\sf\vl}$ is valid in the following generalized
sense. First, $l_i\s>_\ss^{\sf\vl\s}u\s$ whenever
$l_i>_\ss^{\sf\vl}u$, $\dom(\s)\sle\FV(\vl)$ and $\s$ is away from
$\FV(u)-\FV(\vl)$. Second, if $l_i>_\ss^{\sf\vl}u$, $l_i:\sB$ is
computable, $\t$ is computable and $\dom(\t)\sle\FV(u)-\FV(\vl)$, then
$u\t:\sB$ is computable and $l_i\succ_\sB u\t$ (see Lemma 18 in
\cite{blanqui02tcs} or Lemma 54 in \cite{blanqui05mscs}). Hence, by
adapting Lemma \ref{lem-cc}, we can provide an instance of Theorem
\ref{thm-cc-rec} able to handle functions defined by induction on the
structural subterm ordering, by using a status $\cF$-quasi-ordering
compatible with the rank ordering (that is defined on terms of the
same computability predicate only):

\begin{dfn}
\label{def-rank-comp}
An AFS $\vphi$ and a map $\stat:\cF\a\{\lex,\mul\}$ compatible with an
equivalence relation $\simeq_\cF$ on $\cF$ are {\em compatible with
  the rank ordering} when the following conditions are satisfied:
\begin{itemize}
\item if $E$ is an equivalence class modulo $\simeq_\cF$ of status
  $\mul$, then there is a constant type $\sB^E$ such that, for all
  $\sf\in E$ with $\sf:\vT\A\sA$ and $\vphi_\sf=k_1\ldots k_n$, we
  have $T_{k_i}=\sB^E$ for every $i\in[1,n]$;
\item if $E$ is an equivalence class modulo $\simeq_\cF$ of status
  $\lex$, then there is a sequence of constant types $\vec\sB^E$ such
  that, for all $\sf\in E$ with $\sf:\vT\A\sA$, we have
  $\vphi_\sf^\cT(\vT)=\vec\sB^E$.
\end{itemize}
\end{dfn}

\begin{thm}
\label{thm-cc-rec-stat}
In a standard inductive system, the relation ${\ab}\cup{\ar}$
terminates on well-typed terms if there are a well-founded
quasi-ordering $\ge_\cF$ on $\cF$, an AFS $\vphi$ and a status map
$\stat$ compatible with $\simeq_\cF$ and the rank ordering such that,
for every rule $\sf\vl\a r\in\cR$, we have $r\in\CC_\sf(\vl)$, where
$\CC$ is the smallest computability closure closed by the operations I
to V with, in (rec),
${>}={(\a^*\ge_\ss^{\sf\vl})_\stat}$.\footnote{Here, we in fact
  consider a family of $\cF$-quasi-orderings indexed by $\sf\vl$.}
\end{thm}

\begin{prf}
  By adapting Lemma \ref{lem-cc}, we can follow the proof of Theorem
  \ref{thm-cc-rec} but proceed by induction on the DLQO $\succ_\stat$
  associated to:
\begin{itemize}
\item the quasi-ordering $\ge_\cF$ on $\cF$;
\item for each equivalence class $E$ modulo $\simeq_\cF$ of status
  $\mul$ (resp. $\lex$ with $|\vec\sB^E|=n$), the quasi-ordering
  $(\succ_{\sB^E})_\mul$
  (resp. $(\succ_{\sB^E_1},\ldots,\succ_{\sB^E_n})_\lex$);
\item for each symbol $\sf$, the function
  $\psi_\sf(\vt)=\vphi_\sf^\cL(\vt)$,
\end{itemize}
\noindent
which is compatible with application and reduction.\cqfd
\end{prf}

Using this theorem, we can prove the termination of the first two
examples given at the beginning of this section, or the rules defining
the recursor on $\sO$. For instance, if we take $\vphi_+=1$ and
$\stat_+=\lex$, then $\{\lim~x,y\}\sle\CC=\CC_+(\lim~x,y)$ by (arg),
$n\in\CC$ by (var) for $n\in\cX-\FV(\lim~x,y)$, $x\in\CC$ by
(subterm-undef), $(x~n)+y\in\CC$ by (rec) for
$\lim~x>_\ss^{(\lim~x)+y}x~n$, $\l n(x~n)+y\in\CC$ by (abs) for
$n\in\cX-\FV(\lim~x,y)$, and $\lim~(\l n(x~n)+y)\in\CC$ by (undef).

This is however not sufficient to orient the rules defining the
function $\fex$ above since the type for continuations is not strictly
positive. To deal with non-strictly positive types, one needs to
consider type constants with size annotations
\cite{abel04ita,barthe04mscs,blanqui06lpar-sbt}.

%%%%%%%%%%%%%%%%%%%%%%%%%%%%%%%%%%%%%%%%%%%%%%%%%%%%%%%%%%%%%%%%%%%%%%%%%%%%%%
\subsection{Handling matching on non-basic defined symbols}
\label{sec-match-def}

We have already seen at the end of Section \ref{sec-rec} that the rule
(subterm-basic) allows to handle matching on {\em basic} defined
symbols and not only {\em undefined} symbols (constructors) as in the
previous section. Consider now the following set of rules on the
strictly-positive type $\sO$ of ordinals:

\begin{center}
$+:\sO\A\sO\A\sO$\\[2mm]
\begin{rew}
\zero+y&y\\
(\suc~x)+y&\suc~(x+y)\\
(\lim~x)+y&\lim~(\l n~(x~n)+y)\\
(x+y)+z&x+(y+z)\\
\end{rew}
\end{center}

For handling the last rule (associativity), we need $x$ and $y$ to be
computable whenever $x+y$ so is. But this does not follow from the
interpretation of types in standard inductive systems which ensures
that all the arguments of a computable term of the form $\sf\vt$ are
computable if $\sf$ is an {\em undefined} symbol (constructor) and
some positivity conditions are satisfied. However, the
introduction-based interpretation of types can be easily extended to
include other symbols as long as the positivity conditions are
satisfied. Moreover, these conditions can be checked for each argument
independently. Hence the following definitions:

\begin{dfn}[Accessible argument]
\label{def-acc}
  Given a well-founded quasi-ordering $\ge_\cB$ on $\cB$, the set of
  {\em accessible positions} of a symbol $\sf:\vT\A\sB$, $\Acc(\sf)$,
  is the set of integers $i\in[1,|\vT|]$ such that, for all $\sC$
  occurring in $T_i$, either $\sC<_\cB\sB$ or else $\sC\simeq_\cB\sB$
  and $\Pos(\sC,T_i)\sle\Pos^+(T_i)$.

  Let $\cM(\cR)$ be the set of symbols $\sf$ that are {\em strict}
  subterms of a left-hand side of a rule and for which $\Acc(\sf)$ is
  not empty ({\em matched} symbols with accessible arguments).
\end{dfn}

Then, for $I(\sO)$, we can take:

\begin{center}
$F_\sO(X)=\{t\in\SN\mid\all\sf\in\cM(\cR),\all\vT,\all\vu,\tau(\sf)=\vT\A\sO\et
  |\vT|=|\vu|\et t\a^*\sf\vu\A$\\\hsp[-20mm]$\all i\in\Acc(\sf),
  u_i\in\I{T_i}^J\}$\footnote{In a standard inductive system,
  all the arguments of a constructor are accessible
  ($\Acc(\sf)=[1,|\vT|]$ for every $\sf\in\cF-\cD(\cR)$). In this
  case, this new definition of $F_\sO$ is equivalent to the
  introduction-based definition given in the previous section if one
  takes $\Acc(\sf)=\vide$ for every $\sf\in\cD(\cR)$, and assumes that
  $\cM(\cR)=\cF-\cD(\cR)$.}
\end{center}

\noindent
where $J(\sO)=X$ and $J(\sN)=I(\sN)$. But, for $F_\sO(X)$ to satisfy
the property \compneutral, we need to exclude from the set of neutral
terms the terms of the form $\sf\vt$ with $\sf\in\cM(\cR)$:

\begin{dfn}[Neutral term - New definition]
\label{def-neutral-acc}
  Given a set $\cR$ of rewrite rules, a term is {\em
    neutral}\footnote{This definition generalizes and replaces the one
    given in Definition \ref{def-comp-core}.} if it is of the form
  $x\vv$, $(\l xt)u\vv$ or $\sf\vv$ with $\sf\in\cD(\cR)-\cM(\cR)$ and
  $|\vv|\ge\af=sup\{|\vl|\mid\ex r,\sf\vl\a r\in\cR\}$.
\end{dfn}

Then, we have the following property:

\begin{lem}
\label{lem-comp-base-type}
A term $a:\sA$ is computable {\em iff} all its reducts are computable
{\em and}, for all $\sf\in\cM(\cR)$, $i\in\Acc(\sf)$ and $\va$ such
that $a=\sf\va$, $a_i$ is computable.
\end{lem}

\begin{prf}
The only-if part directly follows from \compred\ and the definition of
the interpretation. For the if-part, first note that $a\in\SN$ for all
its reducts are computable and $\I\sA\sle\SN$ by \compsn. Now, let
$\sf\in\cM(\cR)$, $i\in\Acc(\sf)$ and $\va$ such that
$a\a^*\sf\va$. If $a=\sf\va$, then $a_i$ is computable by
assumption. Otherwise, there is $a'$ such that $a\a a'\a^*\sf\va$ and,
since $a'$ is computable by assumption, $a_i$ is computable.\cqfd
\end{prf}

\begin{figure}[ht]
\caption{Computability closure operations V\label{fig-cc-acc}}
\begin{center}
\fbox{\begin{tabular}{rl}
(undef)&${\cF-\cD(\cR)}\sle\CC_\sf(\vl)$\\
(subterm-acc)&if $\sg\vt\in\CC_\sf(\vl)$, $\sg\in\cM(\cR)$, $\sg\vt:\sB\in\cB$
and $i\in\Acc(\sg)$, then $t_i\in\CC_\sf(\vl)$\\
\end{tabular}}
\end{center}
\end{figure}

We can then generalize the closure operations of Figure
\ref{fig-cc-std-ind} for standard inductive systems to the closure
operations of Figure \ref{fig-cc-acc} (we omit the proof).

In addition, we can also give a syntactic criterion for the condition
$\I\sB=\SN$ used in (undef-basic) and (subterm-basic) (see Lemma 16 in
\cite{blanqui02tcs} and Lemma 49 in \cite{blanqui05mscs}):

\begin{dfn}[Basic type]
A type constant is {\em basic} if its equivalence class modulo
$\simeq_\cB$ is basic. An equivalence class $E$ is {\em basic} if for
all $\sB\in E$, $\sf\in\cM(\cR)$, $\sf:\vT\A\sB$, $i\in\Acc(\sf)$,
$T_i$ is a type constant $\sC$ such that $\sC\in E$ or else $\sC<_\cB\sB$
and $[\sC]_{\simeq_\cB}$ is basic.
\end{dfn}

In particular, all first-order data types (natural numbers, lists of
natural numbers, trees, etc.) are basic.

\section{Rewriting modulo an equational theory}
\label{sec-rew-mod}

\renewcommand\ae{\a_\cE}
\newcommand\aer{=_\cE\ar}
\newcommand\aeb{=_\cE\ab}

Rewriting theory has been initially introduced as a decision tool for
equational theories \cite{knuth67}. Indeed, an equational theory
$=_\cE$, \ie the smallest congruence containing $\cE$, is decidable if
there is a set $\cR$ of rewrite rules such that $\ar$ terminates, is
confluent, correct (${\cR}\sle{=_\cE}$) and complete
(${\cE}\sle{=_\cR}$). Knuth and Bendix invented a completion procedure
that, in case of success, builds such a set from $\cE$. This procedure
consists in orienting the equations of $\cE$ (and those generated in
the course of the procedure) in order to use them as rewrite rules.

Yet, some equations or sets of equations, like commutativity, or
associativity and commutativity together (associativity alone is
orientable), are not orientable (no orientation leads to a terminating
relation). A solution consists then in reasoning modulo these
unorientable equations $\cE$ and consider class rewriting modulo
$\cE$, \ie the relation\footnote{We use the relation and notation of
  \cite{huet80jacm} and not the relation
  ${\a_{\cR/\cE}}={=_\cE\ar=_\cE}$ used in \cite{jouannaud86siam} for
  it makes proofs simpler, but the two relations are equivalent from
  the point of view of termination.} $t\aer u$ if there is $t'$ such
that $t=_\cE t'$ and $t'\ar u$ \cite{lankford77tr,huet80jacm}.

Another solution, preferred in practice since it makes rewriting more
tractable, consists in considering rewriting with {\em matching
  modulo} $\cE$, \ie the relation $t\a_{\cR,\cE}u$ if there are a
position $p\in\Pos(t)$, a rule $l\a r\in\cR$ and a substitution $\s$
such that $t|_p=_\cE l\s$ and $u=t[r\s]_p$
\cite{peterson81jacm,jouannaud86siam}. Efficient implementations of
rewriting with matching modulo some equational theories like
associativity and commutativity have been developed
\cite{eker96wrla,kirchnerh01jfp} that are for instance used to
simulate and verify systems modeling chemical reactions or
cryptographic protocols.\footnote{Indeed, the order of molecules in a
  chemical formula is irrelevant, and the order in which messages are
  received may be different from the order messages are sent.}

However, we will only consider class rewriting in this paper. But,
since rewriting with matching modulo is included in class rewriting,
the termination of class rewriting implies the termination of
rewriting with matching modulo. Moreover, many confluence results for
rewriting with matching modulo relies on termination of class
rewriting \cite{jouannaud86siam}.

\medskip

We now show how the notions of computability and computability closure
can be adapted to prove the termination of the relation
${\a}={{\ab}\cup{\aer}}$ for an important class of equational theories
$=_\cE$.

First note that, if there is a {\em non-regular}\footnote{$l=r$ is
  {\em regular} if $\FV(l)=\FV(r)$.} equation (\eg $x\times 0=0$),
then the relation $\aer$ does not terminate. Indeed, if there are
$g=d\in\cE$, $x\in\FV(g)-\FV(d)$ and $l\a r\in\cR$, then $d=d_x^l=_\cE
g_x^l\ar^+g_x^r=_\cE d_x^r=d$ \cite{jouannaud86siam}.

Similarly, if there is a regular {\em non-linear}\footnote{$l=r$ is
  {\em linear} is both $l$ and $r$ are linear.} {\em
  collapsing}\footnote{$l=r$ is {\em collapsing} if $l\in\cX$ or
  $r\in\cX$} equation (\eg $x\et x=x$), then $\aer$ does not terminate
either. Indeed, assume that $t=x\in\cE$ and $x$ freely occurs at two
positions $p$ and $q$ in $t$, and let $t'=t[y]_p$ where
$y\notin\FV(t)$. If $l\a r\in\cR$, then $l=_\cE
t(_x^l)=t'(_x^l)(_y^l)\ar t'(_x^l)(_y^r)\ldots$
\cite{jouannaud86siam}.

We will therefore restrict our attention to regular and non-collapsing
equations, thus excluding regular, linear and collapsing equations
like $x+0=x$, which are easily oriented though.

\medskip

We now extend the notion of neutral term by taking equations into
account:

\begin{dfn}[Neutral term modulo equations]
  Given a set $\cR$ of rewrite rules of the form $\sf\vl\a r$ and a
  set $\cE$ of equations of the form $\sf\vl=\sg\vm$, a term is {\em
    neutral} if it is of the form $x\vv$, $(\l xt)u\vv$ or $\sf\vv$
  with
  $\sf\in\cD(\cR\cup\cE\cup\cE^{-1})-\cM(\cR\cup\cE\cup\cE^{-1})$\footnote{See
    Definition \ref{def-acc}.} and $|\vv|\ge\af=sup\{|\vl|\mid\ex r,\sf\vl\a
  r\in\cR\cup\cE\cup\cE^{-1}\}$. An equation $l=r$ is neutral if $l$
  is of the form $\sf\vl$, $r$ is of the form $\sg\vm$, and both $l$
  and $r$ are neutral. A set of equations $\cE$ is neutral if every
  equation of $\cE$ is neutral.
\end{dfn}

Note that this definition generalizes Definition \ref{def-neutral-acc}
for they are identical if $\cE=\vide$. Note also that, if $\sf\vl
l=r\in\cE$, then $\sf\vl$ is {\em not} neutral.

\medskip

Next, we need the set of neutral terms to be stable by $=_\cE$. It is
not sufficient to require that, for each equation $l=r\in\cE$, $l$ is
neutral iff $r$ is neutral, as shown by the following counter-example:
for each equation $l=r\in\cE=\{\sf=\sg,\sf~x=\sh,\sg~x~y=\sk\}$, $l$
is neutral iff $r$ is neutral ($\sf$ and $\sg$ are not neutral,
$\sf~x$ and $\sh$ are neutral, and $\sg~x~y$ and $\sk$ are neutral),
but $\sf~x=_\cE\sg~x$, $\sf~x$ is neutral and $\sg~x$ is not neutral
because $\af=1$, $\ag=2$ and $\al_\sh=\al_\sk=0$. However, it is
sufficient to require $\cE$ to be neutral:

\begin{lem}
  If $\cE$ is neutral, then the set of neutral terms is stable by
  $=_\cE$.
\end{lem}

\begin{prf}
  Note that $=_\cE$ is the reflexive and transitive closure of
  ${\aa_\cE}={{\ae}\cup{\la_\cE}}$ (the symmetric closure of
  $\ae$). We can therefore proceed by induction on the number of
  $\aa_\cE$ steps, and prove that the set of neutral terms is stable
  by $\aa_\cE$. So, let $t$ be a neutral term and assume that
  $t\aa_\cE t'$. We check that $t'$ is neutral:
\begin{itemize}
\item $x\vv\aa_\cE t'$. Since equations are of the form $\sf\vl=\sg\vm$,
  $t'$ is of the form $x\vv'$ with $\vv(\aa_\cE)_\prod\vv'$.
\item $(\l xt)u\vv\aa_\cE t'$. Since equations are of the form
  $\sf\vl=\sg\vm$, $t'$ is of the form $(\l xt')u'\vv'$ with
  $tu\vv(\aa_\cE)_\prod t'u'\vv'$.
\item $\sf\vv\aa_\cE t'$ with $\sf\in\cD(\cR\cup\cE\cup\cE^{-1})$ and
  $|\vv|\ge\al_\sf$. Either $t'=\sf\vv'$ and $\vv(\aa_\cE)_\prod\vv'$,
  or there are $\vw$, $\sf\vl=\sg\vm\in\cE$ and $\s$ such that
  $\vv=\vl\s\vw$ and $t'=\sg\vm\s\vw$. Since $\cE$ is neutral,
  $|\vl|\ge\af$ and $|\vm|\ge\ag$. Thus, $t'$ is neutral.\cqfd
\end{itemize}
\end{prf}

\comment{
However, it seems possible to complete a set of equations $\cE$ such
that, for each equation $l=r\in\cE$, $l$ is neutral iff $r$ is
neutral, into a set $\cE'\sge\cE$ such that ${=_\cE}={=_{\cE'}}$ and
the set of neutral terms wrt $\cE'$ is stable by $=_\cE$, by adding
{\em extension} rules wrt application (in a way similar to extension
rules for associative and commutative symbols in
\cite{peterson81jacm}): let $\cE'$ be the smallest set of equations
containing $\cE$ and such that, if $\sf\vl=r\in\cE'\cup{\cE'}^{-1}$
and $|\vl|<\af$, then $\sf\vl x=rx\in\cE'$ for some
$x\notin\FV(\sf\vl)\cup\FV(r)$ (note that $\af$ actually depends on
the set of equations but it is easy to see that, here, $\af$ is in
fact invariant).
}

Finally, we need $\SN(\a)$ and thus $\SN(\ab)$ to be stable by
$=_\cE$. This can be achieved by requiring $=_\cE$ to commute with
$\ab$. Putting every thing together, we get:

\begin{dfn}[Admissible theory]
  A set of equations $\cE$ is {\em admissible} if $\cE$ is made of
  regular, non-collapsing and neutral equations only, and $=_\cE$
  commutes with $\ab$.
\end{dfn}

In particular, $=_\cE$ commutes with $\ab$ if:

\begin{lem}
\label{lem-com}
  Given a set of equations $\cE$ of the form $\sf\vl=\sg\vm$, $=_\cE$
  commutes with $\ab$ if $\cE$ satisfies all the following conditions:

\begin{itemize}
\item $\cE$ is linear: $\all l=r\in\cE$, $l$ and $r$ are linear;
\item $\cE$ is regular: $\all l=r\in\cE$, $\FV(l)=\FV(r)$;
\item $\cE$ is algebraic: $\all l=r\in\cE$, $l$ and $r$ are
  algebraic\footnote{They contain no subterm of the form $\l xt$ or
    $xt$.}.
\end{itemize}
\end{lem}

\begin{prf}
We proceed by induction on the number of $\aa_\cE$-steps and show
that, if $t\at[-1]{p}\la_\cE u\at[3]{q}\ab v$, then $t\ab=_\cE v$. The
case ${\ae\ab}\sle{\ab=_\cE}$ is similar for conditions on equations
are symmetric.
\begin{itemize}
\item $p\#q$. Then, $t\ab\la_\cE v$.
\item $p<q$. Then, there are $l\a r$ and $\s$ such that $u|_p=l\s$ and
  $t=u[r\s]_p$. Since $r$ is algebraic and linear, there is
  $x\in\FV(l)$ such that $v=u[l\s']_p$, $x\s\ab x\s'$ and, for all
  $y\neq x$, $y\s'=y\s$. Since $r$ is regular and linear, $t\ab
  u[r\s']\la_\cE v$.
\item $p=q$. Not possible for equations are of the form $\sf\vl=\sg\vm$.
\item $p>q$. There are $x,a,b$ such that $u|_q=(\l xa)b$ and
  $v=u[a_x^b]_q$. Since equations are of the form $\sf\vl=\sg\vm$:
\begin{itemize}
\item Either there is $a'$ such that $a\ae a'$ and $t=u[(\l
  xa')b]_q$. Then, $t\ab u[{a'}_x^b]_q\la_\cE v$.
\item Or there is $b'$ such that $b\ae b'$ and $t=u[(\l
  xa)b']_q$. Then, $t\ab u[a_x^{b'}]_q\la_\cE^* v$.\cqfd
\end{itemize}
\end{itemize}
\end{prf}

The condition of algebraicity could be slightly relaxed. For instance,
the commutation of quantifiers necessary for ensuring the confluence
of the rewrite rules computing the prenex normal form of a formula
\cite{mayr98tcs} commutes with $\ab$:

\begin{rewc}[~~=~~]
\all(\l x\all(\l yPxy))&\all(\l y\all(\l xPxy))\\
\end{rewc}

Now, we generalize the notion of computability to rewriting modulo
some admissible theory:

\begin{dfn}[Computability predicates for rewriting modulo equations]
  Given an admissible set of equations $\cE$ and a type $T$, let
  $\Red_{\cR/\cE}^T$ be the set of all the sets $P\sle\cL^T$ such
  that:

\begin{enumii}{R}
\item
$P\sle\SN(\a)$ where ${\a}={{\ab}\cup{\aer}}$;
\item
$P$ is stable by $\a\cup=_\cE$;
\item
if $t:T$ is neutral and $\a\!(t)\sle P$, then $t\in P$.
\end{enumii}
\end{dfn}

Note that $\Red_{\cR/\vide}^T=\Red_\cR^T$. We now check that the
family $(\Red_{\cR/\cE}^T)_{T\in\cT}$ has all the required properties:

\begin{lem}
  If $\cE$ is an admissible set of equations and $T\in\cT$, then
  $\Red_{\cR/\cE}^T$ is stable by non-empty intersection and admits
  $\SN^T$ as greatest element. Moreover, for all $T,U\in\cT$,
  $P\in\Red_{\cR/\cE}^T$ and $Q\in\Red_{\cR/\cE}^U$,
  $\propto\!(P,Q)\in\Red_{\cR/\cE}^{T\A U}$.
\end{lem}

\begin{prf}
  The proof is similar to the one of Lemma \ref{lem-cand}. We only
  detail the cases that are different. We have
  $\SN^T\in\Red_{\cR/\cE}^T$ for $=_\cE$ commutes with $\ab$ and thus
  with $\a$. For the stability by $\propto$, we only detail
  \compneutral. Let $T,U\in\cT$, $P\in\Red_{\cR/\cE}^T$ and
  $Q\in\Red_{\cR/\cE}^U$, $v:T\A U$ neutral with
  ${\a\!(v)}\sle{\propto\!(P,Q)}$, and $t\in P$. We prove that $vt\in
  Q$ by well-founded induction on $t$ ordered by $\a$ ($t\in\SN$ by
  \compsn). Since $vt$ is neutral, by \compneutral, it suffices to
  prove that, for all $w'$ such that $vt\a w'$, we have $w'\in Q$.

We first prove (a): there are $v'$ and $t'$ such that $w'=v't'$ with
either $v\a v'$ and $t=_\cE t'$, or $v=_\cE v'$ and $t\a t'$. We
proceed by case on $vt\a w'$:
\begin{itemize}
\item $vt\ab w'$. Since $v$ is neutral, it is not an abstraction and
  either $w'=v't$ with $v\ab v'$, or $w'=vt'$ with $t\ab t'$. Hence,
  (a) is satisfied.

\item $vt=_\cE w\ar w'$. We prove (a) by induction on the number of
  $\aa_\cE$-steps. If $vt=w$, then we are done. Assume now that,
  $vt\aa_\cE w=_\cE\ar w'$.

The term $w$ can neither be a variable nor an abstraction for
equations are of the form $\sf\vl=\sg\vm$.

Assume that there are $\sf\vl=\sg\vm\in\cE$ and $\s$ such that
$w=\sf\vl\s$ and $vt=\sg\vm\s$. Since $vt=\sg\vm\s$, there are $\vk$
and $k$ such that $\vm=\vk k$ and $v=\sg\vk\s$. But, then, $v$ cannot
be neutral for $|\vk|<|\vm|\le\ag$.

Therefore, there are $a$ and $b$ such that $w=ab$, $t=_\cE a$ and
$u=_\cE b$. Now, by the induction hypothesis, there are $v'$ and $t'$
such that $w'=v't'$ with either $a\ar v'$ and $b=_\cE t'$, or $a=_\cE
v'$ and $b\ar t'$. Hence, (a) holds.
\end{itemize}
\noindent
If $v\a v'$ and $t=_\cE t'$, then $w'=v't'\in Q$ for
$v'\in{\propto\!(P,Q)}$ by assumption and $t'\in P$ by \compred.
Otherwise, $v=_\cE v'$ and $t\a t'$. Then, $t'\in P$ by \compred, and
$v'$ is neutral since neutral terms are stable by $=_\cE$. Assume now
that $v'\a v''$. Since $\cE$ is admissible, $=_\cE$ commutes with
$\ab$ and thus with $\a$. Hence, there is $e$ such that $v\a e=_\cE
v''$, and $v''\in{\propto\!(P,Q)}$ for $e\in{\propto\!(P,Q)}$ by
assumption and ${\propto\!(P,Q)}$ satisfies \compred. Therefore,
${\a\!(v')}\sle{\propto\!(P,Q)}$ and, by the induction hypothesis on
$t'$, $w'=v't'\in Q$.\cqfd
\end{prf}

\begin{figure}[ht]
\caption{Computability closure operations I'\label{fig-cc-mod}}
\begin{center}
\fbox{\begin{tabular}{rl}
(mod)&if $t\in\CC_\sf(\vl)$ and $t=_\cE u$, then $u\in\CC_\sf(\vl)$
\end{tabular}}
\end{center}
\end{figure}

We now show how to extend Theorem \ref{thm-cc-rec}.

\begin{thm}
\label{thm-cc-mod}
Given a set of rules $\cR$ and an admissible set of equations $\cE$,
the relation ${\a}={{\ab}\cup{\aer}}$ terminates on well-typed terms
if there are $I\in\Pi_{\sB\in\cB}\Red_{\cR/\cE}^\sB$ and a valid
$\cF$-quasi-ordering $\ge$ containing $(=_\cE)_\prod$ such that:

\begin{itemize}
\item every non-basic undefined symbol is computable;
\item for every equation $\sf\vl=\sg\vm\in\cE$, $\vm\in\CC_\sf(\vl)$,
  $\vl\in\CC_\sg(\vm)$ and $(\sf,\vl)\simeq(\sg,\vm)$;
\item for every rule $\sf\vl\a r\in\cR$, $r\in\CC_\sf(\vl)$;
\end{itemize}

\noindent
where $\CC$ is the smallest computability closure closed by the
operations I to IV, and I'.
\end{thm}

\begin{prf}
  We proceed as for Theorem \ref{thm-cc-rec} and show that, for all
  $(\sf,\vt)\in\S_\max$, every reduct $t$ of $\sf\vt$ is computable,
  by induction on ${>}\cup{\a_\prod}$. There are two cases:

\begin{enumerate}
\item
$t=\sf\vu$ with $\vt~(\ab)_\prod~\vu$. By \compred, $\vu$ is
  computable. Therefore, by the induction hypothesis, $t$ is computable.
\item
Otherwise, $\sf\vt=_\cE u\ar t$. We first prove by induction on the
number of equational steps between $\sf\vt$ and $u$, that $u$ is of
the form $\sg\vu$ with $\vu$ computable and
$(\sf,\vt)\simeq(\sg,\vu)$. If there is no equational step, this is
immediate. So, assume that $\sf\vt=_\cE u'\aa_\cE u$. By the induction
hypothesis, $u'$ is of the form $\sg\vu$ with $\vu$ computable and
$(\sf,\vt)\simeq(\sg,\vu)$. The conditions on rules being symmetric,
the case of $\la_\cE$ is similar to the one of $\ae$ for which there
are two cases:

\begin{enumerate}
\item
$u=\sg\vv$ with $\vu~(\ae)_\prod~\vv$. By \compred, $\vv$ is computable
  and $(\sg,\vu)\simeq(\sg,\vv)$ since $\simeq$ contains
  $(=_\cE)_\prod$. Therefore, by transitivity,
  $(\sf,\vt)\simeq(\sg,\vv)$.
\item There are $\sg\vl=\sh\vm\in\cE$, $\s$ and $\vw$ such that
  $\vu=\vl\s\vw$ and $u=\sh\vm\s\vw$. By assumption,
  $\vm\in\CC_\sg(\vl)$ and $(\sg,\vl)\simeq(\sh,\vm)$. Since $\simeq$
  is stable by substitution, $(\sg,\vl\s)\simeq(\sh,\vm\s)$. Since
  $\simeq$ is compatible with application,
  $(\sg,\vl\s\vw)\simeq(\sh,\vm\s\vw)$ and, by transitivity,
  $(\sf,\vt)\simeq(\sh,\vm\s\vw)$. Now, since $>$ is stable by
  substitution, $\CC$ is stable by substitution and
  $\vm\s\in\CC_\sg(\vl\s)$. Hence, by Lemma \ref{lem-cc} and induction
  hypothesis, $\vm\s$ is computable.
\end{enumerate}

Now, for $t$, there are two possibilities:

\begin{enumerate}
\item $t=\sg\vv$ with $\vu~(\ar)_\prod~\vv$. By \compred, $\vv$ is
  computable and, by the induction hypothesis, $t$ is computable.
\item There are $\sg\vl\a r\in\cR$, $\s$ and $\vw$ such that
  $\vu=\vl\s\vw$ and $t=r\s\vw$. By assumption,
  $r\in\CC_\sg(\vl)$. Since $\CC$ is stable by substitution, we have
  $r\s\in\CC_\sg(\vl\s)$. Hence, by Lemma \ref{lem-cc} and induction
  hypothesis, $r\s$ is computable.\cqfd
\end{enumerate}
\end{enumerate}
\end{prf}

%%%%%%%%%%%%%%%%%%%%%%%%%%%%%%%%%%%%%%%%%%%%%%%%%%%%%%%%%%%%%%%%%%%%%%%%%%%%%%
\subsection{$\cF$-quasi-ordering compatible with permutative theories}
\label{sec-f-qo-mod}

We now define an $\cF$-quasi-ordering satisfying the previous conditions
for a general class of equational theories
including {\em permutative}\footnote{An equation $l=r$ is {\em
    permutative} if every variable or function symbol has the same
  number of occurrences in $l$ than it has in $r$. Such equations
  appear in algebra (permutative semi-groups), category theory (middle
  four exchange rule of Mac Lane), linear logic, the calculus of
  structures (medial rule) \cite{strassburger07rta}, automated
  deduction, \ldots} axioms like associativity and commutativity
together \cite{lankford77tr}. It is based on the notion of {\em alien
  subterm} used when studying the preservation (modularity) of
properties like confluence and termination of the disjoint union of
two rewrite systems
\cite{gramlich91tr,gramlich94aaecc,fernandez94adt}.

\begin{dfn}[Alien subterms]
Let $\cM=\bM(\SN)$ be the set of finite multisets on $\SN$. Given a
set $E\sle\cF$, the {\em $E$-alien subterms} ($E$-aliens for short) of
a multiset $M\in\cM$, $\Aliens_E(M)$, is the multiset of terms defined
by induction on $\tgt_\cM$ as follows:
\begin{itemize}
\item $\Aliens_E(\vide)=\vide$;
\item $\Aliens_E(M+N)=\Aliens_E(M)+\Aliens_E(N)$;
\item $\Aliens_E(\mset{t})=\Aliens_E(\mset\vt)$ if $t=\sf\vt$ and $\sf\in E$,
\item $\Aliens_E(\mset{t})=\mset{t}$ otherwise.
\end{itemize}
\noindent
Given an equivalence $\simeq_\cF$ on $\cF$, a set of equations $\cE$
is {\em compatible with $\simeq_\cF$-aliens} if every equation of
$\cE$ is of the form $\sf\vl=\sg\vm$ with $\sf\simeq_\cF\sg$ and
$\Aliens_{[\sf]_{\simeq_\cF}}(\vl)=\Aliens_{[\sg]_{\simeq_\cF}}(\vm)$.
\end{dfn}

Note that $\mset{t}\tge_\cM\Aliens_E(t)$. For instance,
$\Aliens_{\{+\}}((x+y)+(z\times(t+u)))=\mset{x,y,z\times(t+u)}$.

Note also that, for all $\cF$-quasi-orderings $\ge$, if $\cE$ is
compatible with $\simeq_\cF$-aliens then, for all equations
$\sf\vl=\sg\vm\in\cE$, $(\sf,\vl)\simeq(\sg,\vm)$, as required in
Theorem \ref{thm-cc-mod}.

We now prove some properties of aliens:

\begin{lem}
\label{lem-aliens-phi}
If $\t$ is a substitution, then
$\Aliens_E(M\t)=\vphi_E^\t(\Aliens_E(M))$, where
$\vphi_E^\t(M)$ is defined by induction on $\tgt_\cM$ as follows:
\begin{itemize}
\item $\vphi_E^\t(\vide)=\vide$;
\item $\vphi_E^\t(M+N)=\vphi_E^\t(M)+\vphi_E^\t(N)$;
\item
  $\vphi_E^\t(\mset{x\vu})=\Aliens_E(\mset\vt)+\vphi_E^\t(\Aliens_E(\mset\vu))$
  if $x\in\cX$, $x\t=\sf\vt$ and $\sf\in E$;
\item $\vphi_E^\t(\mset{a})=\mset{a\t}$ otherwise.
\end{itemize}
\end{lem}

\begin{prf}
By induction on $M$ with $\tgt_\cM$ as well-founded relation.\qed
\end{prf}

In the following, we assume given a quasi-ordering $\ge_\cF$ on $\cF$,
a set of equations $\cE$ compatible with $\simeq_\cF$-aliens, and an
equivalence class $E$ modulo $\simeq_\cF$.

\begin{lem}
\label{lem-aliens-eq}
If $M\,(=_\cE)_\cM\, N$, then $\Aliens_E(M)\,(=_\cE)_\cM\,\Aliens_E(N)$.
\end{lem}

\begin{prf}
We proceed by induction on $M$ with $\tgt_\cM$ as well-founded relation:
\begin{itemize}
\item $M=N=\vide$. Then, $\Aliens_E(M)=\vide=\Aliens_E(N)$.
\item $M=P+\mset{a}$, $N=Q+\mset{b}$, $P\,(=_\cE)_\cM\, Q$ and $a=_\cE
  b$. By the induction hypothesis, $\Aliens_E(P)\,{(=_\cE)_\cM}$
  $\Aliens_E(Q)$. We now prove that
  $\Aliens_E(\mset{a})\,(=_\cE)_\cM\,\Aliens_E(\mset{b})$, by
  induction on the number of $\aa_\cE$ steps. And since conditions on
  equations are symmetric, it sufficient to prove that, if $a\ae b$,
  then $\Aliens_E(\mset{a})\,(=_\cE)_\cM\,\Aliens_E(\mset{b})$:
\begin{itemize}
\item $a=x\vu$. Since equations are of the form $\sf\vl=\sg\vm$, there
  is $\vv$ such that $b=x\vv$. Therefore,
  $\Aliens_E(\mset{a})=\mset{a}\,(=_\cE)_\cM\,\mset{b}=\Aliens_E(\mset{b})$.
\item $a=(\l xs)\vu$. Since equations are of the form $\sf\vl=\sg\vm$,
  there are $t$ and $\vv$ such that $b=(\l xt)\vv$. Therefore,
  $\Aliens_E(\mset{a})=\mset{a}\,(=_\cE)_\cM\,\mset{b}=\Aliens_E(\mset{b})$.
\item $a=\sf\vu$, $b=\sf\vv$ and $\vu\,(\ae)_\prod\,\vv$.
\begin{itemize}
\item $\sf\in E$. By the induction hypothesis,
  $\Aliens_E(\mset\vu)\,(=_\cE)_\cM\,\Aliens_E(\mset\vv)$. Therefore,
  $\Aliens_E(\mset{a})$ ${(=_\cE)_\cM}\,\Aliens_E(\mset{b})$.
\item $\sf\notin E$. Then,
  $\Aliens_E(\mset{a})=\mset{a}\,(=_\cE)_\cM\,\mset{b}=\Aliens_E(\mset{b})$.
\end{itemize}
\item There are $\vw$, $\sf\vl=\sg\vm\in\cE$ and $\s$ such that
  $a=\sf\vl\s\vw$ and $b=\sg\vm\s\vw$. Since $\cE$ is compatible with
  $\simeq_\cF$-aliens, $\sf\simeq_\cF\sg$ and
  $\Aliens_E(\vl)=\Aliens_E(\vm)$.
\begin{itemize}
\item $\sf\in E$. Then, $[\sf]_{\simeq_\cF}=E$,
  $\Aliens_E(\mset{a})=\Aliens_E(\mset{\vl\s})+\Aliens_E(\vw)$ and
  $\Aliens_E(\mset{b})=\Aliens_E(\mset{\vm\s})+\Aliens_E(\vw)$. By
  Lemma \ref{lem-aliens-phi},
  $\Aliens_E(\mset{\vl\s})=\vphi_E^\s(\Aliens_E(\mset\vl))$ and
  $\Aliens_E(\mset{\vm\s})=\vphi_E^\s(\Aliens_E(\mset\vm))$. Therefore,
  $\Aliens_E(\mset{a})=\Aliens_E(\mset{b})$.
\item $\sf\notin E$. Then, $\sg\notin E$ and
  $\Aliens_E(\mset{a})=\mset{a}\,(=_\cE)_\cM\,\mset{b}=\Aliens_E(\mset{b})$.\cqfd
\end{itemize}
\end{itemize}
\end{itemize}
\end{prf}

\begin{lem}
\label{lem-phi-eq}
If $M\,(=_\cE)_\cM\, N$, then $\vphi_E^\t(M)\,(=_\cE)_\cM\,\vphi_E^\t(N)$.
\end{lem}

\begin{prf}
We proceed by induction on $M$ with $\tgt_\cM$ as well-founded relation:
\begin{itemize}
\item $M=N=\vide$. Then, $\vphi_E^\t(M)=\vide=\vphi_E^\t(N)$.
\item $M=P+\mset{a}$, $N=Q+\mset{b}$, $P\,(=_\cE)_\cM\, Q$ and $a=_\cE
  b$. By the induction hypothesis,
  $\vphi_E^\t(P)\,(=_\cE)_\cM\,\vphi_E^\t(Q)$. We now prove that
  $\vphi_E^\t(a)\,(=_\cE)_\cM\,\vphi_E^\t(b)$:
\begin{itemize}
\item Assume that $a=x\vu$, $x\t=\sf\vw$ and $\sf\in E$. Since
  equations are of the form $\sf\vl=\sg\vm$, there is $\vv$ such that
  $b=x\vv$ and $\vu(=_\cE)_\prod\vv$. Hence,
  $\mset\vu\,(=_\cE)_\cM\,\mset\vv$ and, by Lemma \ref{lem-aliens-eq},
  $\Aliens_E(\mset\vu)\,{(=_\cE)_\cM}\,\Aliens_E(\mset\vv)$. By the induction
  hypothesis,
  $\vphi_E^\t(\Aliens_E(\mset\vu))\,(=_\cE)_\cM\,\vphi_E^\t(\Aliens_E(\mset\vv))$.
  Therefore, we have
  $\vphi_E^\t(a)=$\\$\Aliens_E(\mset\vw)+\vphi_E^\t(\Aliens_E(\mset\vu))\,(=_\cE)_\cM\,\Aliens_E(\mset\vw)+\vphi_E^\t(\Aliens_E(\mset\vv))=\vphi_E^\t(b)$.
\item Otherwise,
  $\vphi_E^\t(a)=\mset{a}\,(=_\cE)_\cM\,\mset{b}=\vphi_E^\t(b)$.\cqfd
\end{itemize}
\end{itemize}
\end{prf}

The ordering on terms that compares the alien subterms with
$(\tgt_\ss)_\cM$ is not stable by substitution as shown by the
following example:
$\Aliens_{\{\sf\}}(\mset{xy})=\mset{xy}\,(\tgt_\ss)_\cM\,\mset{y}=\Aliens_{\{\sf\}}(\mset{y})$
and $\Aliens_{\{\sf\}}(\mset{\sf y})=\mset{y}$. Therefore, we consider
the following restriction of $\tgt_\ss$:

\newcommand\tgtalg{\tgt_\ss^{\ms{alg}}}
\newcommand\tgealg{\tge_\ss^{\ms{alg}}}

\begin{dfn}
Let $\tgtalg$ be the smallest sub-ordering of $\tgt_\ss$ such that,
for all $\sf:\vT\A U$, $\vt:\vT$ and $i\in[1,|\vt|]$, $\sf\vt\tgtalg
t_i$. Let $\tgealg$ be its reflexive closure.
\end{dfn}

\begin{lem}
\label{lem-dlqo-mod}
Let $\ge_\cF$ be a quasi-ordering on $\cF$ and $\cE$ a set of
equations such that:
\begin{itemize}
\item $\cE$ is admissible and compatible with $\simeq_\cF$-aliens;
\item in each equivalence class modulo $=_\cE$, the size of terms is
  bounded.
\end{itemize}
\noindent
Then, the DLQO $\dot\ge$ associated to:
\begin{itemize}
\item the quasi-ordering $\simeq_\cF$ on $\cF$;
\item for each equivalence class $E$ modulo $\simeq_\cF$, the
  quasi-ordering $(=_\cE\tgealg)_\cM$;
\item for each symbol $\sf$, the function
  $\psi_\sf(\vt)=\Aliens_{[\sf]_{\simeq_\cF}}\!(\mset\vt)$ if $\sf$ is
  maximally applied in $\sf\vt$;
\end{itemize}
is a valid $\cF$-quasi-ordering containing $(=_\cE)_\prod$.
\end{lem}

\begin{prf}
The relation $=_\cE\!\tgtalg$ is well-founded since $\tgtalg$ commutes
with $=_\cE$ (for $=_\cE$ is monotone) and, in each equivalence class
modulo $=_\cE$, the size of terms is bounded (see the proof of
Proposition 15 in \cite{jouannaud86siam}).

Therefore, the strict part of ${\ge}={(=_\cE\tgealg)}$ is
${>}={(=_\cE\!\tgtalg)}$, which is well-founded, and its associated
equivalence relation is $=_\cE$.

Let $\dot{>}$ be the strict part of $\dot\ge$ and $\dot\simeq$ be its
associated equivalence relation.

\begin{itemize}
\item Compatibility of $\dot{>}$ with application. The relation
  $\dot{>}$ is compatible with application for it only compares pairs
  $(\sf,\vt)$ such that $\sf$ is maximally applied in $\sf\vt$.
 
\item Compatibility of $\dot{>}$ with reduction. The
  relation $\tgtalg$ commutes with $\a$ for $\a$ is monotone. The
  relation $=_\cE$ commutes with $\ab$ for $\cE$ is admissible. The
  relation $=_\cE$ trivially commutes with $\aer$. Therefore, $>$
  commutes with $\a$. Since both $>$ and $\a$ are well-founded on
  $\SN$, ${>}\cup{\a}$ is well-founded on $\SN$. Hence,
  ${\dot{>}}\cup{\a_\prod}$ is well-founded on $\S_\max$.

\item Stability of $\dot\simeq$ by substitution. It follows from the
  lemmas \ref{lem-aliens-phi}, \ref{lem-aliens-eq} and
  \ref{lem-phi-eq}.

\item Stability of $\dot{>}$ by substitution. Let $E$ be an
  $\simeq_\cF$-equivalence class, and assume that
  $\Aliens_E(\mset\vt)>_\cM\Aliens_E(\mset\vu)$. Then, there are $M$,
  $P\neq\vide$, $N$ and $Q$ such that $\Aliens_E(\mset\vt)=M+P$,
  $\Aliens_E(\mset\vu)=N+Q$, $M\,(=_\cE)_\cM\, N$ and, (*) for all
  $q\in Q$, there is $p\in P$ such that $p>q$. Now, let $\t$ be a
  substitution. By Lemma \ref{lem-aliens-phi},
  $\Aliens_E(\mset{\vt\t})=\vphi_E^\t(M)+\vphi_E^\t(P)$ and
  $\Aliens_E(\mset{\vu\t})=\vphi_E^\t(N)+\vphi_E^\t(Q)$. By Lemma
  \ref{lem-phi-eq}, $\vphi_E^\t(M)\,(=_\cE)_\cM\,\vphi_E^\t(N)$. We
  now prove that $\vphi_E^\t(P)>_\cM\vphi_E^\t(Q)$. To this end, it
  suffices to prove that, for all $q\in Q$, there is $p\in P$ such
  that $\vphi_E^\t(\mset{p})>_\cM\vphi_E^\t(\mset{q})$, that is,
  $\Aliens_E(\mset{p\t})>_\cM\Aliens_E(\mset{q\t})$. So, let $q\in
  Q$. After (*), there is $p\in P$ such that $p>q$. By definition of
  $>$, there are $\vw$ and $i\in[1,|\vw|]$ such that $p=_\cE\sk\vw$
  and $w_i\tgealg q$. Since equations are of the form $\sf\vl=\sg\vm$,
  there are $\sh$ and $\vv$ such that $p=\sh\vv$. Since $p$ is an
  $E$-alien, $\sh\notin E$ and
  $\Aliens_E(\mset{p\t})=\mset{p\t}$. Since $>$ is stable by
  substitution, $p\t>q\t$ and thus $\mset{p\t}>_\cM\mset{q\t}$. By
  definition of aliens,
  $\mset{q\t}\,(\tgtalg)_\cM\,\Aliens_E(\mset{q\t})$. Therefore, by
  transitivity,
  $\Aliens_E(\mset{p\t})>_\cM\Aliens_E(\mset{q\t})$.\cqfd
\end{itemize}
\end{prf}

Note that the terms of an equivalence class modulo $\cE$ are of
bounded size if, for instance, the equivalence classes modulo $\cE$
are of finite cardinality. This is in particular the case of
associativity and commutativity together.

%%%%%%%%%%%%%%%%%%%%%%%%%%%%%%%%%%%%%%%%%%%%%%%%%%%%%%%%%%%%%%%%%%%%%%%%%%%%%%
\subsection{Example of termination proof}

As an example, we check that the conditions of Theorem
\ref{thm-cc-mod} are satisfied by the set $\cR$ of rules defining the
addition on Peano integers given at the beginning of Section
\ref{sec-rec}, and the following set $\cE$ of equations (associativity
and commutativity):

\begin{rulc}
(x+y)+z&=&x+(y+z)\\
x+y&=&y+x\\
\end{rulc}

\noindent
by taking the identity relation for $\simeq_\cF$ and the
$\cF$-quasi-ordering $\dot\ge$ of Lemma \ref{lem-dlqo-mod}.

The set of equations $\cE$ is neutral. By Lemma \ref{lem-com}, $=_\cE$
commutes with $\ab$ since $\cE$ is linear, regular and
algebraic. Therefore, $\cE$ is admissible.

The set of equations $\cE$ is compatible with $\simeq_\cF$-aliens
since, for the associativity equation, we have $+\simeq_\cF +$ and
$\Aliens_{\{+\}}(\mset{x+y,z})=\mset{x,y,z}=\Aliens_{\{+\}}(\mset{x,y+z})$,
and for the commutativity equation, we have $+\simeq_\cF +$ and
$\Aliens_{\{+\}}(\mset{x,y})=\mset{x,y}=\Aliens_{\{+\}}(\mset{y,x})$.

Hence, by Lemma \ref{lem-dlqo-mod}, $\dot\ge$ is a valid
$\cF$-quasi-ordering containing $(=_\cE)_\prod$ for, in each
equivalence class modulo $=_\cE$, the size of terms is bounded.

We now check the conditions on rules and equations:\footnote{As
  already remarked, the condition $(\sf,\vl)\,\dot\simeq\,(\sg,\vm)$
  for every equation $\sf\vl=\sg\vm$ follows from compatibility with
  $\simeq_\cF$-aliens.}

\begin{itemize}
\item For the first rule defining addition, we have $x\in\CC_+(0,x)$
  by (arg).
\item For the second rule defining addition, we have
  $x+y\in\CC_+(x,\suc~y)$ by (rec) since
  $\Aliens_{\{+\}}(x,\suc~y)=\mset{x,\suc~y}>_\cM\Aliens_{\{+\}}(x,y)=\mset{x,y}$,
  and thus $\suc~(x+y)\in\CC_+(x,\suc~y)$ by (undef) and (app).
\item For the commutativity equation, we have
  $\{y,x\}\sle\CC_+(x,y)$ and $\{x,y\}\sle\CC_+(y,x)$ by (arg).
\item Finally, for the associativity equation, we have
  $x\in\CC_+(x+y,z)$ by (arg) and (subterm-acc), and
  $y+z\in\CC_+(x+y,z)$ by (rec) since
  $\Aliens_{\{+\}}(x+y,z)=\mset{x,y,z}>_\cM\mset{y,z}=\Aliens_{\{+\}}(y,z)$. Similarly,
  we have $x+y\in\CC_+(x,y+z)$ and $z\in\CC_+(x,y+z)$.
\end{itemize}

\section{Rewriting with matching modulo $\b\eta$}
\label{sec-hopm}

\newcommand\bo{{\b_0}}
\newcommand\boe{{\bo\eta}}

\renewcommand\ae{\a_\eta}
\newcommand\abo{\a_\bo}
\newcommand\aboe{\a_\boe}
\newcommand\arbe{\a_{\cR,\be}}

In this section, we extend the results of Section \ref{sec-core} to
rewriting with matching modulo $\be$. Consider the following rewrite
rule used for defining a formal derivation operator:

\begin{center}
$\sin,\cos:\sR\A\sR;\quad\times:\sR\A\sR\A\sR;
\quad\sD:(\sR\A\sR)\A(\sR\A\sR)$\\[2mm]
\begin{rew}
\sD~(\l x~\sin~(F~x)) & \l x~(\sD~F~x)\times(\cos~(F~x))
\end{rew}
\end{center}

Using matching modulo $\al$-equivalence only, this rule can be applied
neither to $\sD(\sin)$ nor to $\sD(\l x\,\sin\,x)$. But it can be
applied to $\sD(\l x\,\sin\,x)$ if we use matching modulo
$\b$-equivalence, since $x\la_\b(\l xx)x$,\footnote{In contrast with a
  common practice (Barendregt's variable convention
  \cite{barendregt92chapter}), we often use the same variable name for
  both a bound and a free variable. Although it may be confusing at
  first sight, it has the advantage of avoiding some variable
  renamings.} and to $\sD(\sin)$ if we use matching modulo
$\beta\eta$-equivalence, since $\sin\la_\eta\l x\,\sin\,x$.

Although matching modulo $\b\eta$ is decidable \cite{stirling09lmcs},
it is of non-elementary complexity \cite{statman79tcs} (while
unification modulo $\b\eta$ \cite{huet76hdr} and matching modulo $\b$
are both undecidable \cite{loader03jigpal}). There is however an
important fragment for which the complexity is linear: the class of
$\b$-normal $\eta$-long terms in which every free variable is applied
to distinct bound variables, introduced by Miller for $\l$Prolog
\cite{miller91jlc,qian93tapsoft}. For instance, $\l x\,\sin(Fx)$, $\l
x\l yFyx$ and $\l xx(Fx)$ are patterns (if they are in $\eta$-long
form), while $Fx$ and $\l xFxx$ are not patterns. However, in this
paper, we will consider a slightly different class of terms:

\begin{dfn}[Patterns]
  A term $t$ is a {\em pattern} if $t\in\cP_{\FV(t)}$ where $\cP_V$ is
  defined as follows:
\begin{itemize}
\item if $t\in\cP_V$, then $\l xt\in\cP_{V-\{x\}}$;
\item if $\sf\in\cF$, $\sf:\vT\A U$, $\vt:\vT$ and $\vt\in\cP_V$,
  then $\sf\vt\in\cP_V$;
\item if $x\in V$, $x:\vT\A U$, $\vt:\vT$ and $\vt$ $\eta$-reduces to
  pairwise distinct variables not in $V$, then $x\vt\in\cP_V$.
\end{itemize}
\end{dfn}

Our definition excludes Miller patterns where a bound variable is
applied to a free variable like $\l xx(Fx)$, which is not very common
in practice. On the other hand, our patterns do not need to be in
$\eta$-long form.

To apply the computability closure technique to rewriting with
matching modulo $\b\eta$, we need to prove that, if
$\sf\vt=_\be\sf\vl\s\ar r\s$ and $\vt$ are computable, then $\vl\s$ is
computable, so that $r\s$ is computable if $r\in\CC_\sf(\vl)$. By
confluence of $\abe$ \cite{pottinger78ndj} and $\eta$-postponement
(${\abe^*}\sle{\ab^*\ae^*}$) \cite{curry58book,takahashi95ic}, for
each $i\in[1,|\vl|]$, there is $u_i$ such that
$t_i\ab^*u_i=_\eta\la_\b^*l_i\s$. By \compred, $u_i$ is
computable. Therefore, we are left to prove that, if $u_i$ is
computable and $u_i=_\eta\la_\b^*l_i\s$, then $l_i\s$ is
computable. While computability is preserved by $\eta$-equivalence
(see Lemma \ref{lem-eta} below), it cannot be the case for arbitrary
$\b$-expansions because $\b$-expansion may introduce non-terminating
subterms.

In \cite{miller91jlc}, Section 9.1, Miller remarks that, if $t=_\be
l\s$, $l$ is a pattern \ala Miller, $t$ and $\s$ are in $\b$-normal
$\eta$-long form, then $t=_\boe l\s$, where $\abo$ is the restriction
of $\ab$ to redexes of the form $(\l xt)x$ (or, by $\al$-equivalence,
of the form $(\l xt)y$ with $y\in\cX$ and $\tau(x)=\tau(y)$). So, when
the left-hand sides of rules are patterns, matching modulo $\be$
reduces to matching modulo $\boe$. We now check that $\aboe$
terminates and is {\em strongly} confluent:

\begin{lem}
\label{lem-ae-confl}
$\ae$ terminates and is strongly confluent.
\end{lem}

\begin{prf}
  The relation $\ae$ terminates for it makes the size of terms
  decrease. Assume that $t\at[-1]{p}\la_\eta u\at{q}\ae v$.

\begin{itemize}
\item $p\#q$. Then, $t\ae\la_\eta v$.
\item $p=q$. Then, $t=v$.
\item $p>q$. Then, there are $a$ and $a'$ such that $u|_q=\l xax$,
  $x\notin\FV(a)$, $v=u[a]_q$, $a\ae a'$ and $t=u[\l xa'x]_q$. Since
  $\FV(a')\sle\FV(a)$, $x\notin\FV(a')$ and $t\ae u[a']_q\la_\eta v$.
\item $p<q$. By symmetry, $t\ae\la_\eta v$.\cqfd
\end{itemize}
\end{prf}

\begin{lem}
\label{lem-aboe-confl}
$\aboe$ terminates and is strongly confluent on well-typed terms.
\end{lem}

\begin{prf}
  The relation $\aboe$ terminates on well-typed terms for it is a
  sub-relation of $\abe$ which terminates on well-typed terms
  \cite{pottinger78ndj}. Assume that $t\at[-1]{p}\la_\boe
  u\at[4]{q}\aboe v$. If $p\#q$, then $t\aboe u\la_\boe v$.

\begin{itemize}
\item $t\at[-1]{p}\la_\eta u\at{q}\ae v$. Then, $t=v$ or $t\ae\la_\eta
  v$ by Lemma \ref{lem-ae-confl}.

\item $t\at[-0.5]{p}\la_\eta u\at{q}\abo v$.
\begin{itemize}
\item $p=q$. Not possible.
\item $p>q$. There is $a$ such that $u|_q=(\l xa)x$ and $v=u[a]_q$.
\begin{itemize}
\item $p=q0$. There is $d$ such that $a=dx$, $x\notin\FV(d)$ and
  $t=u[dx]_q$. Thus, $t=v$.
\item $p>q0$. There is $a'$ such that $a\ae a'$ and $t=u[(\l
  xa')x]_q$. Thus, $t\abo u[a']_q\la_\eta v$.
\item $p>q1$. Not possible.
\end{itemize}
\item $p<q$. There is $a$ such that $u|_p=\l xax$, $x\notin\FV(a)$ and
  $t=u[a]_p$.
\begin{itemize}
\item $p0=q$.\footnote{This is exactly the situation of Nederpelt's
    counter-example to the confluence of $\abe$ on untyped
    Church-style $\l$-terms \cite{nederpelt73phd}, which is in fact a
    counter-example to the confluence of $\aboe$ on untyped
    Church-style $\l$-terms.} There is $b$ such that $a=\l yb$ and
  $v=u[a_y^x]_p$. Since $u$ is well-typed, $\tau(x)=\tau(y)$ and, by
  $\al$-equivalence, we can assume wlog that $y=x$. Thus, $t=v$.
\item $p0<q$. There is $a'$ such that $a\abo a'$ and $v=u[\l
  xa'x]_p$. Thus, $t\abo u[a']_p\la_\eta v$.
\end{itemize}
\end{itemize}

\item $t\at[-0.5]{p}\la_\bo u\at{q}\abo v$.
\begin{itemize}
\item $p=q$. Then, $v=t$.
\item $p<q$. There are $a$ and $a'$ such that $u|_p=(\l xa)x$,
  $t=u[a]_p$, $a\abo a'$ and $v=u[(\l xa')x]_p$. Thus, $t\abo
  u[a']_p\la_\bo v$.
\item $p>q$. By symmetry, $t\abo\la_\bo v$.\cqfd
\end{itemize}
\end{itemize}
\end{prf}

Hence, if $t=_\boe l\s$, then $t\aboe^*\la_\boe^*l\s$\footnote{Note
  that $\ae$ cannot be postponed after $\abo$ as shown by the
  following example: $(\l xa)(\l yxy)\ae(\l xa)x\abo a$.}. Therefore,
we could try to prove that computability is preserved by
$\bo$-expansion, all the more so since that, for matching modulo
$\al$-equivalence, computability is preserved by {\em
  head}-$\bo$-expansion as shown by Lemma \ref{lem-comp-beta} (a
result that also holds with pattern matching modulo $\be$ under some
conditions on $\cR$ as we will see it in Lemma
\ref{lem-comp-beta-hopm} below). But this does not seem easy to prove
in general for two reasons.

First, a proof that $u$ is computable whenever $t\la_\bo u$ and $t$ is
computable, by induction on the size of $t$ does not seem to go
through. Indeed, assume that $u=\l xs$. Then, $t=\l xr$ and $r\la_\bo
s$. But $\l xs$ is computable if, for all $e\in\I{\tau(x)}$, $s_x^e$
is computable. Of course, $r_x^e$ is computable but we generally do
not have $r_x^e\la_\bo s_x^e$. We therefore need to consider not
$\bo$-expansion but a restricted form of $\b$-expansion that is stable
by instantiation of the bound variables of a pattern:

\newcommand\LPos{\mr{LPos}}
\renewcommand\sk{\mr{sk}}

\begin{dfn}[Leaf-$\b$-expansion]
  The set $\LPos(t)$ of the (disjoint) {\em leaf positions} of a term
  $t$ is defined as follows:
\begin{itemize}
\item $\LPos(t)=
\{0^{n-1}1p|p\in\LPos(t_1)\}\cup\ldots\cup\{1p|p\in\LPos(t_n)\}$
  if $t=\sf\,t_1\ldots t_n$ and $\sf\in\cF$;
\item $\LPos(t)=\{0p|p\in\LPos(u)\}$ if $t=\l xu$;
\item $\LPos(t)=\{\vep\}$ otherwise.
\end{itemize}
\noindent
Given a term $v$ and a leaf position $p\in\LPos(v)$, let the relation
of {\em $\b$-leaf-expansion} wrt $v$ at position $p$ be the relation
$t\la_{\b,v,p}u$ if there are $\vt,a,x,e,\vb$ such that
$t=v[\vt]_\vq[a_x^e\vb]_p$ and $u=v[\vt]_\vq[(\l xa)e\vb]_p$, where
$\vq$ are all the leaf positions of $v$ distinct from $p$.
\end{dfn}

Second, since we do not consider rewriting on terms in $\b$-normal
form, $l\s$ can contain some arbitrary $\b$-redex $(\l xa)b$ which,
after some $\boe$-reductions, becomes a $\bo$-redex because
$b\aboe^*x$. However, if $l$ is a pattern then such a $\b$-redex can
only occur in $\s$. Therefore, it is not needed to reduce it for
checking that $t$ matches $l$ modulo $\be$. For the sake of
simplicity, we will enforce this property in the definition of
rewriting itself by using the notion of {\em valuation} used for
defining rewriting in CRSs\footnote{In CRSs, $\ar$ is defined as the
  closure of $\cR$ by context and valuation (extended to all terms).}
\cite{klop93tcs}:

\begin{dfn}[Valuation]
\label{def-val}
A substitution $\s$ is {\em valid} wrt a term $t$ if, for all
$p\in\LPos(t)$, $x$ and $t_1,\ldots,t_n$ such that $t|_p=xt_1\ldots
t_n$, there are pairwise distinct variables $y_1,\ldots,y_n$ and a
term $a$ such that $x\s=\l y_1\ldots\l y_na$.  Let the {\em valuation}
of a term $t$ by a substitution $\s$, written $\hat\s(t)$, be the
term:
\begin{itemize}
\item $\l x\hat\s(u)$ if $t=\l xu$ and $\s$ is away from $\{x\}$;
\item $\sf~\hat\s(t_1)\ldots\hat\s(t_n)$ if $t=\sf\,t_1\ldots t_n$;
\item $a\{y_1\to t_1,\ldots,y_n\to t_n\}$ if $t=xt_1\ldots t_n$,
  $x\s=\l y_1\ldots\l y_n a$ and $y_1,\ldots,y_n$ are pairwise
  distinct variables.
\end{itemize}
\end{dfn}

\begin{lem}
\label{lem-val}
If $l$ is a pattern and $p_1,\ldots,p_n$ are the leaf positions of $l$
then, for all substitutions $\s$ valid wrt $l$, we have
$\h\s(l)\la_{\b,l,p_1}^*\ldots\la_{\b,l,p_n}^*l\s$.
\end{lem}

\begin{prf}
  Let $p$ be a leaf position of $l$. By definition of patterns, there
  are terms $\vt$ and pairwise distinct variables $x,\vy$ such that
  $l|_p=x\vt$, $x\in\FV(l)$ and $\vt\ae^*\vy$. Since $\s$ is valid wrt
  $l$, there is $a$ such that $x\s=\l\vy a$ and $\h\s(l)|_p=a\{y_1\to
  t_1,\ldots,y_n\to t_n\}\la_{\b,l,p}^*(\l\vy a)\vt=l\s|_p$.\cqfd
\end{prf}

Hence, valuation preserves typing: $\tau(\h\s(t))=\tau(t)$.

We now introduce our definition of rewriting with matching modulo
$\be$:

\begin{dfn}[Rewriting with pattern matching modulo $\be$]
  Given a set $\cR$ of rewrite rules of the form $\sf\vl\a r$ with
  $\sf\vl$ a pattern, let $t\arbe u$ if there are $p\in\Pos(t)$, $l\a
  r\in\cR$ and $\s$ such that $\tau(t|_p)=\tau(l)$, $\s$ is valid wrt
  $l$, $t|_p=_\eta\h\s(l)$ and $u=t[r\s]_p$.
\end{dfn}

\begin{lem}
The relation $\arbe$ is monotone and stable by substitution.
\end{lem}

\begin{prf}
Monotony is straightforward. We check that it is stable by
substitution. Assume that $t\arbe u$ and let $\t$ be a
substitution. There are $p\in\Pos(t)$, $l\a r\in\cR$ and $\s$ such
that $\tau(t|_p)=\tau(l)$, $\s$ is valid wrt $l$, $t|_p=_\eta\h\s(l)$
and $u=t[r\s]_p$. We have $\tau(t\t|_p)=\tau(t|_p)=\tau(l)$, $\s\t$
valid wrt $l$ and $t\t|_p=_\eta\h\s(l)\t$. We now prove that
$\h\s(l)\t=_\eta\h{\s\t}(l)$. Let $q\in\LPos(l)$. Since $l$ is a
pattern, $l|_q=x\vt$ where $x$ and $\vt\!\ad_\eta$ are pairwise
distinct variables and $\{\vt\!\ad_\eta\}\sle\BV(l,q)$. Since $\s$ is
valid wrt $l$, there is $a$ such that $x\s=\l\vy a$ and
$\h\s(l|_q)=_\eta a$. Wlog we can assume that $\t$ is away from
$\{\vy\}$. Therefore, $x\s\t=\l\vy a\t$ and $\h\s(l|_q)\t=_\eta
a\t=_\eta \h{\s\t}(l|_q)$. Therefore, $t\t\arbe u\t$.\cqfd
\end{prf}

%%%%%%%%%%%%%%%%%%%%%%%%%%%%%%%%%%%%%%%%%%%%%%%%%%%%%%%%%%%%%%%%%%%%%%%%%%%%%%
\subsection{Definition of computability}

Computability is straightforwardly extended to this new form of
rewriting as follows:

\begin{dfn}[Computability predicates for rewriting with matching modulo $\be$]
  Given a set $\cR$ of rewrite rules of the form $\sf\vl\a r$ with
  $\sf\vl$ a pattern, a term is {\em neutral} if it is of the form
  $x\vv$, $(\l xt)u\vv$ or $\sf\vv$ with
  $\sf\in\cD(\cR)-\cM(\cR)$\footnote{See Definition \ref{def-acc}.}
  and $|\vv|\ge\af=sup\{|\vl|\mid\ex r,\sf\vl\a r\in\cR\}$. Given a type
  $T$, let $\Red_{\cR,\be}^T$ be the set of all the sets $P\sle\cL^T$
  such that:

\begin{enumii}{R}
\item
$P\sle\SN(\a)$ where ${\a}={\ab\cup\arbe}$;
\item
$P$ is stable by $\a$;
\item
if $t:T$ is neutral and $\a\!(t)\sle P$, then $t\in P$.
\end{enumii}
\end{dfn}

\begin{lem}
\label{lem-red-boe}
For all type $T$, $\Red_{\cR,\be}^T$ is stable by non-empty
intersection and admits $\SN^T$ as greatest element. Moreover, for
all $T,U\in\cT$, $P\in\Red_{\cR,\be}^T$ and $Q\in\Red_{\cR,\be}^U$,
$\propto\!(P,Q)\in\Red_{\cR,\be}^{T\A U}$.
\end{lem}

\begin{prf}
  The proof is similar to the one of Lemma \ref{lem-cand}. One can
  easily check the stability by non-empty intersection and the fact
  that $\SN^T\in\Red_{\cR,\be}^T$. For the stability by $\propto$,
  there is no change for \compsn\ and \compred. We now detail
  \compneutral. Let $T,U\in\cT$, $P\in\Red_{\cR,\be}^T$,
  $Q\in\Red_{\cR,\be}^U$, $v:T\A U$ neutral such that
  ${\a\!(v)}\sle{\propto\!(P,Q)}$ and $t\in P$. We now show that
  $vt\in Q$ by well-founded induction on $t$ with $\a$ as well-founded
  relation ($t\in\SN$ by \compsn). Since $vt$ is neutral, by
  \compneutral, it suffices to prove that every reduct $w$ of $vt$ is
  in $Q$:
\begin{itemize}
\item $w=v't$ with $v\a v'$. By assumption,
  $v'\in{\propto\!(P,Q)}$. Therefore, $w\in Q$.
\item $w=vt'$ with $t\a t'$. By the induction hypothesis, $w\in Q$.
\item There are $\sf\vl\a r\in\cR$ and $\s$ such that
  $vt=_\eta\h\s(\sf\vl)=\sf\,\h\s(\vl)$ and $w=r\s$. By confluence of
  $\ae$, $(vt)\!\!\ad_\eta$ is of the form $\sf\vm$ with
  $|\vm|=|\vl|$. Since $v$ is neutral, $v$ is of the form $x\vt$, $(\l
  xa)b\vt$ or $\sg\vm$ with $\ag\le|\vm|$. We discuss these cases in
  turn:
\begin{itemize}
\item $v=x\vt$. Then, $v\!\!\ad_\eta$ is of the form $x\vu$. So, this
  case is not possible.

\item $v=(\l xa)b\vt$. Then, $a=_\eta cx$ with $x\notin\FV(c)$ and
  $cb\vt u=_\eta\h\s(\sf\vl)$. Hence, $v\ab v'=a_x^b\vt$, $v't=_\eta
  (cx)_x^b\vt u=cb\vt u=_\eta\h\s(\sf\vl)$ and $v't\a w$. Therefore,
  $w\in Q$ since ${v'}\in{\propto\!(P,Q)}$, $t\in P$ and $Q$ satisfies
  \compred.

\item $v=\sg\vm$ with $\ag\le|\vm|$. Then, $\sg=\sf$ and $|\vm|<|\vm
  u|=|\vl|$. Since $v$ is neutral, $\af\le|\vm|$. By definition of
  $\af$, $|\vl|\le\af$. So, this case is not possible.\cqfd
\end{itemize}
\end{itemize}
\end{prf}

\newcommand\asbe{\a_{\cS,\be}}

We now check that Lemma \ref{lem-comp-beta} still holds if the
following condition is satisfied:

\begin{dfn}
  A set $\cR$ of rules is {\em $\b$-complete} if, for all rules $l\a
  r\in\cR$ and types $T,U$ such that $l:T\A U$, there is
  $x\in\cX-\FV(l)$ such that $\tau(x)=T$ and:
\begin{itemize}
\item $lx\a s_y^x\in\cR$ if $r=\l ys$;\footnote{This case is not
  necessary for Lemma \ref{lem-comp-beta-hopm} to hold but avoids
  adding rules whose right-hand sides are $\b$-redexes.}
\item $lx\a rx\in\cR$ otherwise.
\end{itemize}
\end{dfn}

For instance, $\cR$ is $\b$-complete if, for every rule $l\a r\in\cR$,
$l$ is of base type. On the other hand, the set $\cR=\{\sf\a\l xx\}$
is not $\b$-complete since $\sf x\a x\notin\cR$.

\begin{lem}
\label{lem-comp-beta-hopm}
Assume that $\cR$ is $\b$-complete. Given $T\in\cT$ and
$P\in\Red_{\cR,\be}^T$, $(\l xt)u\vv\in P$ iff $(\l xt)u\vv:T$,
$t_x^u\vv\in P$ and $u\in\SN$.
\end{lem}

\begin{prf}
  Assume that $(\l xt)u\vv\in P$. By \compred, $t_x^u\vv\in P$. By
  \compsn, $(\l xt)u\vv\in\SN$. Therefore, $u\in\SN$.

  Assume now that $t_x^u\vv\in P$ and $u\in\SN$. By \compsn,
  $t_x^u\vv\in\SN$. Therefore, $\vv\in\SN$, $t_x^u\in\SN$ and
  $t\in\SN$. We now prove that, for all $t,u,\vv\in\SN$, $(\l
  xt)u\vv\in P$, by induction on $\a_\prod$. Since $(\l xt)u\vv$ is
  neutral, by \compneutral, it suffices to prove that every reduct $w$
  of $(\l xt)u\vv$ belongs to $P$. Since rules are of the form
  $\sf\vl\a r$, there are three possible cases:
\begin{itemize}
\item $w=t_x^u\vv$. Then, $w\in P$ by assumption.
\item $w=(\l xt')u'\vv'$ and $tu\vv\a_\prod t'u'\vv'$. Then, $w\in P$
  by the induction hypothesis.
\item There are $\sf\vl\a r\in\cR$ and $\s$ such that $\l
  xt=_\eta\h\s(\sf\vl)$ and $w=r\s u\vv$. By confluence of $\ae$,
  there is $a$ such that $t\ae^*ax$, $x\notin\FV(a)$ and
  $a=_\eta\h\s(\sf\vl)$. Wlog we can assume that $x\notin\FV(l)$ and
  $\s$ is away from $\{x\}$. Hence, $t=_\eta\h\s(\sf\vl x)$. Since
  $\cR$ is $\b$-complete, there are two cases:
\begin{itemize}
\item $r=\l ys$ and $\sf\vl x\a s_y^x\in\cR$. Then, $t\arbe
  s_y^x\s$. By monotony and stability by substitution,
  $t_x^u\vv\arbe(s_y^x\s)_x^u\vv$. Hence, $(s_y^x\s)_x^u\vv\in P$ by
  \compred. Therefore, by the induction hypothesis, $(\l
  xs_y^x\s)u\vv\in P$. Wlog we can assume that $\s$ is away from
  $\{y\}$. Hence, $(\l xs_y^x\s)u\vv=_\al r\s u\vv$.
\item $r$ is not an abstraction and $\sf\vl x\a rx\in\cR$. Then,
  $t\arbe (rx)\s=r\s x$. By monotony and stability by substitution,
  $t_x^u\vv\arbe r\s u\vv$. Hence, $r\s u\vv\in P$ by \compred.\cqfd
\end{itemize}
\end{itemize}
\end{prf}

\begin{cor}
\label{cor-comp-lam-hopm}
Assume that $\cR$ is $\b$-complete. Given $T,U\in\cT$,
$P\in\Red_{\cR,\be}^T$ and $Q\in\Red_{\cR,\be}^U$, ${\l
  xt}\in{\propto\!(Q,P)}$ iff $\l xt:U\A T$ and, for all $u\in Q$,
$t_x^u\in P$.
\end{cor}

\begin{prf}
  Assume that $\l xt\in\propto\!(Q,P)$ and $u\in Q$. Then, by
  definition of $\propto$, $(\l xt)u\in P$. Therefore, by \compred,
  $t_x^u\in P$. Assume now that, for all $u\in P$, $t_x^u\in P$. By
  definition of $\propto$, $\l xt\in\propto\!(Q,P)$ if, for all $u\in
  Q$, $(\l xt)u\in P$. By \compsn, $u\in\SN$. Therefore, by Lemma
  \ref{lem-comp-beta-hopm}, $(\l xt)u\in P$.\cqfd
\end{prf}

But $\b$-completeness is not a real restriction from the point of view
of termination since:

\begin{lem}
\label{lem-beta-completion}
  For every (finite) set of rules $\cS$, there is a (finite)
  $\b$-complete set of rules $\cR\sge\cS$ such that $\sf\vl\a
  r\in\CC_\sf(\vl)$ for every $\sf\vl\a r\in\cR$ if $\sf\vl\a
  r\in\CC_\sf(\vl)$ for every $\sf\vl\a r\in\cS$.
\end{lem}

\begin{prf}
  Let $F_\b$ be the function on the powerset of $\cT^2$ such that, for
  all $\cR\sle\cT^2$, $F_\b(\cR)$ is the smallest set such that
  $\cR\sle F_\b(\cR)$ and, for all $l\a r\in\cR$ and $T,U$ such that
  $l:T\A U$, there is $x\in\cX-\FV(l)$ such that $\tau(x)=T$, $lx\a
  s\in F_\b(\cR)$ if $r=\l xs$, and $lx\a rx\in F_\b(\cR)$
  otherwise. Since $F_\b$ is extensive (\ie $\cR\sle F_\b(\cR)$), by
  Hessenberg's fixpoint theorem \cite{hessenberg09crelle}, $F_\b$ has
  a fixpoint $\cR$ such that $\cS\sle\cR$. Since $\cR=F_\b(\cR)$,
  $\cR$ is $\b$-complete.

  Now, if $\cS=\{l_1\a r_1,\ldots,l_n\a r_n\}$ and, for every
  $i\in[1,n]$, $l_i:\vT^i\A\sA_i$ with $\sA_i\in\cB$, then
  $\card(\cR)\le n+\S_{i=1}^n|\vT^i|$.

  Assume now that $\sf\vl\a r\in\CC_\sf(\vl)$ for every $\sf\vl\a
  r\in\cS$, and that there are $\sf\vl\a r\in\cR$ and $T,U$ such that
  $\sf\vl:T\A U$. By assumption, $r\in\CC_\sf(\vl)$. Let now
  $x\in\cX-\FV(l)$. Wlog, we can assume that $x\notin\BV(r)$. Hence,
  $r\in\CC_\sf(\vl x)$. By (arg), $x\in\CC_\sf(\vl x)$. Therefore, by
  (app), $rx\in\CC_\sf(\vl x)$. Now, if $r=\l ys$, then
  $s_y^x\in\CC_\sf(\vl x)$ by (red).\cqfd
\end{prf}

Note moreover that ${\cS}\sle{\cR}\sle{\a_\cS\abo^=}$. Therefore,
$\ab\cup\arbe$ and $\ab\cup\asbe$ have the same normal forms and, if
$=_{\cR\b}$ (resp. $=_{\cS\b}$) is the smallest congruence containing
$\ab$ and $\cR$ (resp. $\cS$), then ${=_{\cR\b}}$ is equal to
${=_{\cS\b}}$.

\subsection{Preservation of computability by $\eta$-equivalence}

In this section, we prove that computability is preserved by
$\eta$-equivalence if ${\la_\eta\arbe}\sle{\arbe=_\eta}$. Then, we
give sufficient conditions for this commutation property to hold.

\begin{lem}
\label{lem-typ-ord}
Let $>_\cT$ be the smallest transitive relation on types containing
$>_\cB$ and such that $T\A U>_\cT T$ and $T\A U>_\cT U$. The relation
$>_\cT$ is well-founded.
\end{lem}

\begin{prf}
  Wlog we can assume that the symbol $\A$ is not a type
  constant. Then, let $\succ$ be the smallest transitive relation on
  $\cB\cup\{\A\}$ containing $>_\cB$ and such that ${\A}\succ{\sA}$
  for all $\sA\in\cB$. The relation $\succ$ is well-founded for
  $>_\cB$ is well-founded. Hence, $>_\cT$ is well-founded for it is
  included in the recursive path ordering (RPO) built over $\succ$
  \cite{dershowitz79focs}.\cqfd
\end{prf}

\begin{lem}
\label{lem-eta}
Let $\cR$ be a set of rules such that
${\la_\eta\arbe}\sle{\arbe=_\eta}$, and assume that types are
interpreted as in Section \ref{sec-match-def}. If $t:T$ is computable,
$t=_\eta u$ and $u:T$, then $u$ is computable.
\end{lem}

\begin{prf}
Note that, by Lemma \ref{lem-ae-confl}, $t\ae^*\la_\eta^*u$. Since $t$
and $u$ are well-typed and $\ae$ preserves typing, all terms between
$t$ and $u$ are of type $T$.

We then proceed by induction on (1) the type of $t$ ordered with
$>_\cT$ (well-founded by Lemma \ref{lem-typ-ord}), (2) the rank of $t$
(see Definition \ref{def-rank}) if $t$ is of base type, (3) $t$
ordered by $\a$ ($t\in\SN$ by \compsn), and (4) the number of
$\aa_\eta$-steps between $t$ and $u$.

If $T=V\a T'$, then $u$ is computable if, for all computable $v:V$,
$uv:T'$ is computable. By monotony, $tv=_\eta uv$. Since $tv:T'$ and
$T>_\cT T'$, $uv$ is computable by the induction hypothesis.

If $t=u$, then $u$ is computable. Assume now that $t=_\eta t'\aa_\eta
u$. By the induction hypothesis, $t'$ is computable. Therefore, we are
left to prove the lemma when $=_\eta$ is replaced by $\aa_\eta$.
 
Assume now that $T$ is a type constant $\sA$. By Lemma
\ref{lem-comp-base-type}, a term $a:\sA$ is computable iff all its
reducts are computable and, for all $\sf\in\cM(\cR)$, $i\in\Acc(\sf)$
and $\va$ such that $a=\sf\va$, $a_i$ is computable.

We first prove that, for all $\sf\in\cM(\cR)$, $i\in\Acc(\sf)$ and
$\vu$ such that $u=\sf\vu$, $u_i:\vV\A\sB$ is computable. Since $t$ is
of base type and $t\aa_\eta u=\sf\vu$, there are $\vt$ such that
$t=\sf\vt$ and $\vt~(\aa_\eta)_\prod~\vu$. Now, $u_i$ is computable
if, for all computable $\vv:\vV$, $u_i\vv$ is computable. By monotony,
$t_i\vv\aa_\eta^=u_i\vv$ and $t_i\vv$ has a type or a rank smaller
than the type or rank of $\sf\vt$ (for $i\in\Acc(\sf)$). Therefore,
$u_i\vv$ is computable by the induction hypothesis.

We now prove that all the reducts $v$ of $u$ are computable.

\begin{itemize}
\item $t\at{p}\ae u\at{q}\ab v$.\footnote{This case could be
    simplified and dealt with by \compred\ if $\ae$ was included in
    $\a$. But, then, we would have to check Lemma \ref{lem-red-boe}
    again. The present proof shows that this is not necessary.} We now
  prove that there is $t'$ such that $t\ab^+t'\ae^*v$, so that we can
  conclude by the induction hypothesis:
\begin{itemize}
\item $p\#q$. In this case, $t\ab\ae v$.

\item $p\le q$. There are $a$ and $a'$ such that $t|_p=\l xax$,
  $x\notin\FV(a)$, $u=t[a]_p$, $a\ab a'$ and $v=t[a']_p$. Thus, $t\ab
  t[\l xa'x]_p\ae v$.
\item $p>q$. There are $a$ and $b$ such that $u|_q=(\l xa)b$ and
  $v=u[a_x^b]_q$.
\begin{itemize}
\item $p\ge q1$. There is $d$ such that $t=u[(\l xa)d]_q$ and $d\ae
  b$. Thus, $t\ab u[a_x^d]\ae^*v$.
\item $p=q0$. Then, $t=u[(\l x(\l xa)x)b]_q$. Thus, $t\abo u\ab v$.
\item $p>q0$. There is $d$ such that $t=u[(\l xd)b]_q$ and $d\ae
  a$. Thus, $t\ab u[d_x^b]\ae v$.
\end{itemize}
\end{itemize}

\item $t\at{p}\ae u\at[9]{q}\arbe v$. We now prove that there is $t'$
  such that $t\arbe t'\ae^=v$, so that we can conclude by the
  induction hypothesis.
\begin{itemize}
\item $p\#q$. Then, $t\arbe t'\ae v$.
\item $p\ge q$. Then, $t\arbe v$.
\item $p<q$. There are $a$ and $a'$ such that $t|_p=\l xax$,
  $x\notin\FV(a)$, $u=t[a]_p$, $a\arbe a'$ and $v=t[a']_p$. Thus,
  $t\arbe t[\l xa'x]_p\ae v$.
\end{itemize}

\item $t\at[-0.5]{p}\la_\eta u\at[9]{q}\arbe v$. By assumption, there
  is $t'$ such that $t\a t'=_\eta v$, so that we can conclude by the
  induction hypothesis.

\item $t\at[-0.5]{p}\la_\eta u\at{q}\ab v$. We now prove that, either
  $v=t$ and $v$ is computable for $t$ is computable, or there is $t'$
  such that $t\ab t'\la_\eta^*v$ and we can conclude by the induction
  hypothesis:
\begin{itemize}
\item $p\#q$. Then, $t\ab t'\la_\eta v$.
\item $p=q$. Not possible.
\item $p>q$. There are $a$ and $b$ such that $u|_q=(\l xa)b$ and
  $v=u[a_x^b]_q$.
\begin{itemize}
\item $p=q0$. There is $d$ such that $a=dx$, $x\notin\FV(d)$ and
  $t=u[db]_q$. Thus, $t=v$.
\item $p>q0$. There is $a'$ such that $a\ae a'$ and $t=u[(\l
  xa')b]_q$. Thus, $t\ab u[{a'}_x^b]_q\la_\eta v$.
\item $p\ge q1$. There is $b'$ such that $b\ae b'$ and $t=u[(\l
  xa)b']_q$. Thus, $t\ab u[a_x^{b'}]_q\la_\eta^*v$.
\end{itemize}
\item $p<q$. There is $a$ such that $u|_p=\l xax$, $x\notin\FV(a)$ and
  $t=u[a]_p$.
\begin{itemize}
\item $p0=q$. There is $b$ such that $a=\l yb$ and $v=u[\l
  xb_y^x]_p$. As already mentioned in Lemma \ref{lem-aboe-confl},
  since $u$ is well-typed, we can assume wlog that $y=x$. Thus, $t=v$.
\item $p0<q$. There is $a'$ such that $a\ab a'$ and $v=u[\l
  xa'x]_p$. Thus, $t\ab u[a']_p\la_\eta v$.\cqfd
\end{itemize}
\end{itemize}
\end{itemize}
\end{prf}

In the previous proof, we have seen that
${\ae\arbe}\sle{\arbe\ae^=}$. Hence, if we also have
${\la_\eta\arbe}\sle{\arbe=_\eta}$, then
${\aa_\eta\arbe}\sle{\arbe^+=_\eta}$, a property that, after
\cite{jouannaud84cade}, we call:

\begin{dfn}
A relation $R$ {\em locally $\eta$-commutes} if ${\aa_\eta
  R}\sle{R^+=_\eta}$.
\end{dfn}

We now provide sufficient conditions for this property to hold:

\begin{dfn}
  A set $\cR$ of rules is {\em $\eta$-complete} if, for all $l,k,r,x$
  such that $lk\a r\in\cR$, $k\ae^*x$ and $x\in\cX-\FV(l)$, we have:
\begin{itemize}
\item $l\a s\in\cR$ if $r=sk'$, $k'\ae^*x$ and
  $x\notin\FV(s)$;\footnote{This case is not necessary for Lemma
  \ref{lem-eta} to hold but avoids adding rules whose right-hand sides
  are $\eta$-redexes.}
\item $l\a\l xr\in\cR$ otherwise.
\end{itemize}
\end{dfn}

\begin{lem}
If $\cR$ is $\eta$-complete, then ${\la_\eta\arbe}\sle{\arbe=_\eta}$
and $\arbe$ locally $\eta$-commutes.
\end{lem}

\begin{prf}
  Assume that $t\at[0]{p}\la_\eta u\at[9]{q}\arbe v$.
\begin{itemize}
\item $p\#q$. Then, $t\arbe\la_\eta v$.
\item $p\ge q$. Then, $t\arbe v$.
\item $p<q$. There is $a$ such that $u|_p=\l xax$, $x\notin\FV(a)$ and
  $t=u[a]_p$.
\begin{itemize}
\item $p01\le q$. Not possible since the rules are of the form
  $\sf\vl\a r$.
\item $p00\le q$. There is $a'$ such that $v=u[\l xa'x]_p$ and
  $a\arbe a'$. Then, $t\arbe u[a']_p\la_\eta v$.
\item $p0=q$. There are $\sf\vl\a r\in\cR$ and $\s$ such that
  $ax=_\eta\h\s(\sf\vl)$ and $v=u[\l xr\s]_p$. By confluence of $\ae$,
  there are $\vm$ and $k$ such that $\vl=\vm k$, $a=_\eta\h\s(\sf\vm)$
  and $x=_\eta\h\s(k)$. Since $k$ is a pattern, there is $y\in\cX$
  such that $k\ae^*y$ and $y\s\ae^*x$. Wlog we can assume that
  $y=x$. Let $\t$ be the restriction of $\s$ on $\FV(\sf\vm)$. Since
  $x\notin\FV(a)$ and the set of free variables of a term is invariant
  by $=_\eta$, we have $x\notin\FV(\vm)$ and $\t$ away from
  $\{x\}$. Now, since $\cR$ is $\eta$-complete, there are two cases:
\begin{itemize}
\item $r=sk'$, $k'\ae^*x$, $x\notin\FV(s)$ and $\sf\vm\a
  s\in\cR$. Then, $a\arbe s\t$ and $\FV(s\t)\sle\FV(a)$. Since
  $x\notin\FV(a)$, $x\notin\FV(s\t)$ and $s\t\la_\eta \l xs\t
  x$. Since $x\la_\eta^*x\s$ and $x\la_\eta^*k'$, we have
  $x\la_\eta^*k'\s$. Therefore, $t\arbe\la_\eta^*u[\l xs\t k'\s]=v$.
\item Otherwise, $\sf\vm\a\l xr\in\cR$. Hence, $a\arbe(\l
  xr)\t$. Since $\t$ is away from $\{x\}$, $(\l xr)\t=\l xr\t$. Since
  $x=x\t\la_\eta^*x\s$, $r\t\la_\eta^*r\s$. Therefore,
  $t\arbe\la_\eta^*v$.\cqfd
\end{itemize}
\end{itemize}
\end{itemize}
\end{prf}

For instance, $\cR=\{\sf x\a x\}$ is not $\eta$-complete since
$\sf\a\l xx\notin\cR$ and, indeed, the relation $\la_\eta$ does not
commute with $\arbe$ because of the non-joinable critical pair
$\sf\la_\eta\l x\sf x\ar\l xx$. Adding the rule $\sf\a\l xx$ allows us
to recover commutation.

But $\eta$-completeness is not a real restriction from the point of
view of termination since:

\begin{lem}
\label{lem-eta-completion}
  For every (finite) set of rules $\cS$, there is an $\eta$-complete
  (finite) set of rules $\cR\sge\cS$ such that, using the rules of
  Figure \ref{fig-cc-hopm}, $\sf\vl\a r\in\CC_\sf(\vl)$ for every
  $\sf\vl\a r\in\cR$ if $\sf\vl\a r\in\CC_\sf(\vl)$ for every
  $\sf\vl\a r\in\cS$.
\end{lem}

\begin{prf}
  Let $F_\eta$ be the function on the powerset of $\cT^2$ such that,
  for all $\cR\sle\cT^2$, $F_\eta(\cR)$ is the smallest set such that
  $\cR\sle F_\eta(\cR)$ and, for all $l,k,r,x$ such that $lk\a
  r\in\cR$, $k\ae^*x$ and $x\in\cX-\FV(l)$, $l\a s\in F_\eta(\cR)$ if
  $r=sk'$, $k'\ae^*x$ and $x\notin\FV(s)$, and $l\a\l xr\in
  F_\eta(\cR)$ otherwise.

  Since $F_\eta$ is extensive (\ie $\cR\sle F_\eta(\cR)$), by
  Hessenberg's fixpoint theorem \cite{hessenberg09crelle}, $F_\eta$
  has a fixpoint $\cR$ such that $\cS\sle\cR$. Since
  $\cR=F_\eta(\cR)$, $\cR$ is $\eta$-complete.

  If $\cS=\{l_1\a r_1,\ldots,l_n\a r_n\}$ and, for every $i\in[1,n]$,
  $l_i:\vT^i\A\sA_i$ with $\sA_i\in\cB$, then $\card(\cR)\le
  n+\S_{i=1}^n|\vT^i|$.

  Assume now that $\sf\vl\a r\in\CC_\sf(\vl)$ for every $\sf\vl\a
  r\in\cS$, and that there are $\sf\vl k\a r\in\cR$ and
  $x\in\cX-\FV(\vl)$ such that $k\ae^*x$. By assumption,
  $r\in\CC_\sf(\vl k)$. By (var), $x\in\CC_\sf(\vl)$. Therefore, by
  (eta), $k\in\CC_\sf(\vl)$. Now, since $x\in\cX-\FV(\vl)$, we can get
  $r\in\CC_\sf(\vl)$ by replacing, everywhere in the derivation proof
  of $r\in\CC_\sf(\vl k)$, $\CC_\sf(\vl k)$ by $\CC_\sf(\vl)$, and the
  proofs of $k\in\CC_\sf(\vl k)$ obtained with (arg), by the proof of
  $k\in\CC_\sf(\vl)$ obtained with (var) and (eta). Therefore, by
  (abs), $\l xr\in\CC_\sf(\vl)$. Now, if $r=sk'$ with $k'\ae^*x$ and
  $x\notin\FV(s)$, then $s\in\CC_\sf(\vl)$ by (eta).\cqfd
\end{prf}

The fact that the rules of Figure \ref{fig-cc-hopm} are valid
computability closure operations is proved in \cite{blanqui00rta}.

Note that ${\cS}\sle{\cR}\sle{\la_\eta^*\cS\ae^*}$. Hence, if
$=_{\cR\be}$ (resp. $=_{\cS\be}$) is the smallest congruence
containing $\ae$, $\ab$ and $\cR$ (resp. $\cS$), then ${=_{\cR\be}}$
is equal to ${=_{\cS\be}}$. Moreover, $\arbe$ and $\asbe$ have the
same normal forms on $\eta$-long terms.

We have seen that termination of rewriting with matching modulo $\be$
relies on commutation properties between $\arbe$ and $\aa_\be$. Such
conditions are well-known in first-order rewriting theory: the notion
of compatibility of Peterson and Stickel \cite{peterson81jacm}, the
notion of local $E$-commutation of Jouannaud and Muñoz
\cite{jouannaud84cade} and, more generally, the notion of local
coherence modulo $E$ of Jouannaud and Kirchner
\cite{jouannaud86siam}. Similarly, the addition of {\em extension
  rules} to make a system compatible, locally commute or locally
coherent is also well-known since Lankford and Ballantyne
\cite{lankford77tr}.

%%%%%%%%%%%%%%%%%%%%%%%%%%%%%%%%%%%%%%%%%%%%%%%%%%%%%%%%%%%%%%%%%%%%%%%%%%%%%%
\subsection{Preservation of computability by leaf-$\b$-expansion}

We now prove that computability is preserved by leaf-$\b$-expansion,
but for patterns containing undefined symbols {\em only}.

\begin{dfn}
  Let $v$ be a term, $p\in\LPos(v)$ and $\vq$ be the leaf positions of
  $v$ distinct from $p$. We say that a term $t$ is {\em valid} wrt
  $(v,p)$ if there are $\vt$ and $u$ such that $t=v[\vt]_\vp[u]_p$
  and, for all $\vy$, $a$ and $\vb$ such that $u=(\l\vy a)\vb$ and
  $|\vy|=|\vb|$, we have $\vb\in\I{\tau(\vy)}$ and, for all
  $j\in[1,|\vb|]$, either ${b_j\!\!\ad_\eta}\in\BV(l,p)$ or
  $\FV(b_j)\cap\BV(l,p)=\vide$.
\end{dfn}

Note that, if $l$ is a pattern and $\s$ is valid wrt $l$, then every
term $t$ such that $\h\s(l)\la_{\b,l,p_1}^*\ldots\la_{\b,l,p_k}^*t$,
where $p_1,\ldots,p_k$ are leaf positions of $l$, is valid wrt
$(l,p_1),\ldots,(l,p_k)$ (for ${b_j\!\!\ad_\eta}\in\BV(l,p_i)$ in this
case).

\begin{lem}
\label{lem-leaf-beta-exp}
Let $\cR$ be a $\b$ and $\eta$-complete set of rules, and assume that
types are interpreted as in Section \ref{sec-match-def}. Let $l$ be a
term containing undefined symbols only, and let $p$ be a leaf position
of $l$. If $t\in\I{\tau(l)}$, $t\la_{\b,l,p}u$ and $u$ is valid wrt
$(l,p)$, then $u\in\I{\tau(l)}$.
\end{lem}

\begin{prf}
  Let $S=\I{\tau(l)}$. Note that $l$ does not need to be a pattern, a
  property that cannot be preserved when instantiating bound
  variables. In fact, the complete structure of $l$ is not
  relevant. Because we look at leaf-$\b$-expansions, only the top part
  of $l$ that is above the leaf positions is relevant. Hence, let
  $||\_||$ be the measure on terms defined as follows:
\begin{itemize}
\item $||l||=1+||m||$ if $l=\l zm$,
\item $||l||=1+\sup\{||l_1||,\ldots,||l_n||\}$ if $l=\sf l_1\ldots
  l_n$ and $n\ge 1$,
\item $||l||=0$ otherwise.
\end{itemize}

We prove the lemma by induction on (1) $||l||$, (2) $\tau(l)$, (3) $t$
ordered by $\a$ (for $t\in\SN$ by \compsn), and (4) the terms $\vb$
such that $u|_p=(\l\vy a)\vb$ (for $u$ is valid wrt $(l,p)$) ordered
by $\a$ (for $\vb\in\SN$ by \compsn). We proceed by case on $l$:

\begin{itemize}
\item $||l||=0$. Then, there are $a,x,e,\vb$ such that $t=a_x^e\vb$
  and $u=(\l xa)e\vb$. Since $u$ is valid, $e\in\I{\tau(x)}$. Hence,
  $e\in\SN$ by \compsn. Therefore, by Lemma \ref{lem-comp-beta-hopm},
  $u\in S$.

\item $l=\l zm$. Then, there are $r,s,q$ and $M$ such that
  $l:\tau(z)\A M$, $t=\l zr$, $u=\l zs$ and $r\la_{\b,m,q}s$. That is,
  ${S}={\propto\!(\I{\tau(z)},\I{M})}$ and there are $\vt$, $a$, $x$, $e$,
  $\vb$ such that $r=m[\vt]_\vk[a_x^e\vb]_q$ and $s=m[\vt]_\vk[(\l
  xa)e\vb]_q$, where $\vk$ are all the leaf positions of $m$ distinct
  from $q$. By Corollary \ref{cor-comp-lam-hopm}, $u\in S$ if, for all
  $g\in\I{\tau(z)}$, $s_z^g=m[\vt_z^g]_\vk[(\l
  xa)_z^ge_z^g\vb_z^g]_q\in\I{M}$. So, let $g\in\I{\tau(z)}$. By
  Corollary \ref{cor-comp-lam-hopm},
  $r_z^g=m[\vt_z^g]_\vk[(a_x^e)_z^g\vb_z^g]_q\in\I{M}$. Let
  $b_0=e$. Since $u$ is valid and $z\in\BV(l,p)$, for all
  $i\in[0,|\vb|]$, either $b_i\ae^*z$ and $b_i(_z^g)\ae^*g$, or
  $z\notin\FV(b_i)$ and $b_i(_z^g)=b_i$. Therefore, $s_z^g$ is
  valid. Wlog we can assume that $x\neq z$ and $x\notin\FV(g)$. Hence,
  $(\l xa)_z^g=\l xa_z^g$ and
  $(a_x^e)_z^g=(a_z^g)_x^{e_z^g}$. Therefore, $r_z^g\la_{\b,m,q}s_z^g$
  and, by the induction hypothesis (1), $s_z^g\in\I{M}$.

\item $l=\sf\vl$ with $\tau(\sf)=\vT\A U$. We proceed by case on $\tau(l)$:

\begin{itemize}
\item $\tau(l)=V\A W$. By definition of computability, $u\in S$ if,
  for all $v\in\I{V}$, $uv\in\I{W}$. So, let $v\in\I{V}$. Then,
  $tv\in\I{W}$ and $tv\la_{\b,lx,0p}uv$. Moreover, $uv$ is valid wrt
  $(lx,0p)$ and $||lx||=||l||$. Therefore, by the induction hypothesis
  (2), $uv\in\I{W}$.

\item $\tau(l)\in\cB$. Since $t\la_{\b,l,p}u$, there are $i,q,\vt,\vu$
  such that $p=0^{|\vl|-i}1q$, $t=\sf\vt$, $u=\sf\vu$ and
  $t_i\la_{\b,l_i,q}u_i$, that is, there are $a,x,e,\vb$ such that
  $t_i|_q=a_x^e\vb$ and $u_i|_q=(\l xa)e\vb$.

  By Lemma \ref{lem-comp-base-type}, $u\in S$ if all its reducts are
  in $S$ and, if $\sf\in\cM(\cR)$ and $i\in\Acc(\sf)$, then
  $u_i\in\I{T_i}$. Assume that $\sf\in\cM(\cR)$ and
  $i\in\Acc(\sf)$. By Lemma \ref{lem-comp-base-type},
  $t_i\in\I{T_i}$. If $t_i=u_i$ then $u_i\in\I{T_i}$. Otherwise,
  $t_i\la_{\b,l_i,q}u_i$. Therefore, since $u_i$ is valid wrt
  $(l_i,q)$, by the induction hypothesis (1), $u_i\in\I{T_i}$. We now
  prove that, if $u\at[1]{q}\a v$, then $v\in S$:

\begin{itemize}
\item $p\#q$. Then, $t\a t'\la_{\b,l,p}v$. By \compred, $t'\in
  S$. Since $t'$ is valid wrt $(l,p)$, by the induction hypothesis
  (3), $v\in S$.

\item $p>q$. Not possible since $l$ contains undefined symbols only.

\item There are $\sf\vl\a r\in\cR$ and $\t$ such that $\tau(\l
  xa)=\tau(\sf\vl)$, $\l xa=_\eta\h\t(\sf\vl)$ and $v=r\t e\vb$. Wlog
  we can assume that $x\notin\FV(\sf\vl)$ and $\t$ is away from
  $\{x\}$. Then, as already seen in the proof of Lemma
  \ref{lem-comp-beta-hopm}, $a=_\eta\h\t(\sf\vl x)$. Since $\cR$ is
  $\b$-complete, there are two cases:
\begin{itemize}
\item There is $s$ such that $r=\l xs$. Then, $\sf\vl x\a s\in\cR$ and
  $a\arbe s\t$. Hence, $t\a t'=u[(s\t)_x^e\vb]_p\la_{\b,l,p}u[(\l
  xs\t)e\vb]_p=v$. By \compred, $t'\in S$. Since $v$ is valid wrt
  $(l,p)$, by the induction hypothesis (3), $v\in S$.
\item Otherwise, $\sf\vl x\a rx\in\cR$ and $a\arbe r\t x$. Hence, $t\a
  t'=u[(r\t x)_x^e\vb]_p\la_{\b,l,p}u'=u[(\l xr\t x)e\vb]_p$ $\ae
  v$. By \compred, $t'\in S$. Since $u'$ is valid wrt $(l,p)$, by the
  induction hypothesis (3), $u'\in S$. Therefore, by Lemma
  \ref{lem-eta}, $v\in S$.
\end{itemize}

\item There is $a'$ such that $a\a a'$ and $v=u[(\l
  xa')e\vb]_p$. Then, $t\a t'=u[{a'}_x^e\vb]_p\la_{\b,l,p}v$. By
  \compred, $t'\in S$. Since $v$ is valid wrt $(l,p)$, by the
  induction hypothesis (3), $v\in S$.

\item There is $e'$ such that $e\a e'$ and $v=u[(\l
  xa)e'\vb]_p$. Then, $t\a^*t'=u[a_x^{e'}\vb]_p\la_{\b,l,p}v$. By
  \compred, $t'\in S$. Since $u$ is valid wrt $(l,p)$,
  $e\in\I{\tau(x)}$. By \compred, $e'\in\I{\tau(x)}$. Therefore, $v$
  is valid wrt $(l,p)$ and, by the induction hypothesis (4), $v\in S$.

\item There is $\vb'$ such that $\vb\a_\prod\vb'$ and $v=u[(\l
  xa)e\vb']_p$. Then, $t\a t'=u[a_x^e\vb']_p\la_{\b,l,p}v$. Since $u$
  is valid wrt $(l,p)$, $\vb$ are computable. Thus, by \compred,
  $\vb'$ are computable and $v$ is valid. Therefore, by the induction
  hypothesis (3), $v$ is computable.\cqfd
\end{itemize}
\end{itemize}
\end{itemize}
\end{prf}

Finally, we check that $\b$ and $\eta$-completion commute when left
and right-hand sides are $\be$-normal. Hence, any (finite) set of
rules $\cS$ whose left-hand and right-hand sides are $\be$-normal can be
completed into a (finite) $\b$ and $\eta$-complete set of rules
$\cR\sge\cS$.

\begin{lem}
  $\b$-completion (resp. $\eta$-completion) preserves
  $\eta$-completeness (resp. $\b$-completeness when left-hand and right-hand
  sides are $\be$-normal).
\end{lem}

\begin{prf}
\begin{itemize}
\item We will say that $\cR$ is $\be$-normal if, for every rule $l\a
  r\in\cR$, both $l$ and $r$ are $\be$-normal. We first prove that the
  function $F_\eta$ defined in the proof of Lemma
  \ref{lem-eta-completion} preserves $\b$-completeness and
  $\be$-normality. Let $\cR$ be a $\be$-normal and $\b$-complete set
  of rules. We have to prove that $F_\eta(\cR)$ is $\be$-normal and
  $\b$-complete, that is, if there are $l\a r\in F_\eta(\cR)$ and
  $T,U\in\cT$ such that $l:T\A U$, then there is $x\in\cX-\FV(l)$ such
  that $\tau(x)=T$ and, either $r=\l ys$ and $l x\a s_y^x\in
  F_\eta(\cR)$, or $lx\a rx\in F_\eta(\cR)$. Let $l\a r\in
  F_\eta(\cR)-\cR$ and assume that there is $gk\a d\in\cR$ such that
  $k\ae^*x\in\cX-\FV(g)$. Then, either:
\begin{itemize}
\item $d=sk'$, $k'\ae^*x\in\cX-\FV(s)$ and $l\a r=g\a s$. Since $\cR$
  is $\be$-normal, $k=k'=x$ and $r$ is not an abstraction. Therefore,
  $lx\a rx\in F_\eta(\cR)$ since $lx=gk$, $rx=sk'=d$ and $gk\a
  d\in\cR$. Moreover, $l$ is $\be$-normal since $l=g$ and $gk$ is
  $\be$-normal, and $r$ is $\be$-normal since $r=s$ and $d=sk'$ is
  $\be$-normal.

\item $l\a r=g\a\l xd$. Since $\cR$ is $\be$-normal, $k=x$. Therefore,
  $lx\a d\in F_\eta(\cR)$ since $lx=gk$ and $gk\a d\in\cR$. Moreover,
  $l$ is $\be$-normal since $l=g$ and $g$ is $\be$-normal, and $r$ is
  $\be$-normal since $r=\l xd$, $d$ is $\be$-normal and $d$ is not of
  the form $sk'$ with $k'\ae^*x\in\cX-\FV(s)$.
\end{itemize}

\item We now prove that the function $F_\b$ defined in the proof of
  Lemma \ref{lem-beta-completion} preserves $\eta$-completeness. Let
  $\cR$ be an $\eta$-complete set of rules. We have to prove that
  $F_\b(\cR)$ is $\eta$-complete, that is, if $lk\a r\in F_\b(\cR)$
  and $k\ae^*x\in\cX-\FV(l)$ then, either $r=tk'$,
  $k'\ae^*x\in\cX-\FV(t)$ and $l\a t\in F_\b(\cR)$, or $l\a\l xr\in
  F_\b(\cR)$. Let $l\a r\in F_\b(\cR)-\cR$ and assume that there are
  $g\a d\in\cR$ and $T,U\in\cT$ such that $g:T\A U$. Then, there is
  $x\in\cX-\FV(g)$ such that $x:T$ and either:
\begin{itemize}
\item $d=\l ys$ and $lk\a r=gx\a s_y^x$. Wlog we can assume that
  $y=x$. If $r=tk'$ and $k'\ae^*x\in\cX-\FV(t)$, then $d$ is not
  $\be$-normal. Therefore, $l\a\l xr\in F_\b(\cR)$ since $l=g$, $r=s$
  and $g\a\l xs\in\cR$.

\item $lk\a r=gx\a dx$. Therefore, $l\a d\in F_\b(\cR)$ since $l=g$
  and $g\a d\in\cR$.\cqfd
\end{itemize}
\end{itemize}
\end{prf}

%%%%%%%%%%%%%%%%%%%%%%%%%%%%%%%%%%%%%%%%%%%%%%%%%%%%%%%%%%%%%%%%%%%%%%%%%%%%%% 
\subsection{Handling the subterms of a pattern}

We now show that Theorem \ref{thm-cc-rec} extends to rewriting with
pattern matching modulo $\be$:

\begin{figure}[ht]
\caption{Computability closure operations VI\label{fig-cc-hopm}}
\begin{center}
\fbox{\begin{tabular}{rl}
(subterm-abs)&if $\l xt\in\CC_\sf(\vl)$ and $x\in\cX-\FV(\vl)$,
then $t\in\CC_\sf(\vl)$\\
(subterm-app)&if $tx\in\CC_\sf(\vl)$ and $x\in\cX-(\FV(t)\cup\FV(\vl))$,
then $t\in\CC_\sf(\vl)$\\
(eta)&if $t\in\CC_\sf(\vl)$, $t=_\eta u$ and $\tau(t)=\tau(u)$,
then $u\in\CC_\sf(\vl)$\\
\end{tabular}}
\end{center}
\end{figure}

\begin{thm}
\label{thm-cc-hopm}
Given a set of rules $\cR$ that is both $\b$ and $\eta$-complete, the
relation ${\ab}\cup{\arbe}$ terminates on well-typed terms if there is
an $\cF$-quasi-ordering $\ge$ valid wrt the interpretation of Section
\ref{sec-match-def} such that, for every rule $\sf\vl\a r\in\cR$,
$\vl$ are patterns containing undefined symbols only and
$r\in\CC_\sf(\vl)$, where $\CC$ is the smallest computability closure
closed by the operations I to VI.
\end{thm}

\begin{prf}
  We proceed as for Theorem \ref{thm-cc-rec} by showing that, for all
  $(\sf,\vt)\in\S_\max$, every reduct $t$ of $\sf\vt$ is computable,
  by well-founded induction on ${>}\cup{\a_\prod}$. There are two
  cases:

\begin{itemize}
\item There is $\vu$ such that $t=\sf\vu$ and $\vt\a_\prod\vu$. By
  \compred, $\vu$ is computable. Therefore, by the induction
  hypothesis, $\sf\vu$ is computable.

\item There are $\vs$, $\vw$, $\sf\vl\a r\in\cR$ and $\s$ such that
  $\vt=\vs\vw$, $\vs=_\eta\h\s(\vl)$ and $t=r\s\vw$. By Lemma
  \ref{lem-eta}, $\h\s(\vl)$ are computable. Let now $i\in[1,|\vl|]$
  and $p_1,\ldots,p_n$ be the leaf positions of $l_i$. By Lemma
  \ref{lem-val}, we have
  $\h\s(l_i)\la_{\b,l,p_1}^*\ldots\la_{\b,l,p_n}^*l_i\s$. Since all
  the terms between $\h\s(l_i)$ and $l_i\s$ are valid, by Lemma
  \ref{lem-leaf-beta-exp}, $l_i\s$ is computable. Since
  $r\in\CC_\sf(\vl)$ and $\CC$ is stable by substitution (for $>$ is
  stable by substitution), we have $r\s\in\CC_\sf(\vl\s)$. Now, Lemma
  \ref{lem-cc} is easily extended with the rules of Figure
  \ref{fig-cc-hopm} for destructuring patterns
  \cite{blanqui00rta}. Therefore, $r\s$ is computable since, for all
  $(\sg,\vu)\in\S_\max$, if $(\sf,\vt)>(\sg,\vu)$, then $\sg\vu$ is
  computable by the induction hypothesis.\cqfd
\end{itemize}
\end{prf}

For instance, let us check that these conditions are satisfied by the
formal derivation rule given at the beginning of the section. Let
$l=\l x\,\sin(Fx)$ and assume that $\sD>_\cF\times$. By (arg),
$l\in\CC=\CC_\sD(l)$. By (var), $x\in\CC$. By (subterm-abs),
$\sin(Fx)\in\CC$. By (subterm-acc), $Fx\in\CC$. By (subterm-app),
$F\in\CC$. By (undef), $\cos(Fx)\in\CC$. By (rec), $\sD Fx\in\CC$ for
${l}\tgt_\ss{F}$. By (rec), $(\sD Fx)\times(\cos(Fx))\in\CC$ for
$\sD>_\cF\times$. Therefore, by (abs), $\l x(\sD
Fx)\times(\cos(Fx))\in\CC$.

%%%%%%%%%%%%%%%%%%%%%%%%%%%%%%%%%%%%%%%%%%%%%%%%%%%%%%%%%%%%%%%%%%%%%%%%%%%%%%
\subsection{Application to CRSs and HRSs}
\label{sec-hrs}

CRSs \cite{klop80phd,klop93tcs} can be seen as an extension of the
untyped $\l$-calculus with no object-level application symbol but,
instead, symbols of fixed arity defined by rules using a matching
mechanism equivalent to matching modulo $\be$ on Miller patterns.

In HRSs \cite{nipkow91lics,mayr98tcs}, one considers simply-typed
$\l$-terms in $\b$-normal $\eta$-long form with symbols defined by
rules using Miller's pattern-matching mechanism.

Note that, although HRS terms are simply typed, one can easily encode
the untyped $\l$-calculus in it by considering an object-level
application symbol. Similarly, in CRSs, one easily recovers the
untyped $\l$-calculus by considering an object-level application
symbol. Such a CRS is called $\b$-CRS in \cite{blanqui00rta}.

In HALs \cite{jouannaud91lics}, Jouannaud and Okada consider arbitrary
typed $\l$-terms with function symbols of fixed arity defined by
rewrite rules, and computation is defined as the combination of
$\b$-reduction and rewriting.

These three approaches can be seen as operating on the same term
algebra ($\l$-calculus with symbols of fixed arity, which is a
sub-algebra of the one we consider here) with different reduction
strategies wrt $\b$-reduction \cite{oostrom93hoa}: in HALs, there is
no restriction; in CRSs, every rewrite step is followed by a
$\b$-development of the substituted variables (see the notion of
valuation in Definition \ref{def-val}); finally, in HRSs, terms are
$\b$-normalized.

More precisely, in a CRS, a term is either a variable $x$, an
abstraction $\l xt$, or the application of a function symbol $\sf$ to
a fixed number of terms. A CRS term is therefore in $\b$-normal
form. The set of CRS terms is a subset of the set of terms that is
stable by reduction or expansion (if matching substitutions are
restricted to CRS terms). Only rewrite rules can contain terms of the
form $x\vt$, but every rewrite step is followed by a
$\b$-development. Hence, the termination of a CRS can be reduced to
the termination of the corresponding HAL, because a rewrite step in a
CRS is included in the relation $\arbe\cup\ab^*$.

In an HRS, terms are in $\b$-normal $\eta$-long form and, after a
rewrite step, terms are $\b$-normalized and $\eta$-expanded if
necessary \cite{mayr98tcs}. Hence, in an HRS, the reduction relation
is ${\ar\ab^!\a_{\o\eta}^!}$, where $\a_{\o\eta}$ is the relation of
{\em $\eta$-expansion}\footnote{That is, the relation $\la_\eta$
  restricted to terms not of the form $\l xt$ and to contexts not of
  the form $C[[]u]$ \cite{dicosmo96tcs}.} and $R^!$ denotes
normalization wrt $R$. Hence, our results can directly apply to HRSs
if the set of terms in $\eta$-long form is stable by rewriting for, in
this case, no $\eta$-expansion is necessary\footnote{The set of terms
  in $\eta$-long form is stable by $\b$-reduction
  \cite{huet76hdr}.}. This is in particular the case if the right-hand
side of every rule is in $\eta$-long form \cite{huet76hdr}. Otherwise,
one needs to extend our results by proving the termination of
$\a\cup\a_{\o\eta}$ instead (see \cite{dicosmo96tcs} for the case
where ${\a}={{\ab}\cup{\ar}}$ and $\cR$ is a set of algebraic rewrite
rules).

\section{Conclusion}

We have provided a new, more general, presentation of the notion of
computability closure \cite{blanqui02tcs} and how it can be extended
to deal with different kinds of rewrite relations (rewriting modulo
some equational theory and rewriting with matching modulo $\be$) and
applied to other frameworks for higher-order rewriting (Section
\ref{sec-hrs}). In particular, for dealing with recursive function
definitions, we introduced a new more general rule (Figure
\ref{fig-cc-rec}) based on the notion of $\cF$-quasi-ordering
compatible with application (Definition \ref{def-f-quasi-ord}).

Parts of this work have been formalized in the proof assistant Coq
\cite{coq}: {\em pure} $\l$-terms\footnote{Using named variables and
  explicit $\al$-equivalence \cite{curry58book} which is closer to
  informal practice than de Bruijn indices \cite{debruijn72im}.},
computability predicates on (untyped) $\l$-terms, simply-typed
$\l$-terms using typing environments, the interpretation of types
using accessible arguments as in Section \ref{sec-match-def}, and the
smallest computability closure closed by the operations I, II, IV and
V \cite{blanqui13coq-cc}.\footnote{The definitions and theorems
  without their proofs are available on {\tt
    http://color.inria.fr/doc/main.html}. In particular, $\l$-calculus
  is formalized in the files {\tt LTerm.v}, {\tt LSubs.v}, {\tt
    LAlpha.v}, {\tt LBeta.v} and {\tt LSimple.v}; computability is
  formalized in {\tt LComp.v}, {\tt LCompRewrite.v} and {\tt
    LCompSimple.v}; the interpretation of type constants as in Section
  \ref{sec-match-def} is formalized in {\tt LCompInt.v}; the notion of
  $\cF$-quasi-ordering is formalized in {\tt LCall.v}; and the notion
  of computability closure is formalized in {\tt LCompClos.v}. As an
  example, G\"odel system T is proved terminating in {\tt LSystemT.v}
  by using the lexicographic status $\cF$-quasi-ordering
  $(\tge_\ss)_\lex$.} Therefore, the complete formalization of the
results presented in this paper is not out of reach. In particular,
the operations III and VI, and the computability closure for rewriting
modulo some equational theory. On the other hand, the computability
closure for rewriting with matching modulo $\be$ seems more difficult.

For the sake of simplicity, we have presented this work in Church
simply-typed $\l$-calculus \cite{church40jsl} but, at the price of
heavier notations, a special care for type variables, and assuming
that $\af=\sup\{|\vl|\mid\sf\vl\a r\in\cR\}$ is
finite,\footnote{Because, in this case, $\af$ may be infinite if $\cR$
  is infinite, which may be the case if one considers the rewrite
  relation generated by a conditional rewrite system, or applies some
  semantic labeling to a finite rewrite system \cite{zantema95fi}.}
these results can be extended to polymorphic and dependent types, and
type-level rewriting (\eg strong elimination), following the
techniques developed in \cite{blanqui05mscs}.

But the notion of computability closure has other interesting
properties or applications:

\begin{itemize}
\item As shown in \cite{blanqui06wst-hodp}, it has some important
  relationship with the notion of dependency pair \cite{arts00tcs} and
  can indeed be used to improve the static approach to higher-order
  dependency pairs \cite{kusakari09ieice}.

\item The notion of computability closure and Jouannaud and Rubio's
  higher-order recursive path ordering (HORPO)
  \cite{jouannaud99lics,jouannaud07jacm} share many similarities. The
  notion of computability closure is even used in HORPO for
  strengthening it. HORPO is potentially more powerful than CC
  because, when comparing the left-hand side of a rule $\sf\vl$ with
  its corresponding right-hand side $r$, in CC, the subterms of $r$
  must be compared with $\sf\vl$ itself while, in HORPO, the subterms
  of $r$ may be compared with subterms of $\sf\vl$. However, in
  \cite{blanqui06tr}, I showed that HORPO is {\em included} in the
  monotone closure of the least {\em fixpoint} of the monotone
  function $\cR\mapsto\{(\sf\vl,r)\mid
  r\in\CC_\sf(\vl),\tau(\sf\vl)=\tau(r),\FV(r)\sle\FV(l)\}$ (where
  $\CC$ is the smallest computability closure defined by the rules I
  to IV), and that Dershowitz' first-order recursive path ordering
  \cite{dershowitz82tcs} is {\em equal} to this fixpoint (when $\CC$
  is restricted to first-order terms). This and the fact that HORPO
  could not handle the examples of Section \ref{sec-ho-subterm}
  motivated a series of papers culminating in the definition of the
  computability path ordering (CPO) subsuming both HORPO and CC, but
  currently limited to matching modulo $\al$-equivalence
  \cite{blanqui08csl,blanqui15lmcs}.

\item In Section \ref{sec-ho-subterm}, we have seen that, on
  non-strictly positive inductive types, the computability closure can
  handle recursors (by using an elimination-based interpretation of
  types), but cannot handle arbitrary function definitions (\eg the
  function $\fex$). This can however be achieved by extending the type
  system with size annotations (interpreted as ranks) and using an
  $\cF$-quasi-ordering comparing size annotations. This line of
  research was initiated independently by Gim\'enez
  \cite{gimenez96phd} and Hughes, Pareto and Sabry
  \cite{hughes96popl}, and further developed by Xi \cite{xi02hosc},
  Abel \cite{abel04ita}, Barthe {\em et al} \cite{barthe04mscs} and
  myself \cite{blanqui04rta}. By considering explicit quantifications
  and constraints on size annotations, one can even handle conditional
  rewrite rules \cite{blanqui06lpar-sbt}. Moreover, in
  \cite{blanqui09csl}, Roux and I showed that these developments can
  to some extent be seen as an instance of higher-order semantic
  labeling \cite{zantema95fi,hamana07ppdp}, a technique which consists
  in annotating function symbols with the semantics of theirs
  arguments in some model of the rewrite system.

\item In \cite{jouannaud06rta-horpo}, using a complex notion of
  ``neutralization'' that requires the introduction of new function
  symbols, Jouannaud and Rubio provide a general method for building a
  reduction ordering for rewriting with matching modulo $\be$ on {\em
    $\b$-normal} terms from a reduction ordering for rewriting with
  matching modulo $\al$-equivalence on arbitrary terms, if the latter
  satisfies some conditions. Then, they provide a restriction of HORPO
  satisfying the required conditions. A precise comparison between
  this approach and the one developed in Section \ref{sec-hopm}
  remains to be done. It could perhaps shed some light on this notion
  of neutralization.
\end{itemize}

\noindent{\bf Acknowledgements.} The author thanks very much the
anonymous referees for their very careful reading and many
suggestions, and Ali Assaf and Ronan Saillard for their comments on
Section \ref{sec-hopm}.

\renewcommand\ss\latexss
\bibliographystyle{alpha}
%\bibliography{main}%long,abbrev,long-names,mybib}
\newcommand{\etalchar}[1]{$^{#1}$}

\end{document}